\documentclass[twocolumn,tighten]{aastex62}

\usepackage{graphicx}	
\usepackage{amsmath}	
\usepackage{amssymb,bm}	
\usepackage{hyperref}	
\usepackage{color}	
\usepackage{upgreek}	

\newcommand{\bvec}[1]{\mathbf{#1}}

\newcommand{\BBB}{{\bm B}}

\newcommand{\EQ}{\begin{equation}}
\newcommand{\EN}{\end{equation}}
\newcommand{\EQA}{\begin{eqnarray}}
\newcommand{\ENA}{\end{eqnarray}}

\newcommand{\ve}[1]{\boldsymbol{#1}}

{}

\newcommand{\K}{\,{\rm K}}

\newcommand{\msol}{\textup{M}_\odot}

\newcommand{\vect}[1]{{{\mbox{\boldmath $#1$}}}} 
\definecolor{midblue}{rgb}{0.0,0.4,0.7}
\definecolor{midgreen}{rgb}{0.0,0.8,0.3}
\definecolor{mypurple}{rgb}{0.8,0.2,0.8}
\newcommand{\edits}[1]{\textcolor{black}{#1}}

\received{Jun 2019}
\revised{--}
\accepted{--}
\submitjournal{ApJ}

\shorttitle{MHD supernova explosions}
\shortauthors{Evirgen and Gent}

\begin{document}

\title{MHD supernova explosions -- Large-scale magnetic field effects}

\correspondingauthor{Cetin Can Evirgen}
\email{c.c.evirgen@newcastle.ac.uk}

\author[0000-0002-0786-7307]{Cetin C. Evirgen}
\affil{School of Mathematics, Statistics and Physics,\,
Newcastle University\,
Newcastle upon Tyne,\,NE1 7RU,\,UK}

\author{Frederick A. Gent}
\affiliation{ReSoLVE Centre of Excellence,\,
Department of Computer Science, Aalto University,\,
Aalto, PO Box 15400,\,FI-00076,\,Finland}

\begin{abstract}

We examine the effect of uniform ambient magnetic fields on the evolution of
\edits{supernova-driven blast waves into a homogeneous ambient ISM in thermal equilibrium.} 
Using the \edits{\sc{Pencil Code}} we
\edits{simulate} high resolution nonideal magnetohydrodynamic \edits{simulations in 3D}.
We find that supernova blast waves \emph{are} sensitive to plane-parallel magnetic
fields of strength in excess of 1\,$\upmu$G for ambient gas number density
1\,cm$^{-3}$. 
Perpendicular to the field, the \edits{inward} magnetic pressure gradient \edits{induces 
retrograde mass accretion in the wake of the primary shock front.}
 Subsequently, we find that
the primary shockwave expands faster perpendicular to the field, \edits{but with reduced
momentum,} while the remnant \edits{core} is subject to magnetic confinement.
This leads to a decrease in fractional volume of hot gas
but also an increase in the density {and temperature} of hot gas in the magnetically confined remnant.
The magnetic pressure gradient behind the shock front generates enhanced regions
favourable to UV-heating and thus reduces net radiative losses.
Although the presence of a strong uniform magnetic field can reduce
momentum early on, \edits{and hence residual kinetic energy,}
it increases the efficiency of \edits{residual total
energy injection by the SN} into the ISM \edits{by up to 40\% within 1\,Myr}.
\end{abstract}


\section{Introduction} \label{sec:intro}

Supernova explosions (SNe) play a significant role in galactic dynamics{.
They provide pressure support and inject energy, driving turbulence and
large-scale outflows} \citep{MO77,CC85,MK04,MQT05,SILCC2}. 
Isolated SNe can significantly impact their local ambient environment, whether 
evacuating or fragmenting a dense molecular cloud from within, or
expansion into, and advection through, less dense hot gas regions.
Clusters of SN can form larger hot gas structures (superbubbles or chimneys),
spanning typically hundreds of parsec, and feeding the galactic fountain.

Both individual SNe and SNe clusters, such as superbubbles arising from OB
associations, have been studied analytically
\citep{Taylor50,Sedov59,THI81,Cioffi88} and numerically
\citep{Chevalier74,FMZ91,SC92,JN96,Tomisaka98,CK01,HT06,ZP08,KO15,YMSN17}.
SNe evolution is spherically symmetric in a non-magnetized uniform ambient
ISM \citep{Spitzer78,Cioffi88}.
From a 1D model \citet{Chevalier74} concluded that if the postshock magnetic
pressure does not exceed the ram pressure, then the magnetic field does not
affect the SN remnant dynamics, other than the thickness of its shell.
Further, from 3D ideal {magnetohydrodynamic (MHD)} simulations \citet{KO15} conclude that the
final momentum injected by an SNe is insensitve to the the magnetic field, 
although if the ambient plasma-$\beta$ is greater than 1 the late stage
remnant temperature is affected.
\citet{HT06} show that strong plane-parallel magnetic fields can inhibit the expansion
of individual SN remnants perpendicular to the magnetic field, thus altering the aspect ratio
of the remnant and leading to a deviation from the spherical expansion typical of
{hydrodynamic (HD)} remnants.
\citet{CK01} show the expansion profile for an individual SN remnant in a magnetized medium,
with the blast wave expanding {faster} perpendicular to a uniform, plane-parallel magnetic field.
While the results of \citet{HT06} and \citet{CK01} seem to disagree, we aim to show that
they {instead} describe {complimentary} aspects of MHD remnant expansion.

\edits{In this Paper, we focus on understanding the role of a large-scale magnetic field
in simulations of SN remnants in an idealised physical setup. Our initial aim is twofold: to examine the
effects of large-scale magnetic fields in isolation, and to assess the claim that large-scale
magnetic fields do not affect momentum injection in SN remnants. In order to make direct comparison,
we use \edits{similar parameters to numerical models from earlier studies}.}

\edits{
SN explosions are a major component of larger-scale numerical models of
galaxies, evolving turbulence in which SN
remnants are more realistically embedded}.
\edits{However, these models are typically too coarse to resolve SN remnants
self-consistently}.
Remnant-scale modelling \edits{provides necessary constraints on the critical}
aspect of energy injection \edits{to apply in the} larger simulations.
However, a less idealized model \edits{will utlimately be required} for
comparison with \edits{SN remnant} observations.
We \edits{shall} develop such models in an incremental and cumulative manner \edits{in
subsequent studies}.

\edits{In Section\,\ref{sec:model} we present details of the numerical model.
We then discuss magnetic effects on the aspect ratio of SN remamnts in 
Section\,\ref{sec:aspect}.
This is followed examination of the effect of large-scale magnetic fields on
momentum injection by SN remnants in Section\,\ref{sec:momentum}.
In Section\,\ref{sec:therm} we discuss magnetic effects on the thermodynamical
properties of SN remnants.
Residual energy injection by the remnants into the ISM is measured   
in Section\,\ref{sec:energy} and we consider the critical magnetic 
field strength for MHD effects to become important in Section\,\ref{sec:beta}.
We present a summary and discussion of our results in Section\,\ref{sec:summary}.}

\edits{
As a sanity check we compute some shock-tube experiments and and other related
tests, which verify our qualitative results are not sensitive to numerical recipes
or parameters. 
These are external to the focus of the paper so we include them in an Appendix.
In Appendix\,\ref{app:riemann}, we discuss the effect of physical parameters
on 1D HD and MHD shocks. Appendix\,\ref{app:amb_rho} provides a preliminary set of results
on the effects of large-scale magnetic fields on 1D shocks in ambient gas densities
ranging from $n=10^{-2}$\,cm$^{-3}$ to $n=10^{-2}$\,cm$^{2}$. Appendix\,\ref{app:amb_tt}
examines the effect of changing ambient gas temperature in 3D simulations of SN explosions.}

\section{\edits{Numerical methods}} \label{sec:model}

Here we explore further the MHD effects, especially momentum injection, the
anisotropy of the remnant, and effects on the gas and density distribution
within the remnant.
We use the Pencil Code\footnote{\href{https://github.com/pencil-code}{
https://github.com/pencil-code}}, adapted for highly compressible nonideal
MHD turbulence.

\edits{ 
The MHD equations applied include the compressible form of the continuity 
equation
  \begin{eqnarray}
    \label{eq:mass}
     \frac{D\rho}{Dt} &=&+ \dot{\rho}_{\rm SN}- \rho\nabla\cdot \vect{ u} + \zeta_D\nabla^2\rho 
                         + \vect\nabla\zeta_D\cdot\vect\nabla\rho, 
  \end{eqnarray}
  where $\rho$ is the gas density and $\dot{\rho}_{\rm SN}$ is the mass of the
  SN ejecta, which for the purposes of these experiments is set to zero.
  An artificial shock dependant mass diffusion, $\zeta_D$, is required for
  numerical stability in simulations of the SN-driven turbulent ISM.
  $\zeta_{\rm D}\propto f_{\rm shock}$, where 
  \begin{equation}\label{eq:fshock}
  {f_{\rm shock}=\left<\underset{5}{\rm max}\left[
       \left(-\nabla\cdot\bm u\right)_{+}\right]\right>
       \left({\rm min}\left(\Delta x,\Delta y,\Delta z\right)\right)^2.}
  \end{equation}
  This is described in \citet{GMKSH19}, along with shock dependant
  diffusivities included in the following equations.
  The material derivative is
  \[
    \frac{D}{Dt} = \frac{\partial}{\partial t} + \vect{u}\cdot\vect{\nabla}.
  \]
}

\edits{The momentum equation evolving velocity, $\vect{u}$,
  includes an artifical shock dependent viscosity, $\zeta_{\nu}$ and a 
  momentum conserving correction term for $\zeta_D$ from Equation\,\eqref{eq:mass}.
  The pressure force is expressed in terms of specific entropy, $s$, 
  specific heat capacity of the gas at constant pressure, $c_p$, 
  and sound speed, $c_s$, yielding
  \begin{eqnarray}
    \rho\frac{D \vect{u}}{Dt}
    & = & -c_{s}^{2}\rho \vect{\nabla} \left(\frac{s}{c_{p}} +  \textrm{ln} \rho\right) 
    +\vect\nabla \cdot \left( 2\rho \nu\vect{\nabla} \bm{\mathsf{S}} \right)
    + \vect{j}\times\vect{B}\nonumber
    \\
    \label{eq:mom}
    &&+\rho\vect{\nabla}\left(\zeta_{\nu}
     \vect{\nabla}\cdot \vect{u} \right)
    -\vect u \left(\zeta_D\nabla^2\rho + \bm\nabla\zeta_D\cdot\bm\nabla\rho\right).
  \end{eqnarray}
  This also includes the Lorentz force due to the magnetic field, $\vect B$,
  and the current density, $\vect j=\mu_0^{-1}\vect{\nabla}\times\vect{B}$.
  The vacuum magnetic permeability is denoted $\mu_0$.
  Viscous stresses are accounted for through the shear viscosity $\nu$
  within the divergence of the rate of strain tensor
  $\bm{\mathsf{S}}$, defined by
  \[
  \label{eq:str}
   2 \mathsf{S}_{ij}= \frac{\partial u_{i}}{\partial x_{j}}+
      \frac{\partial u_{j}}{\partial x_{i}}
    -\frac{2}{3}\delta_{ij}\vect{\nabla} \cdot\vect{u},\quad 
    {\rm with}\quad  \bm{\mathsf{S}}^2\equiv \mathsf{S}_{ij}\mathsf{S}_{ij}.
  \]
}

\edits{  The energy equation is evolved in the form of the specific entropy with
  \begin{eqnarray}
    \label{eq:ent}
    \rho T\frac{D s}{Dt}&  =
    &\dot\sigma_{\rm{SN}}+\rho\Gamma-\rho^2\Lambda+2 \rho \nu \bm{\mathsf{S}}^{2}
    +\eta\mu_0|\vect{j}|^2
    \\\nonumber
    &&  
    +\vect{\nabla} \cdot\left( c_p[\chi+\zeta_{\chi}] \rho \vect\nabla T\right)  
    \\\nonumber
    &&  
     - c_{\rm{v}}\,T \left(
    \zeta_D\nabla^2\rho + \vect\nabla\zeta_D\cdot\vect\nabla\rho\right),
  \end{eqnarray}
  where $T$ denoted the gas temperature and $c_v$ the specific heat capacity
  of the gas at constant volume. 
  Heat sources and sinks include SN explosion thermal energy, 
  $\dot\sigma_{\rm{SN}}=10^{51}$\,erg at time $t=0$, and viscous and 
  Ohmic heating, in which $\eta$ denotes the resistivity. 
  As applied in \citet{Gent13a}, the radiative cooling, $\Lambda$,
  applies \citet{WHMTB95} at lower temperatures and \citet{SW87} for the hot gas.
  Diffuse UV-heating, $\Gamma$, follows \citet{WHMTB95}.
  Thermal conductivity, $\chi$ and shock dependant $\zeta_\chi$ are applied,
  and an energy conserving correction term due to $\zeta_D$ from 
  Equation\,\eqref{eq:mass}.
}
\begin{figure}
  \centering
  \includegraphics[trim=0.15cm 0.2cm 0.3cm 0.7cm, clip=true,width=0.93\linewidth]{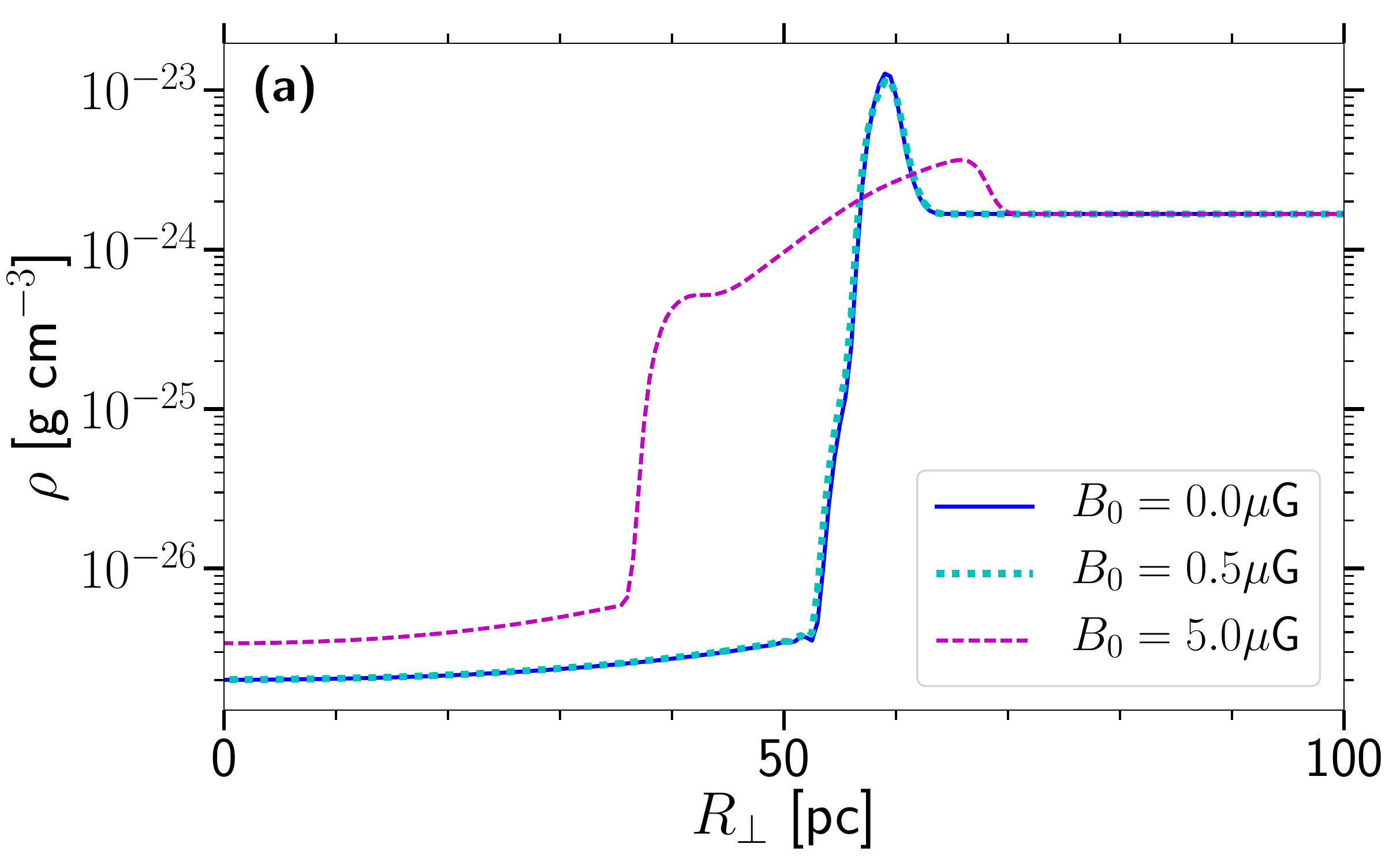}
  \includegraphics[trim=0.15cm 0.2cm 0.3cm 0.7cm, clip=true,width=0.93\linewidth]{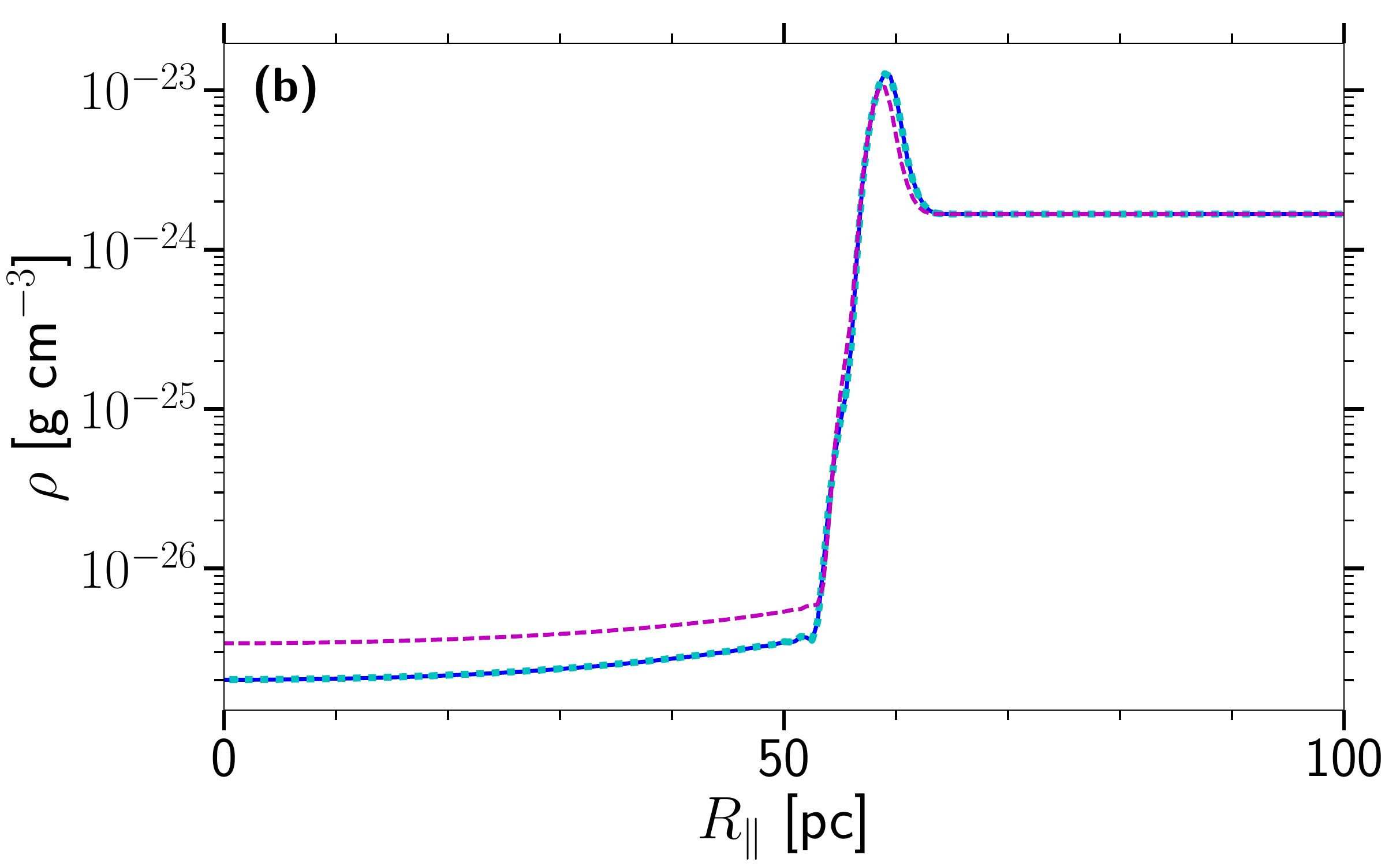}
  \caption{Radial profiles of gas density {\textbf{(a)}}\,perpendicular 
	and {\textbf{(b)}}\,parallel to the magnetic field at $t=1$\,Myr.
  \label{fig:rho_profs}
	}
\end{figure}

\edits{
The induction equation is solved in terms of the vector potential, $\vect{A}$,
which conserves $\vect\nabla\cdot\vect{B}=0$ by design.
In contrast to the previous equations we
 do not include any shock capturing resistivity, as this has the 
unphysical effect of suppressing magnetic field in the SN remnant shell
through excessively rapid diffusion (reconnection).
However, we are interested in investigating the nature of dynamo in
the turbulent ISM, so isotropic resistivity is included to yield
  \begin{eqnarray}
  \label{eq:ind}
    \frac{\partial \vect{A}}{\partial t}   &=
        \vect{u}\times\vect{B}
      +\eta\vect\nabla^2\vect{A}
      +\vect\nabla\cdot\vect{A}\vect\nabla\eta.
  \end{eqnarray}
The system of equations is completed by the ideal gas equation of state
with adiabatic index $\gamma=c_p/c_v=5/3$.
The monatomic gas with hydrogen and helium abundances representative of 
the Solar neighbourhood of the Milky Way has mean molecular weight 0.531
when assumed to be fully ionised.
A comprehensive description of the Pencil Code application to SN driven
ISM turbulence is given in \citet[][Chapter 3]{Gent12}, while the further
enhancement of  
s}hock handling and stability methods for reproducing SN blast waves in HD
are presented in \citet{GMKSH19}.

Here, we consider a plane-parallel uniform magnetic field $\BBB=(0,B_0,0)$,
for $B_0\in[0,5]\,\upmu$G.
With a fiducial gas number density, $n_0$, {for the ambient ISM} of
1\,cm$^{-3}$ and resolution 0.5\,parsec along each side, we use a Cartesian
grid of 528$^2$ by 576 parallel to the field.
For consistent {thermal pressure in} the ambient ISM {throughout the
duration of} the model blastwaves we set {a} thermal equilibrium 
temperature {of} $T\simeq260$\,K.

The numerical model uses constant magnetic resistivity,
$\eta=8\cdot 10^{-4}$\,kpc\,km\,s$^{-1}$ and sound speed dependent viscosity,
\edits{$\nu = \nu_0 c_{\rm s}$}, where \edits{$\nu_0 = \Delta x = 5\cdot 10^{-4}$\,kpc} is the
grid resolution, and $c_{\rm s}$ is the speed of sound.
A test of alternate Prandtl numbers produced qualitatively 
similar solutions, and results were convergent with resolution of 0.5\,pc,
other than a thinner more dense remnant shell.

\section{SN remnant aspect ratio}\label{sec:aspect}

\begin{figure}
  \centering
  \includegraphics[trim=0.2cm 2.1cm 0.1cm 0.4cm, clip=true,width=0.93\linewidth]{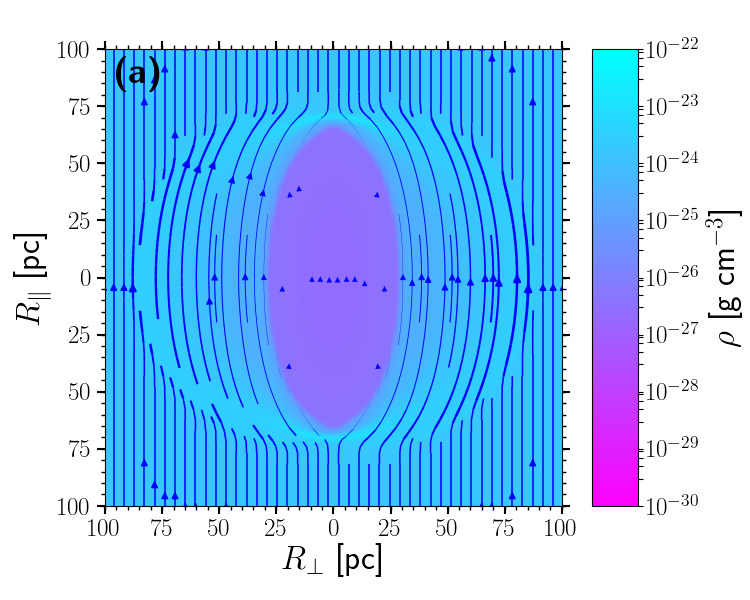}
  \includegraphics[trim=0.2cm 0.6cm 0.1cm 0.4cm, clip=true,width=0.93\linewidth]{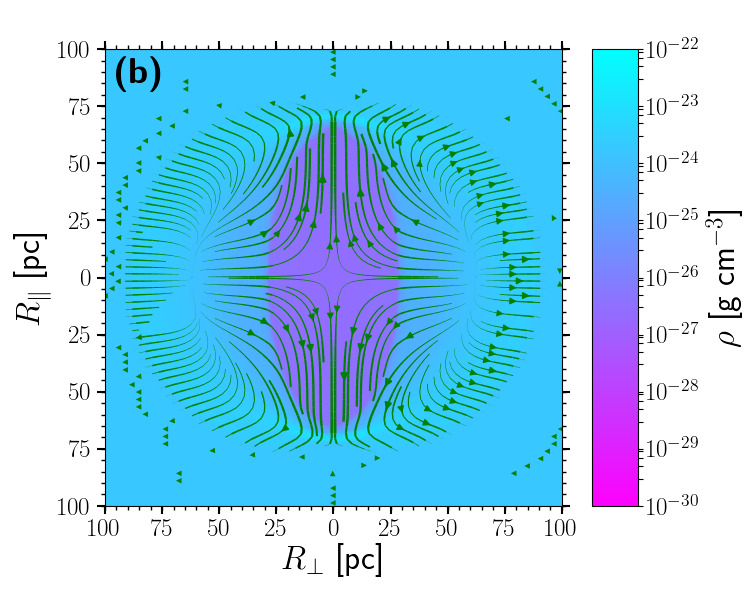}
  \caption{Cross-section of the strongly magnetized remnant at 2\,Myr, with
	{\textbf{(a)}}\,magnetic field lines and {\textbf{(b)}}\,velocity
	stream lines overlaid. Line thickness is proportional to field strength.}
  \label{fig:str_line}
\end{figure}

We show the radial profiles at 1\,Myr of gas density in
Figure\,\ref{fig:rho_profs}, perpendicular to the magnetic field, 
Panel\,{\bf{(a)}}, and parallel, {\bf{(b)}}, for models with $B_0=0,0.5$ and
$5\,\upmu$G.
With a weak magnetic field $B_0\leq0.5\,\upmu$G, the evolution of the remnant
is very similar to the HD model.
With stronger $B_0$ the remnant core is less diffuse, and the slightly less
dense remnant shell parallel to the field is coincident with the HD model shell.
However, perpendicular to the field the mass profile of the remnant is 
substantially altered.
Mass is more confined within the remnant, but Figure\,\ref{fig:rho_profs} also
shows that the SN shockwave propagates faster perpendicular to the magnetic
field {while carrying} less mass at the shock front.

\begin{figure}
  \centering
  \includegraphics[trim=0.4cm 13.5cm 0.8cm  0.5cm, clip=true,width=0.93\linewidth]{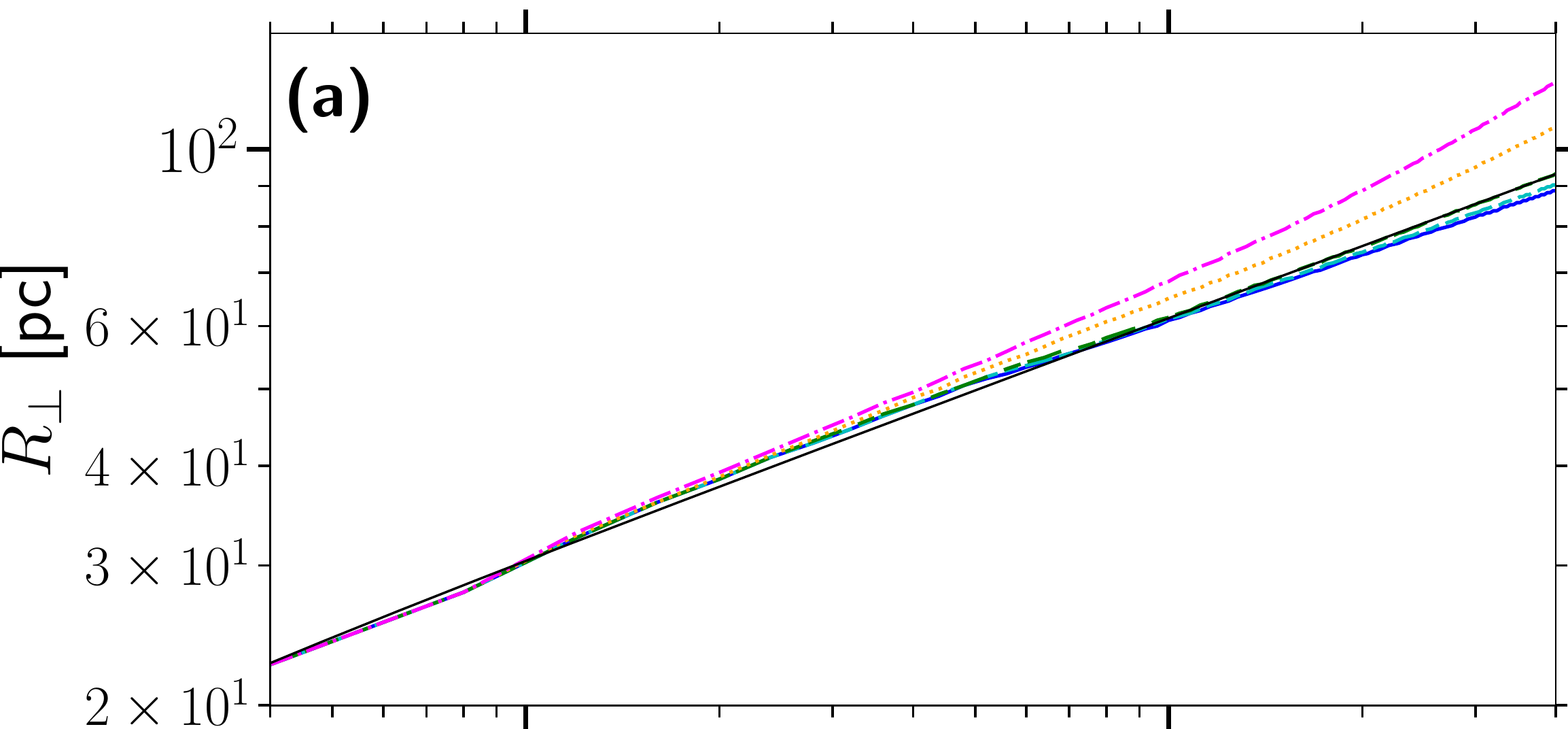}
  \includegraphics[trim=0.4cm  0.3cm 0.8cm 11.5cm, clip=true,width=0.93\linewidth]{fig/sn_radii_perp.pdf}
  \caption{SN remnant radius, $R_\perp$, perpendicular and 
        $R_{\parallel}$, parallel to $B_0$.
	Numerical solutions are compared to the HD analytical solution
	\citep{Cioffi88}.
  \label{fig:sn_rad}}
\end{figure}

Figure\,\ref{fig:str_line}{\textbf{(a)}} shows the shockwave {at 2\,Myr}
compressing magnetic field lines {to form a dense magnetic collar}.
Magnetic field is {initially} evacuated from the core of the remnant.
{It is evident that the MHD remnant core geometry resembles a prolate
spheroid with the polar radius parallel to the magnetic field.
The ratio of polar to equatorial radius increases with field strength.
In contrast, the remnant shock front forms the surface of an oblate
spheroid with equatorial radius perpendicular to the magnetic field.
This ratio of polar to equatorial radius decreases with field strength.}

Figure\,\ref{fig:str_line}{\textbf{(b)}} shows the structure of flows driven
by the remnant in a strongly magnetized ambient ISM.
The shockwave drives outward flows as expected.
We find inward flows of comparable speed to the outward flows.
{There are retrograde shocks towards the core in all the models, but for the 
HD and weakly magnetic models these carry negligible mass.}
{For strong $B_0$ the magnetic forces drive inward gas flow from the shell, perpendicular to
the magnetic field.}
As the flow approaches the thermally dominated core, the outward thermal
pressure gradient channels the flows parallel to the magnetic field,
creating a quadrupolar velocity field between the shockwave and core.
{In the HD and weakly magnetized remnants, the internal gas flows
 of the remnant remain radially outward.}

By 2\,Myr the magnetic field along with some of the gas, has filled almost
half the remnant void behind the shock front.
This forms an inter-shock region between the core of the remnant and its shell. 
At even later times the field strength throughout the inter-shock region is
close to its original strength.
The gas in the core remains hot and diffuse, with magnetic field strength
considerably weaker than outside the core. 

{Figure\,\ref{fig:sn_rad} shows the {radial} evolution of the shockwave
for the HD, weakly magnetized and strongly magnetized remnants.
The HD remnant has a single characteristic
length; the shockwave radius, since the shockwave and core are coupled.
This shows good agreement with the
analytical solution of \citet{Cioffi88} for a spherically symmetric model.
 Shock wave radii are {near} 
identical for all models, parallel to the magnetic field.
The weakly magnetized remnant behaviour {in Figure\,\ref{fig:sn_rad}}
closely resembles that of the HD remnant.
However, perpendicular to the strongly magnetized field the shockwaves
diverge from the other profiles {near} $t=100$\,kyr and propagate
faster.}

{Given the ubiquity of magnetic fields in the ISM, it is reasonable to 
anticipate spheroid morphology to be common in SN remnants. 
This may be useful in understanding the 3D structure of observed remnants.
For example, G351.0-5.4, discovered in \citet{Gasp14} is modelled in first
approximation as spherical.
This may be reasonable for observations of the shell, which from our models 
appears only slightly oblate, but misleading for the remnant core.
{Alternatively, comparisons of gamma-ray and radio data around remnant W44
\citep{Cardillo14} indicate a smaller spheroid in gamma-ray emission with 
polar radius offset from the surrounding spheroid in radio emission, and the
presence of magnetic field in the shock $B\ge10^2\,\upmu$G.
Complex interaction with molecular clouds and the turbulent ISM affect the
morphology of the remnant, but the inter-shock region may also be part of the
explanation for the misalignment in the observational profiles.}

Modelling the local bubble of the Milky Way with a prolate sphere \citet{ABFM18}
find this a reasonable fit even for the magnetic shell.
In contrast to our findings, they determine the remnant field to be vertical,
out of alignment to the neighbourhood galactic field, and highly
anisotropic between North and South.
Perhaps, however, an oblate spheroid shell with polar axis parallel to the 
galactic plane would be an alternative model consistent with  the 
galactic magnetic field.

\edits{Any speculation about how our results relate to observational features,
are of course at this stage highly tentative. 
The large scale alignment of magnetic fields in the ISM will surely not be 
uniform on scales of a few parsecs.
The structure of the turbulent magnetic field, the stratification and
inhomogeneity of the gas density and temperature, and the turbulent motion of
the ambient ISM will be considered in future work, let alone the effects of
chemistry and ionization on the observational signatures.
Anisotropy of SN remants may of course also arise independent of magnetic 
effects.}

\section{Momentum injection} \label{sec:momentum}

In a typical {HD} supernova explosion, the propagation of the shock, and
subsequent momentum injection into the surrounding medium,
is dominated by \edits{the} thermal pressure gradients\edits{, which evacuate mass
from the core into} the blastwave.
In a very similar physical set up to ours,  \citet{KO15}  find 
the SN terminal momentum injection \edits{to be} about $4\times 10^5\,\msol$\,km\,s$^{-1}$,
defined by
\begin{equation}\label{eq:momtot}
Mu_n=\int_{\mathrm{V}}\rho\left(\ve{u}\cdot\ve{\hat{n}}\right) \mathrm{d}V,
\end{equation}
where $\ve{\hat{n}}$ is the radial unit vector from the centre of the SN
explosion\edits{.}

\begin{figure}
  \centering
  \includegraphics[trim=0.1cm 0.0cm 1.0cm 0.7cm, clip=true,width=0.93\linewidth]{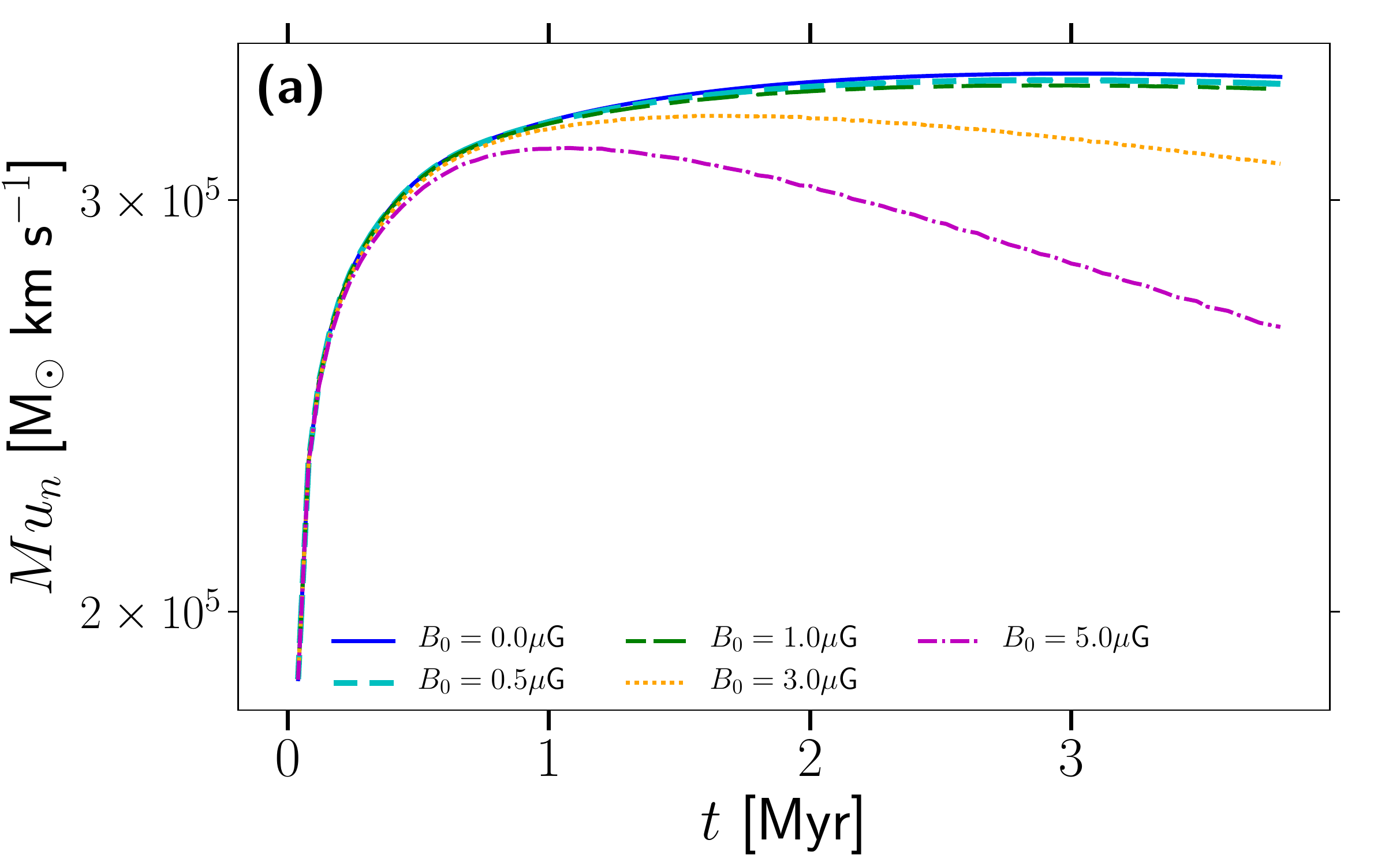}
  \includegraphics[trim=0.1cm 0.0cm 1.0cm 0.7cm, clip=true,width=0.93\linewidth]{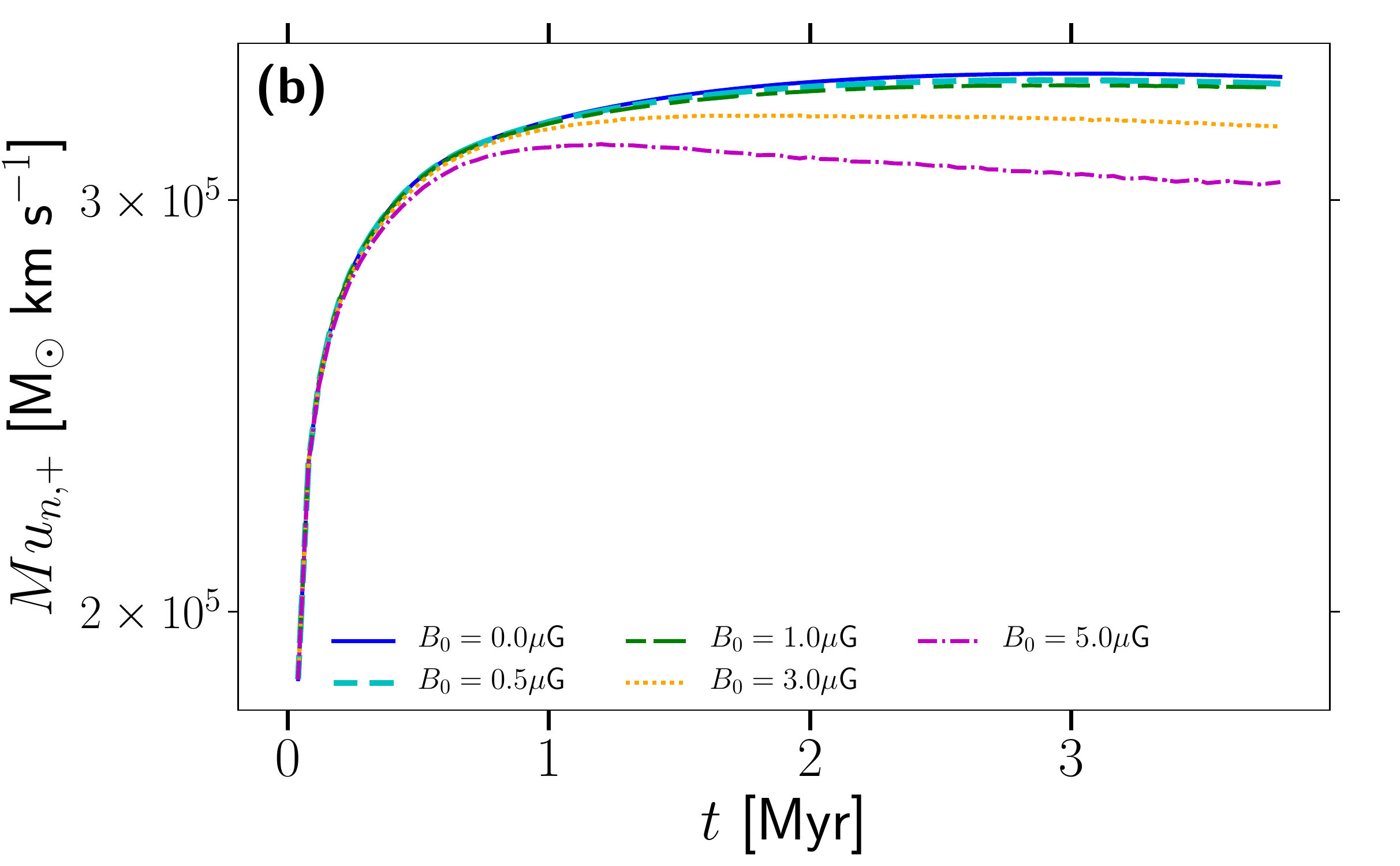}
  \caption{Time evolution of \edits{ {\bf{(a)}}\,total and {\bf{(b)}}\,outward } 
	momentum injection for a number of SN models with magnetic field strengths $B_0\in [0,5]\ \upmu$G.
	Momentum, at a given snapshot, is calculated as
	$\int_{\mathrm{\edits{SNR}}}\rho\left(\ve{u}\cdot\ve{\hat{n}}\right) \mathrm{d}V$,
	where $\ve{\hat{n}}$ is the radial unit vector from the centre of the SN explosion.
	The scatter plots with blue star and magenta circle symbols show outward momentum injection for the HD model
	and strongly magnetized model ($B_0=5\upmu$G), respectively.}
  \label{fig:mom_time}
\end{figure}

\begin{figure*}
  \centering
  \includegraphics[trim=1.65cm 0.5cm 2.4cm 1.0cm, clip=true,width=0.45\linewidth]{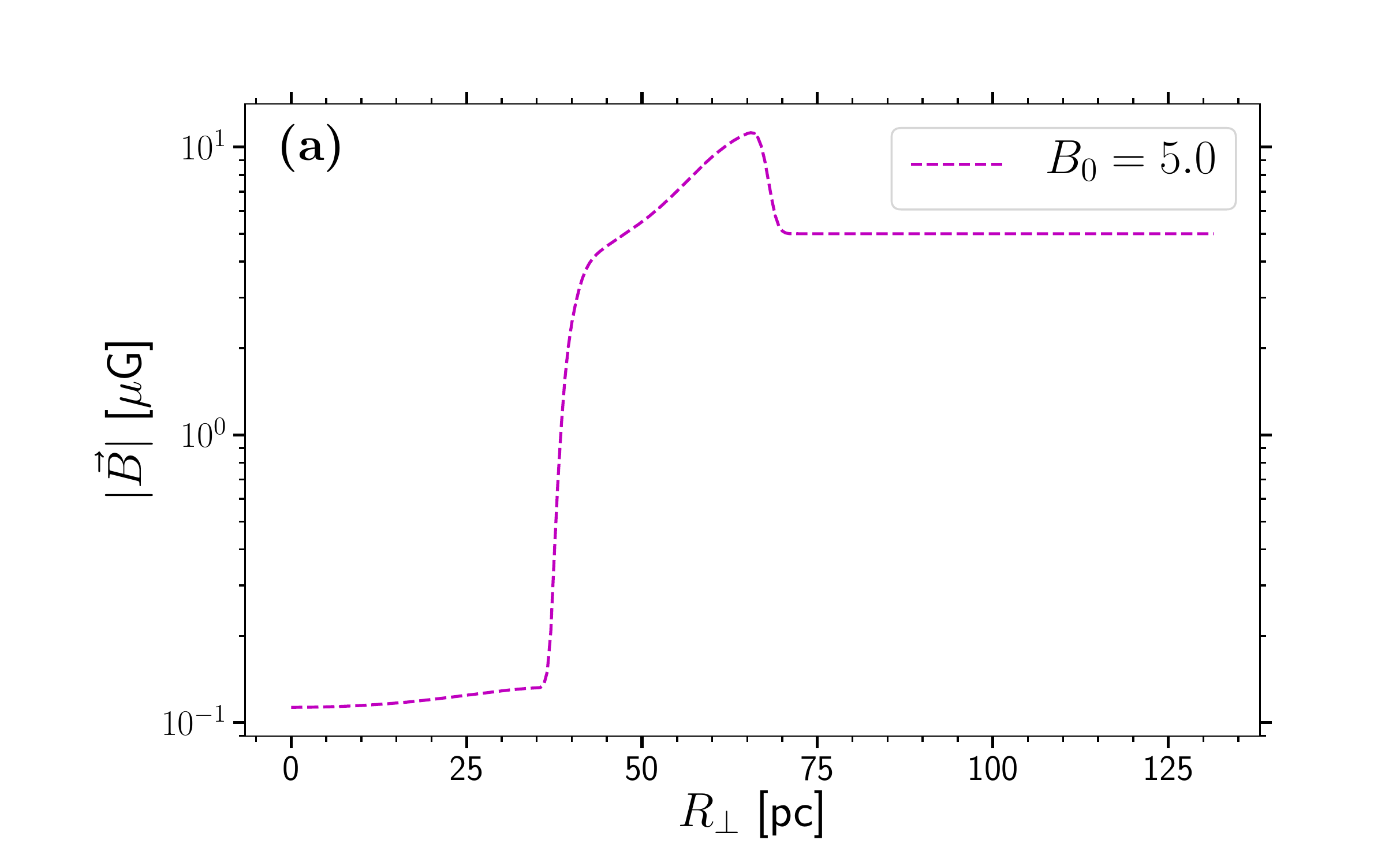}
  \includegraphics[trim=1.65cm 0.5cm 2.4cm 1.0cm, clip=true,width=0.45\linewidth]{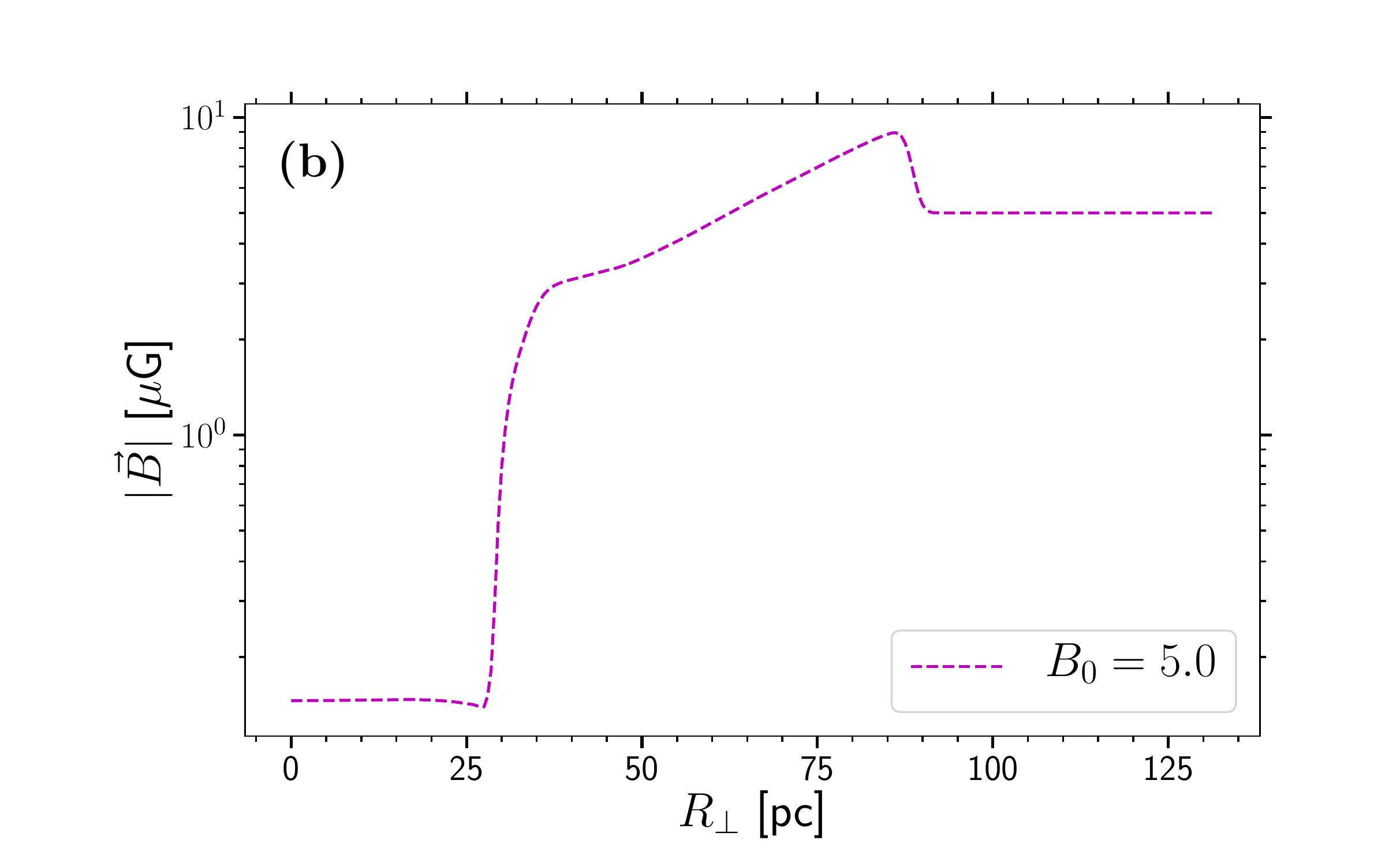}\\
  \includegraphics[trim=1.65cm 0.5cm 2.4cm 1.0cm, clip=true,width=0.45\linewidth]{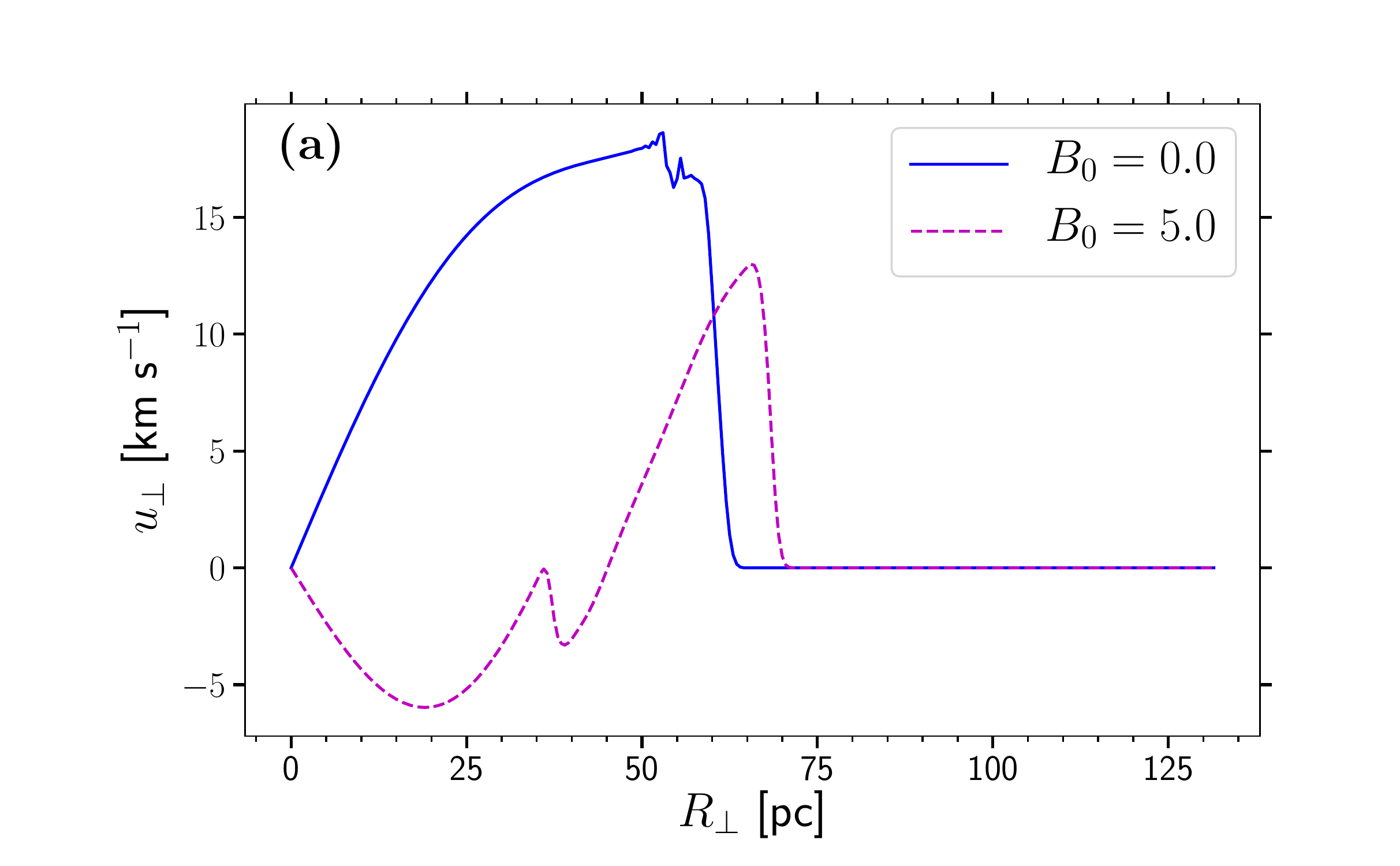}
  \includegraphics[trim=1.65cm 0.5cm 2.4cm 1.0cm, clip=true,width=0.45\linewidth]{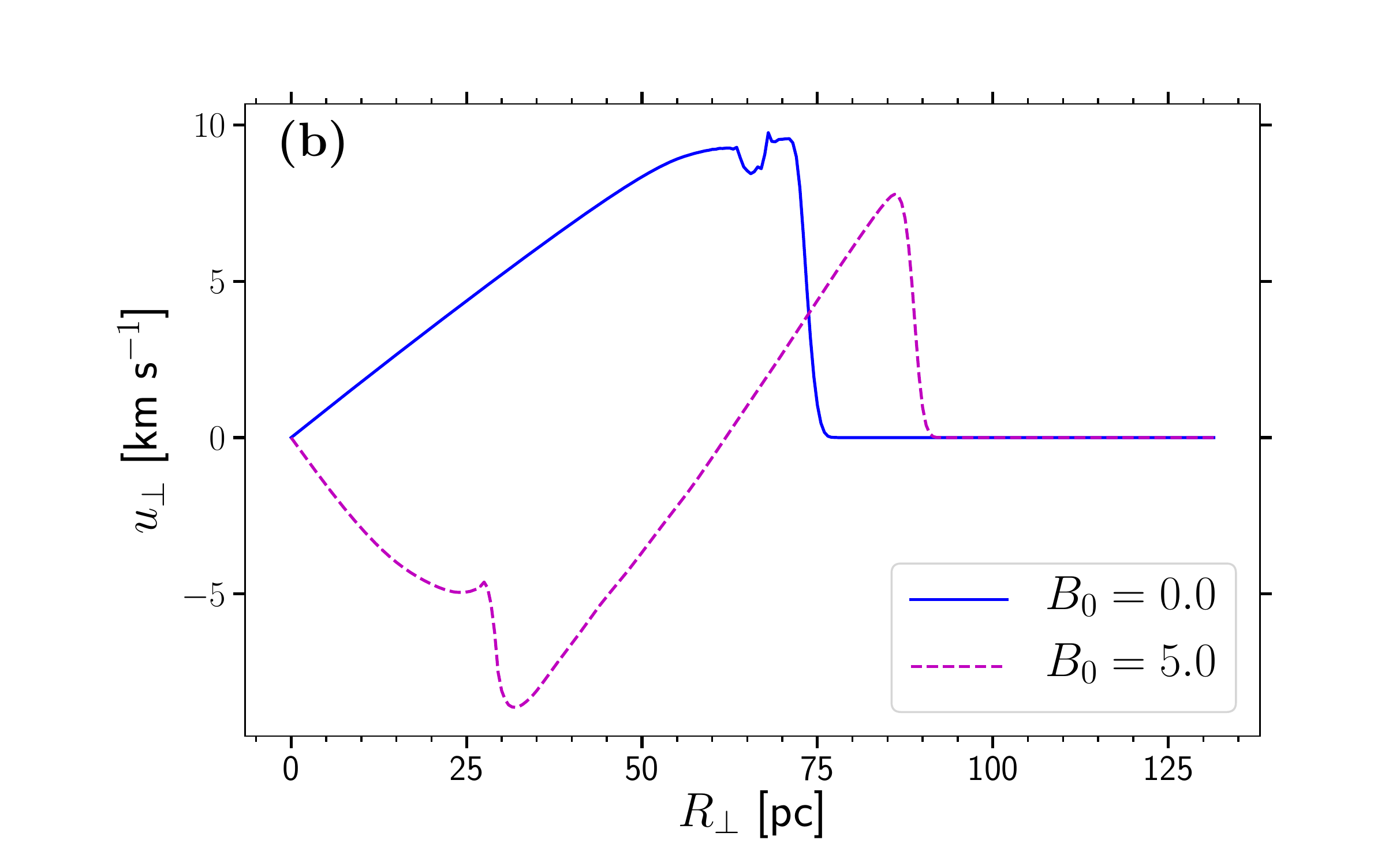}\\
  \includegraphics[trim=1.65cm 0.5cm 2.4cm 1.0cm, clip=true,width=0.45\linewidth]{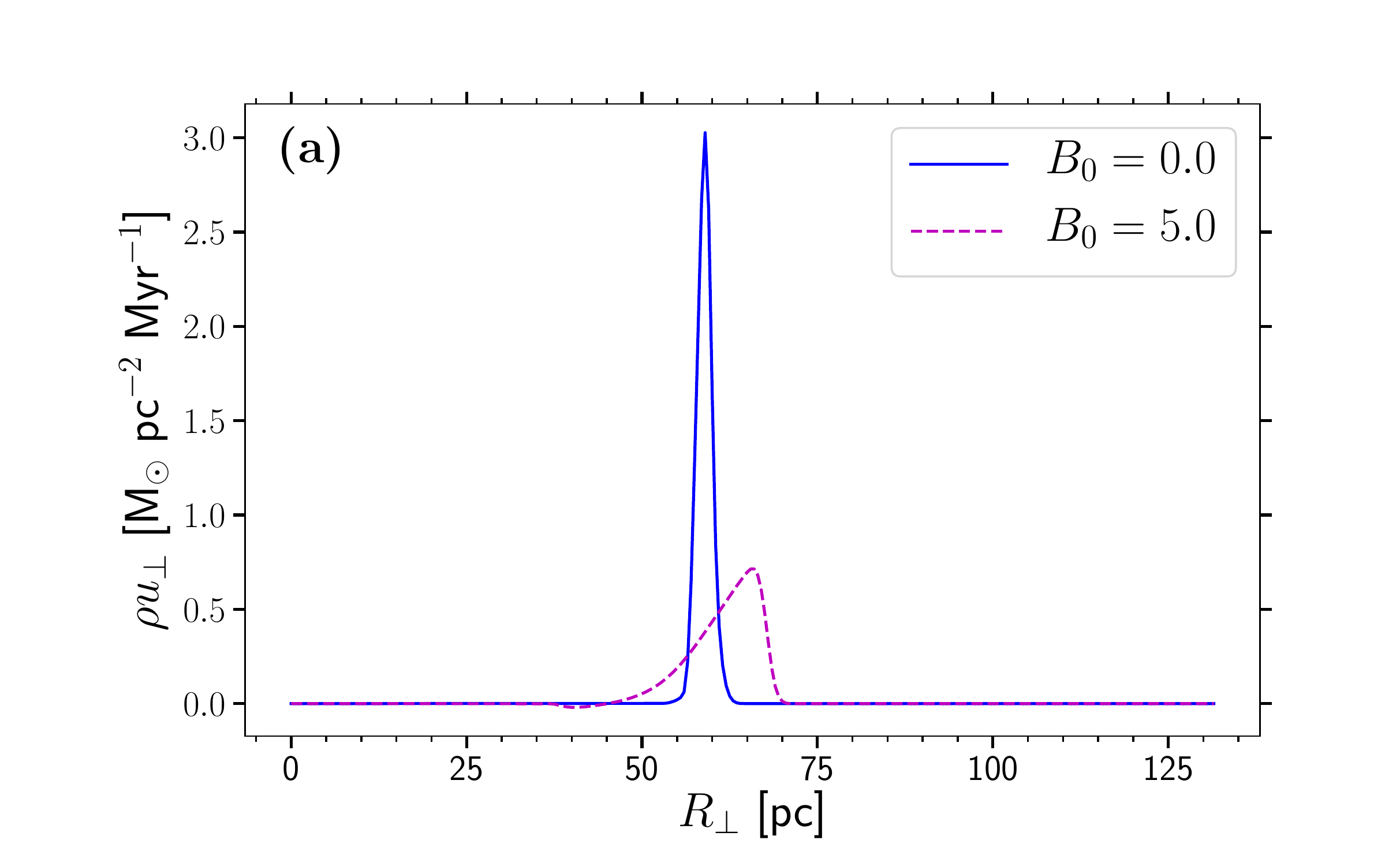}
  \includegraphics[trim=1.65cm 0.5cm 2.4cm 1.0cm, clip=true,width=0.45\linewidth]{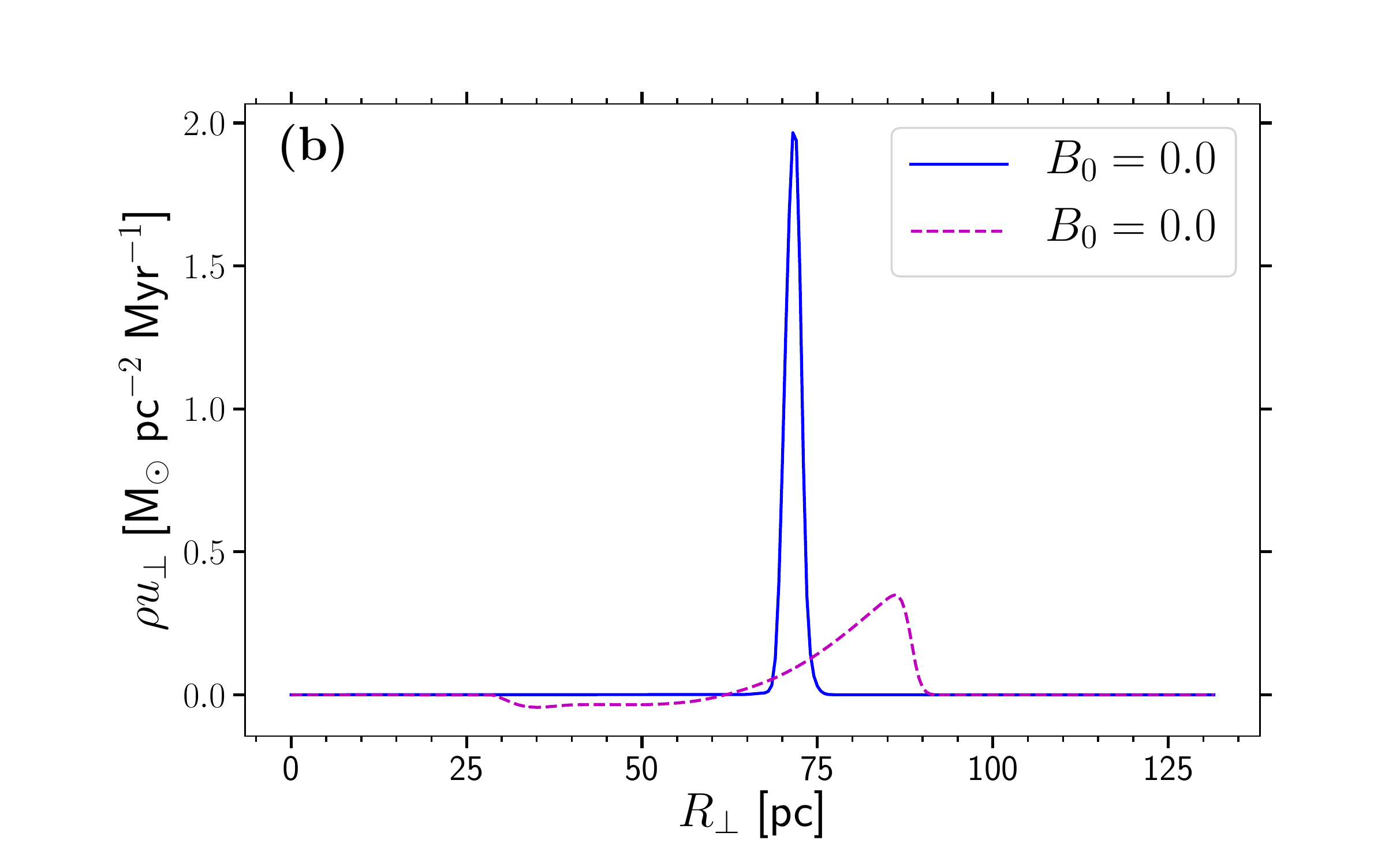}
  \caption{\edits{Radial profiles perpendicular to the magnetic
	\edits{field} at {\bf{(a)}}\,1\,Myr and {\bf{(b)}}\,2\,Myr
        \edits{ of magnetic field strength (top), velocity and momentum 
         (bottom)}.}}
  \label{fig:mom_rad}
\end{figure*}

\edits{We consider this calculation with a modification to account for non-spherical
expansion of the SN remnant. The most general approach would be to adopt
a tri-axial ellipsoid profile. However in this case, the remnants have two distinct
length scales, the SNR radius parallel and perpendicular
to the magnetic field, $R_{\parallel}$ and $R_{\perp}$, respectively. 
This change produces a bi-axial ellipsoidal
remnant, \edits{rather than} the spherical HD remnants. \edits{The} ellipsoidal
surface \edits{is} represented by
\begin{equation}
  g({\mathbf{x}})=\left(\dfrac{x}{R_\perp}\right)^2+\left(\dfrac{y}{R_\parallel}\right)^2+\left(\dfrac{z}{R_\perp}\right)^2,\label{ellipse}
\end{equation}
where $x$ and $z$ are the directions perpendicular to the magnetic field, and $y$ is parallel to
the magnetic field. Here $g(\bvec{x})=1$ represents the remnant shell.\footnote{$g(\bvec{x})=k$, where $0\leq k\leq 1$, represent contours
of the remnant.}
We now define the normal to the shell as 
  \[\hat{\bvec{n}}=\dfrac{\nabla g}{\left|\nabla g\right|},\]
where
  \[\nabla g=\left(\dfrac{2x}{R_\perp^2},\dfrac{2y}{R_\parallel^2},\dfrac{2z}{R_\perp^2}\right).\]
It can be shown that $R_\perp=R_\parallel$ leads to a spherical remnant profile
and $\hat{\bvec{n}}=\hat{\bvec{x}}$.}

\edits{With this definition for $\hat{\bvec{n}}$ applied to Equation\,\eqref{eq:momtot},
we obtain total momentum injection $3.3\times 10^5\,\msol$\,km\,s$^{-1}$
for the HD remnant, similar to \citet{KO15}.
we plot the evolution of total momentum for the HD and a range of MHD remnants in
Figure\,\ref{fig:mom_time}\,{\bf{(a)}}.}
As \edits{the} thermal pressure gradient \edits{reduces} 
in the remnant the pressure driven snowplough
\edits{(PDS)} transitions 
to the momentum conserving snowplough \edits{(MCS)}, as described
by \citet{Cioffi88}.
\edits{It can seen that MCS 
for the HD model initiates a little later than 2\,Myr.
A transient MCS occurs much earlier ($t\simeq1$\,Myr) for the strongly
magnetized model, subsequentially losing momentum thereafter.}
\edits{As it might be argued that only the outward momentum drives the 
ISM turbulence we also calculate the outward momentum injection, $Mu_{n}|_+$,
which is plotted in Figure\,\ref{fig:mom_time}\,{\bf{(b)}}}.

\edits{There is a negligible difference between the HD and weak MHD models, 
but for magnetic field strength above 1\,$\upmu$G the momentumn injection
diverges significantly from the HD.
Contrasting panels {\bf{(a)}} and {\bf{(b)}} it is evident that
although HD remnants feature retrograde shocks, these carry negligible
mass back towards the core, while there is significant inward 
momentum for the strong MHD shocks.}

\begin{figure}
  \centering
  \includegraphics[trim=0.225cm 2.1cm 0.5cm 0.0cm, clip=true,width=0.93\linewidth]{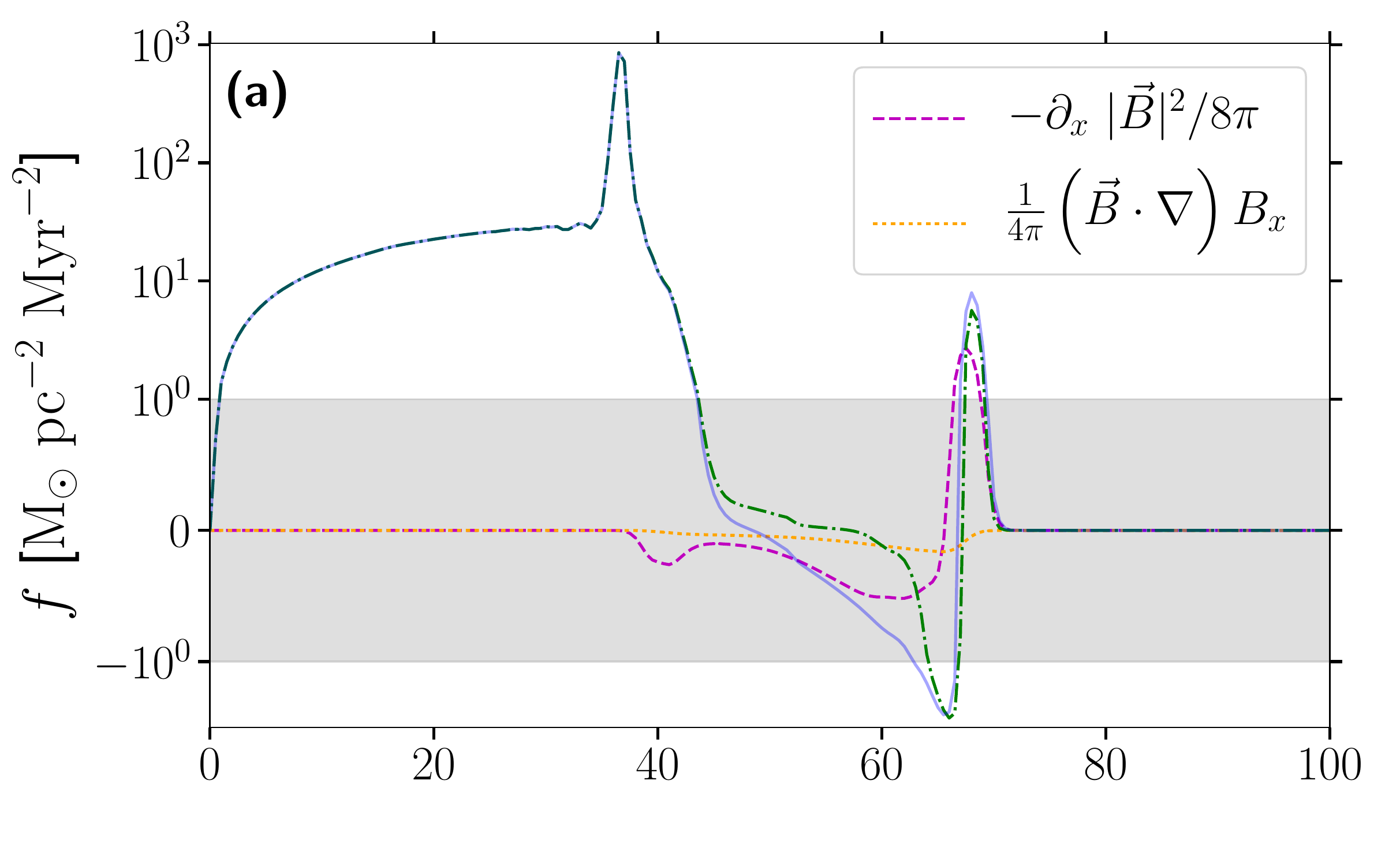}
  \includegraphics[trim=0.225cm 0.1cm 0.5cm 0.6cm, clip=true,width=0.93\linewidth]{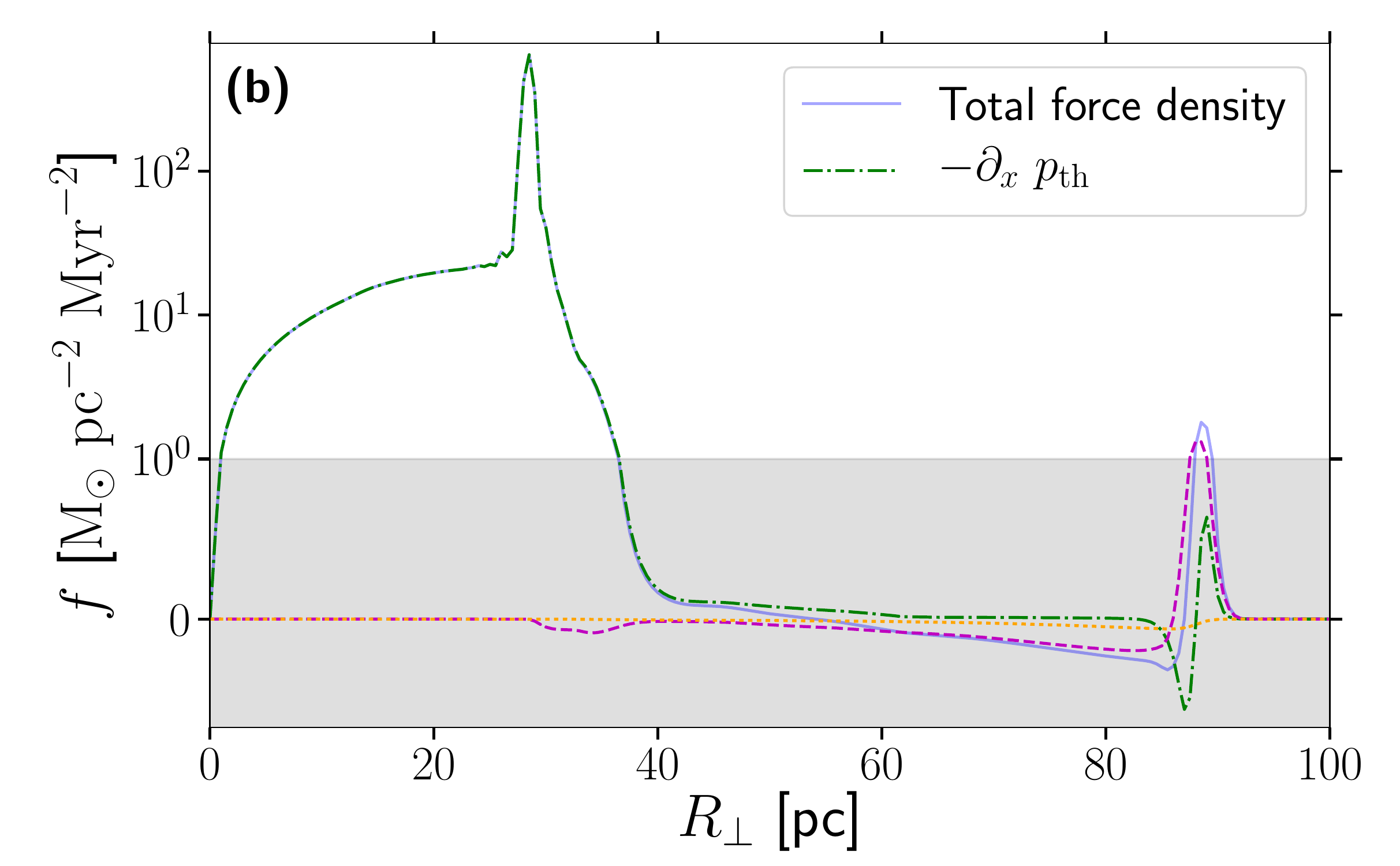}
  \caption{Force density, $f$
	   perpendicular to magnetic field with $B_0=5\upmu$G
           at $t=1$\,Myr {\textbf{(a)}}\,
           and $t=2$\,Myr {\textbf{(b)}}.
	   The vertical scale across the shaded area is linear and
           logarithmic elsewhere.
           }
  \label{fig:force_terms}
\end{figure}  

\citet{KO15} find that momentum injection ceases within two to three
shell-forming
times, before the magnetic energy becomes comparable to the thermal
energy in the shell. Subsequently, they argue that magnetic effects will not
lead to further momentum injection by the remnant.
\edits{However, from both panels in Figure\,\ref{fig:mom_time} we find our
strong MHD solutions ($B_0=3.0,\,5.0\upmu$G) diverge from the HD solution
within 200\,kyr.
These show non-monotonic evolution, reaching a local maximum 
(\edits{$3.1\times\,10^5\,\msol$\,km\,s$^{-1}$}) after 700\,kyr followed by decay.}
\edits{However, outward momentum injection
becomes steady after 2\,Myr and levels off at $3\times 10^5\,\msol$\,km\,s$^{-1}$.
This suggests that momentum injection is reduced by up to $10\%$ in the 
presence of a large-scale plane-parallel uniform magnetic field.
\citet{EGSFB19} show that a $B_{\rm{rms}}\sim 3\upmu$G dynamo-evolved magnetic field,
featuring a strong locally plane-parallel large-scale coherent structure,
leads to a decrease in vertical velocity profile. An important contributing
factor could be suppression of momentum injection \edits{at the SN forcing scale},
as seen here.}

\edits{To illustrate what is changing due to the strong MHD we show radial profiles
perpendicular to the magnetic field at 1\,Myr and 2\,Myr in panels (a) and (b),
respectively in Figure\,\ref{fig:mom_rad} of the magnetic field strength (top),
velocity (centre) and momentum (bottom).}

Momentum in the shell of the \edits{HD} remnant is a factor of 4--6 higher \edits{and
centered narrowly about the shock front}.
The MHD shock \edits{in all profiles} is \edits{broader, projectinq both in advance of the HD shock 
position and including an extended weak inward momentum behind the 
shock front.}
\edits{Even slowing 
below 10\,km s$^{-1}$ by 2\,Myr the MHD blast wave
still remains supersonic.}
We find that \edits{the blast wave remains supersonic even up to 4\,Myr.
We see from that the magnetic field is swept up into the remnant shell, but
that some of this relaxes back towards the core over time.}

In order to explore {what are the magnetic effects that increase the 
outward \edits{velocity, but inhibit the outward} momentum of the MHD remnant in
the latter stage}, we consider the momentum equation perpendicular to $B_0$,
\begin{equation}
\rho \edits{\frac{D u_\perp}{Dt}} =
-\edits{\frac{\partial}{\partial R_\perp}}\left(p+\dfrac{|\vect{B}|^2}{2\upmu_0}\right)+\dfrac{1}{\mu_0}\left(\vect{B}\cdot\vect{\nabla}\right)B_\perp + \mathcal{D},
\end{equation}
where $p$ denotes thermal pressure and 
$\mathcal{D}$ refers to diffusive terms in the equation. For our analysis,
we are interested in the interaction between the pressure gradient terms and the
magnetic tension term on the right hand side of the equation.

\edits{For the strong MHD model} Figure\,\ref{fig:force_terms} shows \edits{the
forces applying perpendicular to $B_0$ at $t=1$\,Myr, 
Panel\,{\bf{(a)}} and $t=2$\,Myr {\bf{(b)}}.
Interestingly, the} very large thermal pressure gradient \edits{extending
through the inner core to about 40\,pc at 1\,Myr is confined to only
30\,pc by 2\,Myr.
A negative thermal pressure gradient does evolve int he wake of the shock,
and this is true also for the HD remnant.
However this is too remote from the core to account for its confinement
late on.
}
Magnetic effects gradually become dominant at the blastwave, 
primarily the magnetic pressure gradient\edits{s},
\edits{where the outward} magnetic pressure gradient is comparable to {the} thermal
gradient at $t=1$\,Myr, and 2--3 \edits{times} greater at $t=2$\,Myr.
\edits{However, the negative pressure gradient in the wake of the shock 
appears to be the most interesting. 
The negative force applies throughout the \emph{intershock region}, 
between the primary blast wave and the inner thermal pressure front.
The magnetic pressure force peels mass away from the remnant shell
and drives it inwards along with some of the magnetic field.
This broadens the mass shell profile we see from 
Figure\,\ref{fig:rho_profs} and the magnetic field profile from
Figure\,\ref{fig:mom_rad} (top panels) and provides the 
substantive forces confining the remnant core compared to HD.}
{For most of} the $60<R_{\perp}<90$\,pc region {by 2\,Myr the}
magnetic pressure gradient dominates {the} thermal pressure gradient.
Magnetic tension is subdominant perpendicular to the magnetic field.

\begin{figure}
  \centering
  \includegraphics[trim=0.2cm 2.2cm 0.1cm 0.45cm, clip=true,width=0.93\linewidth]{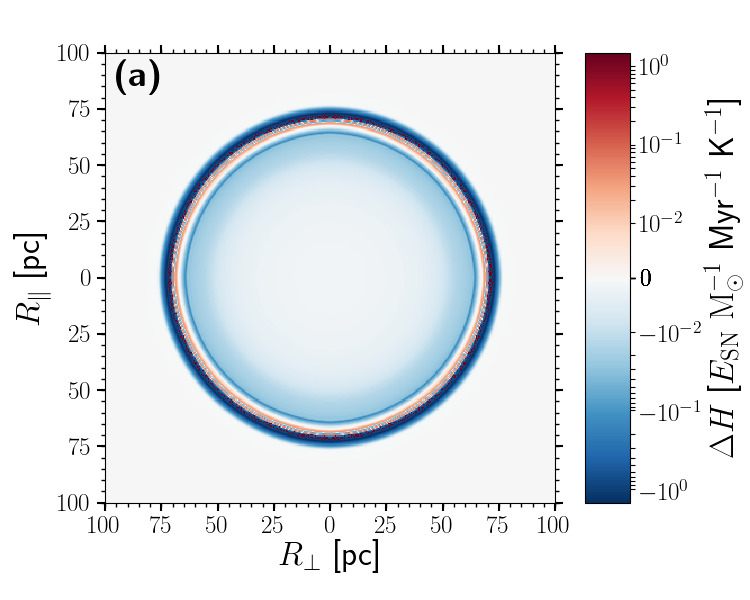}
  \includegraphics[trim=0.2cm 0.6cm 0.1cm 0.45cm, clip=true,width=0.93\linewidth]{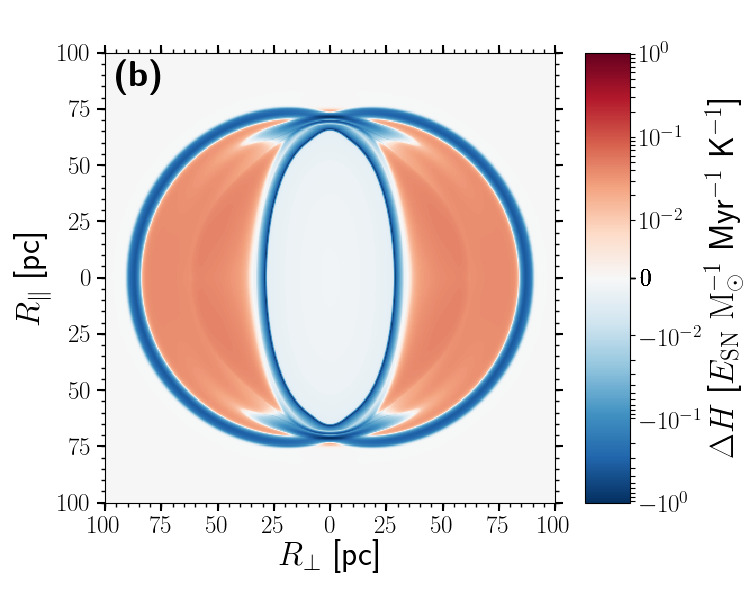}
  \caption{{Cross sections at 2\,Myr for the HD model,
          {\bf{(a)}}, and the model with $B_0=5\upmu$G, {\bf{(b)}},
          of the net heating, $\Delta H=T^{-1}(\Gamma-\rho\Lambda)$.}}
  \label{fig:netheat}
\end{figure}

\section{\edits{Thermodynamics of MHD remnants}}\label{sec:therm}

\edits{The effect of confinement of the remnant core by the inward magnetic
pressure gradient is to reduce adiabatic cooling compared to HD.
The system is non-adiabatic, so how does the redistribution of mass 
in the strong MHD case affect its thermal properties?}
Snapshots of the net heating, $\Delta H=T^{-1}(\Gamma-\rho\Lambda)$, are 
presented in Figure\,\ref{fig:netheat} for the HD model, Panel\,{\bf{(a)}}, and
strong MHD model, {\bf{(b)}}.
In the HD and weak MHD models, cooling dominates everywhere in the remnant,
except for a thin layer just behind the cooling shell.
In the strong MHD model cooling dominates in the confined remnant inner core
and shell, but in the {inter-shock region} perpendicular to the field UV-heating
\edits{exceeds radiative losses}.
In Figure\,\ref{fig:rho_profs}\,{\bf{(a)}} this applies for
$35\lesssim R_\perp\lesssim60$\,pc.

\begin{figure}
  \centering
  \includegraphics[trim=0.5cm 0.2cm 0.8cm 0.4cm, clip=true,width=0.93\linewidth]{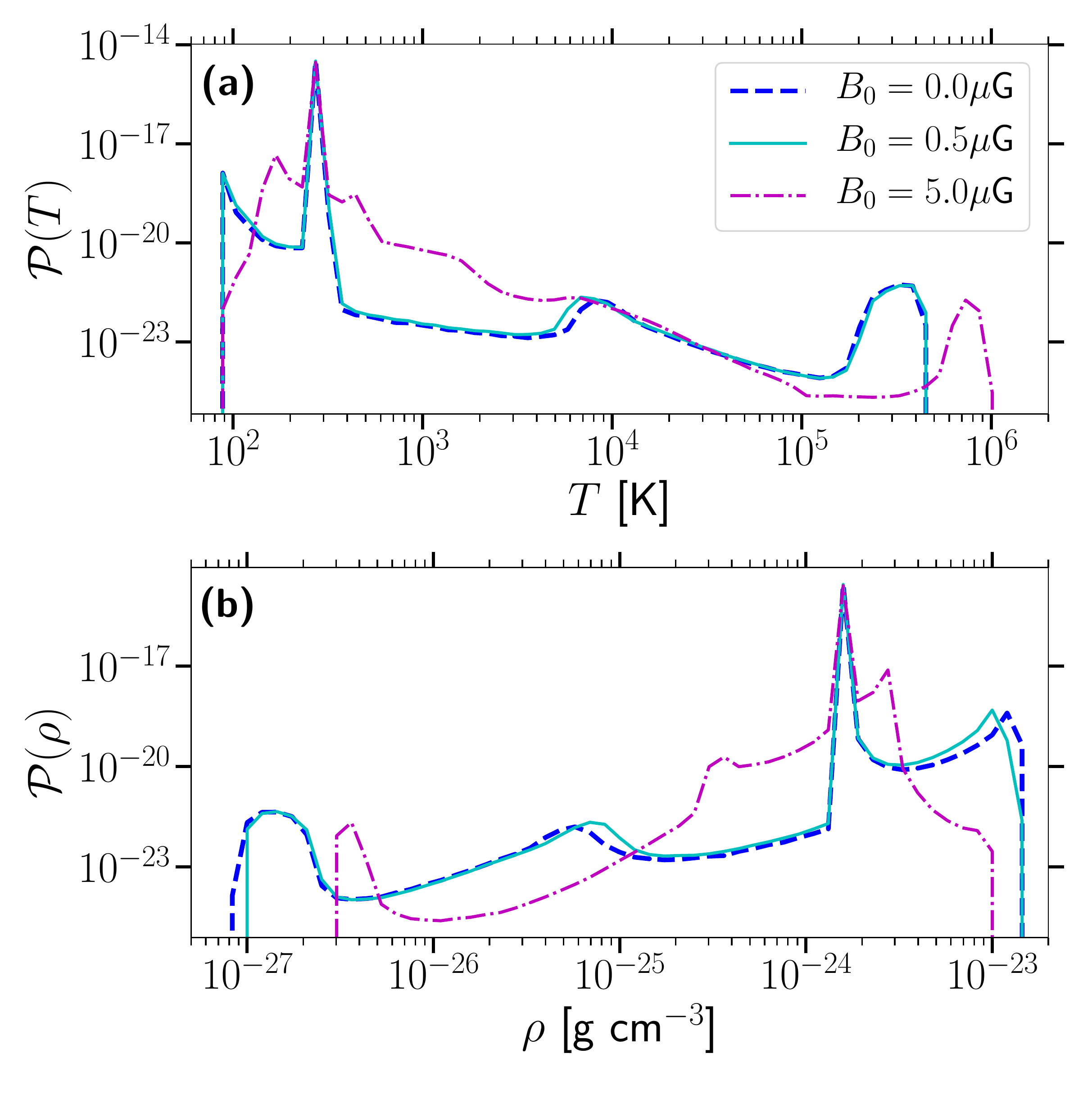}
  \caption{Mass-weighted probability density functions (PDFs) of 
		{\textbf{(a)}} gas temperature and {\textbf{(b)}} gas density at $2.0$\,Myr
		for HD (dashed blue) and MHD models
		with $B_0=0.5\upmu$G (solid cyan) and $B_0=5.0\upmu$G (dash-dotted magenta)
		 models.}
  \label{fig:sn_pdfs}
\end{figure}

In Figure\,\ref{fig:sn_pdfs} we show the probability density
functions (PDFs) of temperature, {\bf(a)}, and gas density, {\bf(b)}.
Each panel features the respective PDFs for models with $B_0=0$, 0.5 and
5\,$\upmu$G at $t=2$\,Myr.
All models have identical initial temperature and gas density distributions.
The peaks at $T\simeq260$\,K and $\rho\simeq1.67$\,g\,cm$^{-3}$ identify
the ambient ISM.
For the HD model and model with $B_0=0.5\upmu$G, the differences in the 
PDFs are otherwise also negligible.
{The other local maxima identify the cold remnant shell ($n>10$\,cm$^{-3}$,\,$T<100$\,K), 
the hot diffuse core ($n<0.01$\,cm$^{-3}$,\,$T>10^5$\,K), and the
accumulation of thermally stable warm gas ($n\simeq0.1$\,cm$^{-3}$,\,$T\simeq10^4$\,K).}

However, with the strong magnetic field the peak temperatures representing the
remnant shell and the remnant cores are hotter and more dense.
{Its magnetically confined hot gas has a smaller fractional volume.
Hot gas in all models cools slowly, due to the low density, but the strong
{MHD hot gas cools due to radiative losses relatively faster due to its slightly higher density.
It nevertheless remains hotter, because the adiabatic cooling is reduced.}
The strong MHD model has almost identical temperature distribution in the
range $10^4<T<10^5$\,K, although more dense.

\begin{figure}
  \centering
  \includegraphics[trim=0.00cm 2.2cm 0.2cm 0.15cm, clip=true,width=0.93\linewidth]{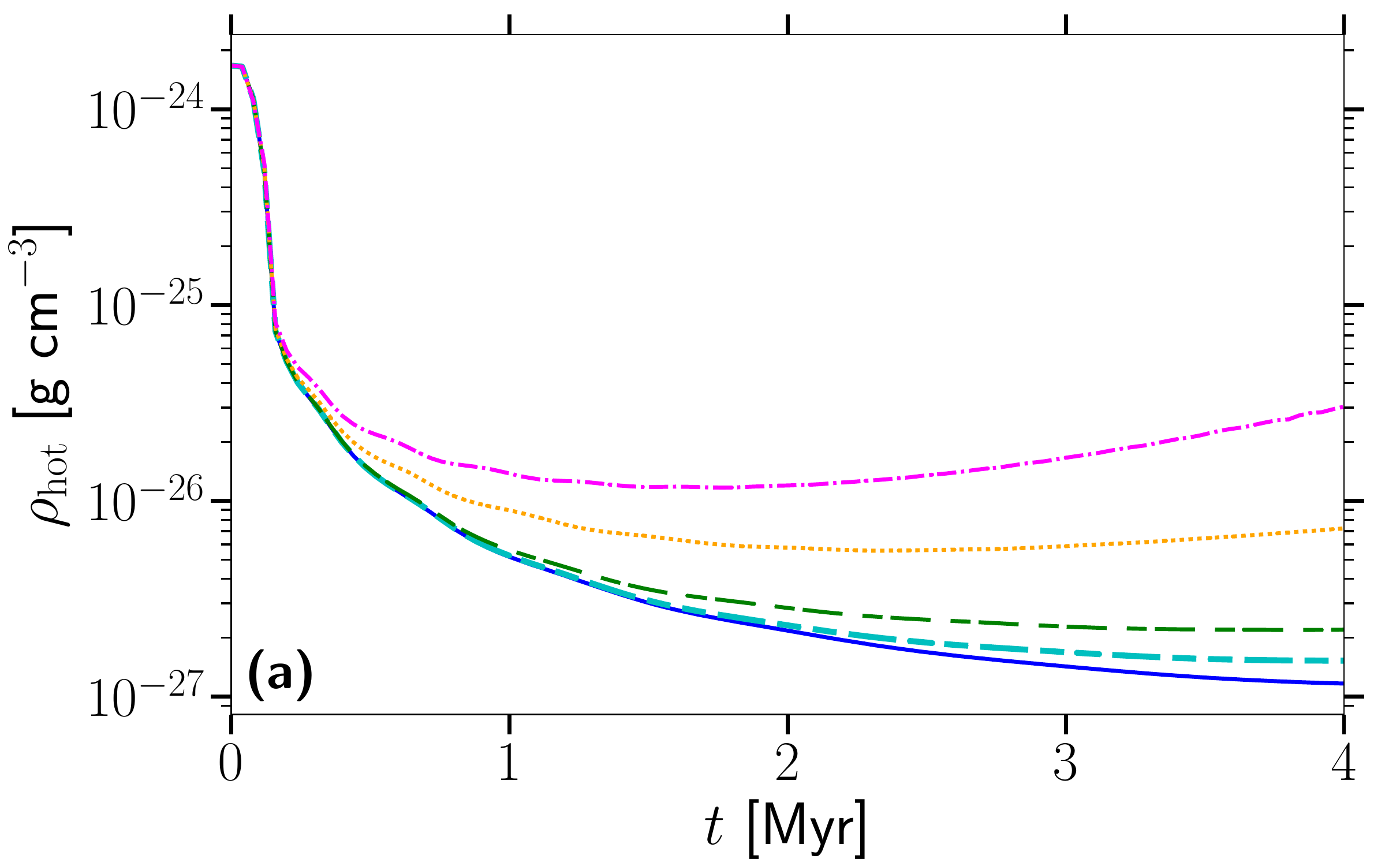}
  \includegraphics[trim=0.00cm 2.2cm 0.2cm 0.15cm, clip=true,width=0.93\linewidth]{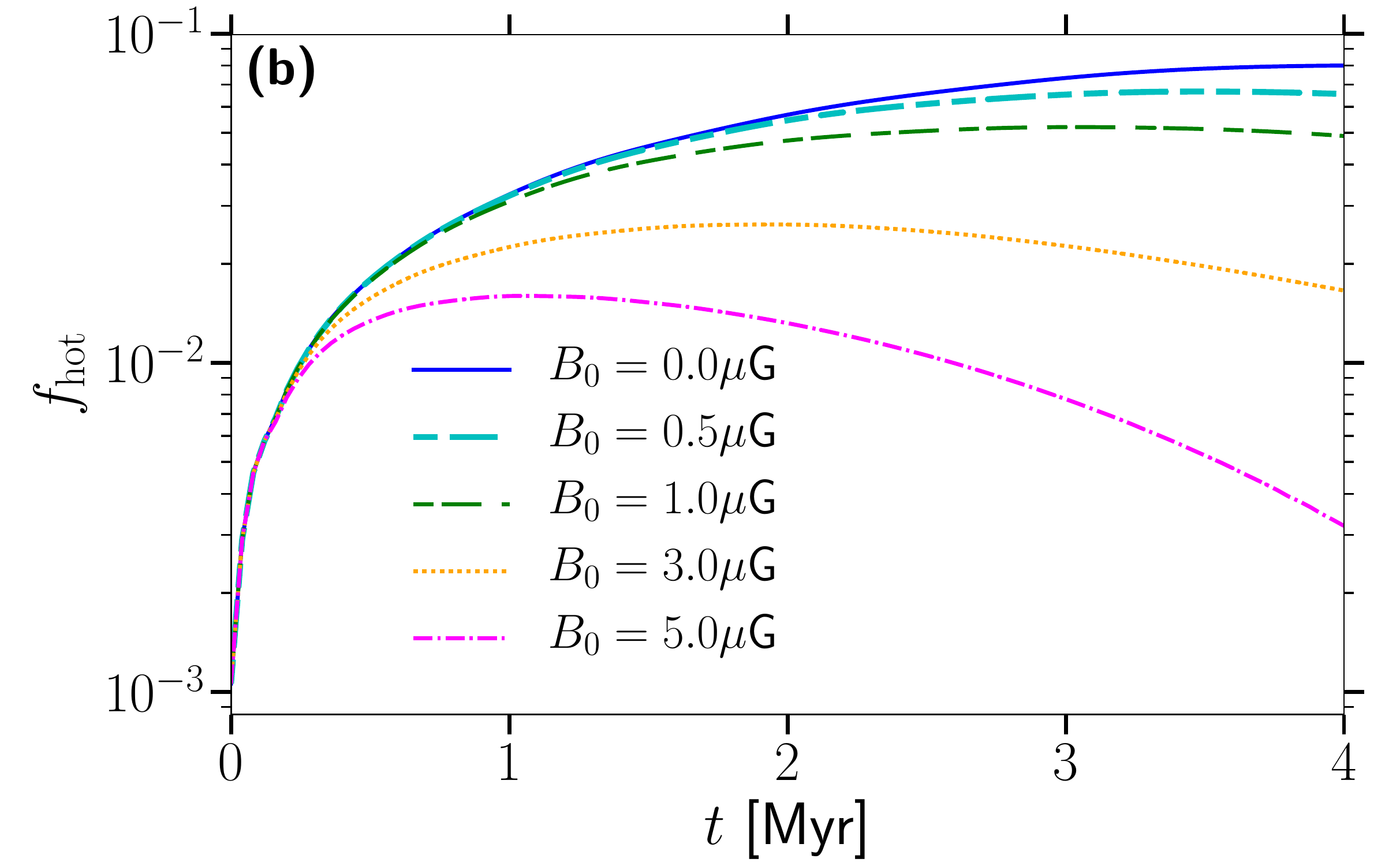}
  \includegraphics[trim=0.30cm 0.1cm 0.2cm 0.15cm, clip=true,width=0.93\linewidth]{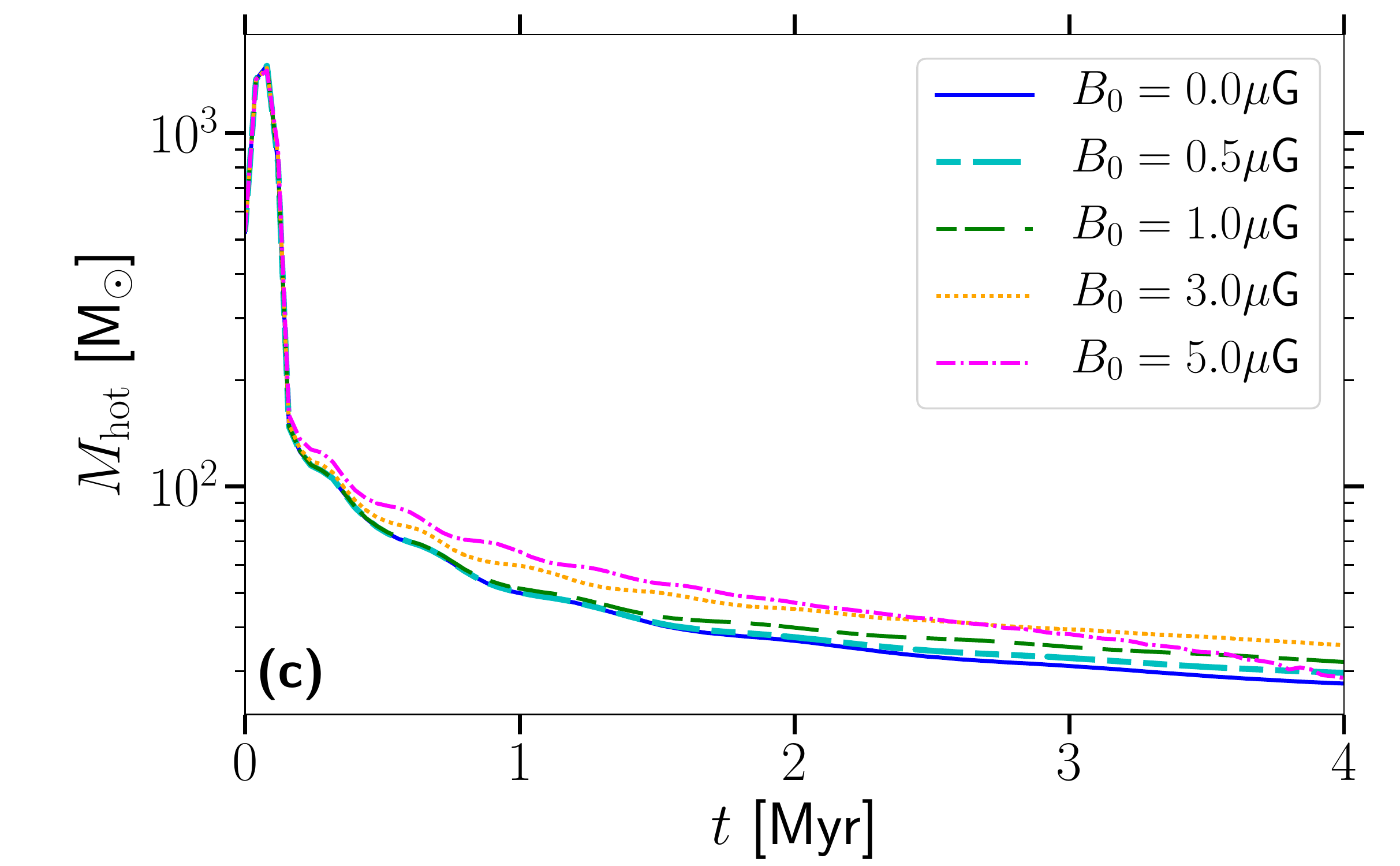}
  \caption{Time evolution of {\textbf{(a)}}\,mean density of hot gas 
	{\textbf{(b)}}\,fractional volume of hot gas 
	\edits{{\textbf{(c)}}\,mass of hot gas} for the HD (solid blue)
  and MHD models with $B_0=0.5\upmu$G. Hot gas is defined as $T>2\cdot 10^4$\,K \citep{KO15}.}
  \label{fig:rho_hot}
\end{figure}

\begin{figure*}
  \centering
  \includegraphics[trim=1.5cm 2.1cm 2.4cm 1.4cm, clip=true,width=0.45\linewidth]{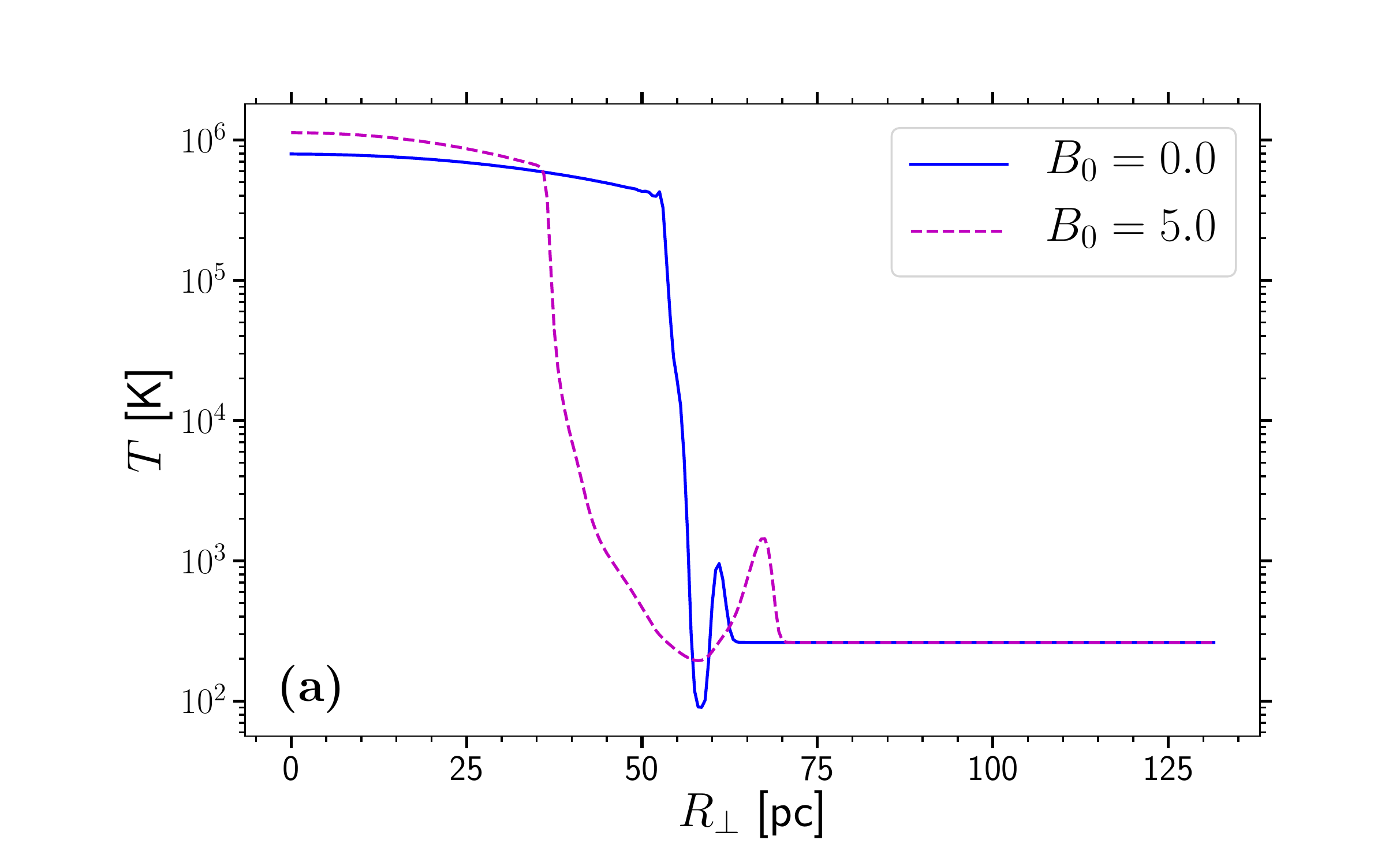}
  \includegraphics[trim=1.5cm 2.1cm 2.4cm 1.4cm, clip=true,width=0.45\linewidth]{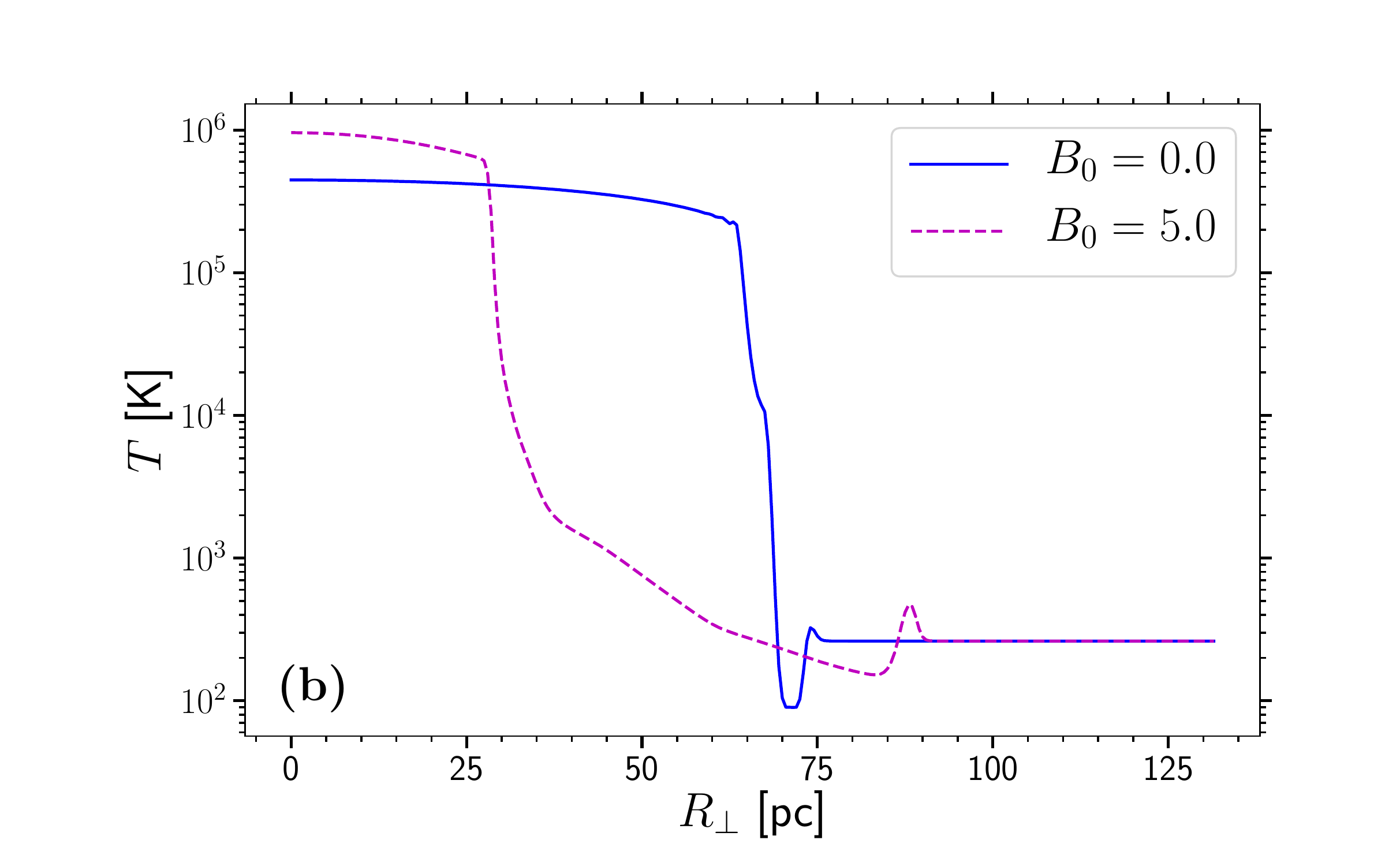}\\
  \includegraphics[trim=1.5cm 0.0cm 2.4cm 1.4cm, clip=true,width=0.45\linewidth]{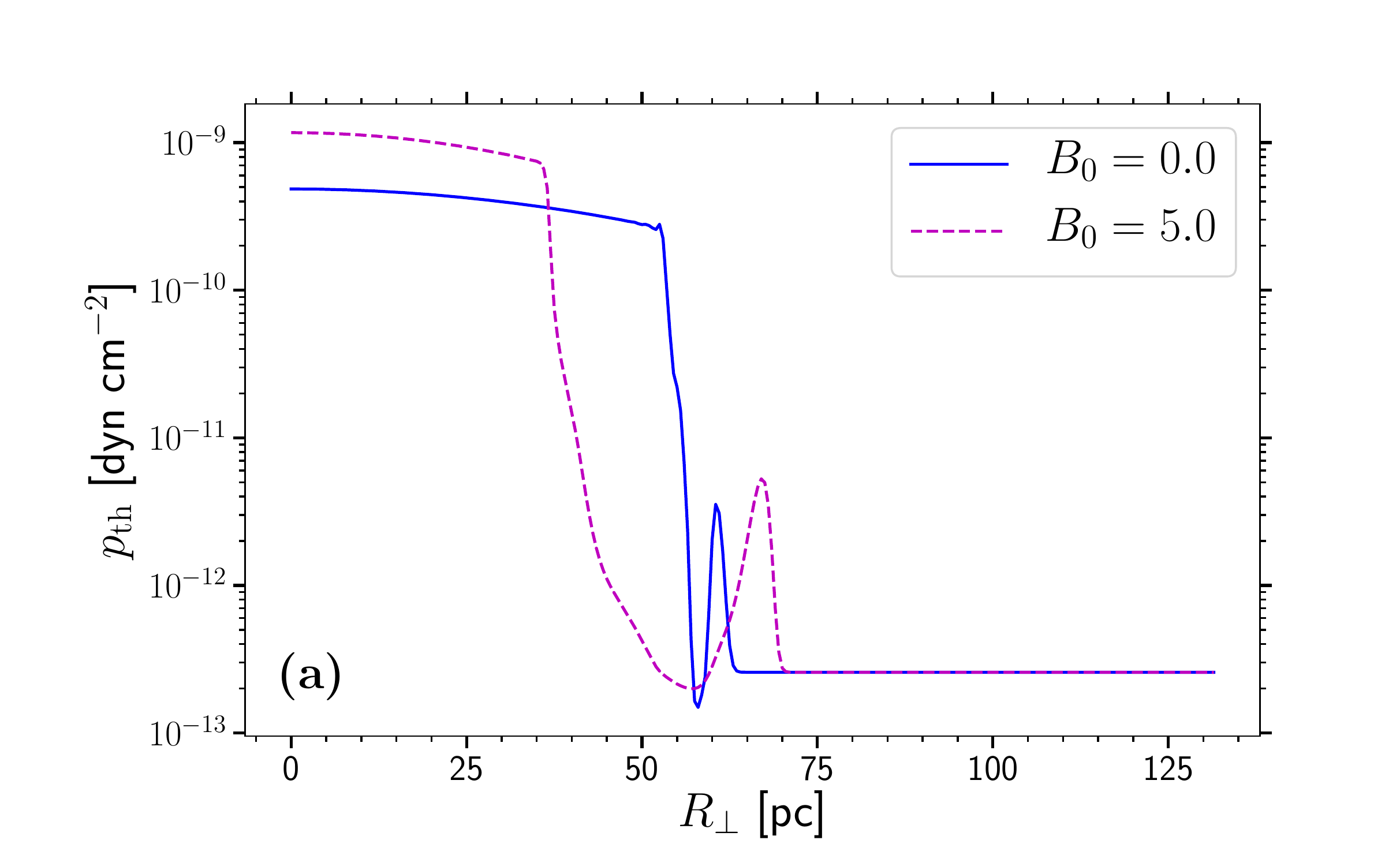}
  \includegraphics[trim=1.5cm 0.0cm 2.4cm 1.4cm, clip=true,width=0.45\linewidth]{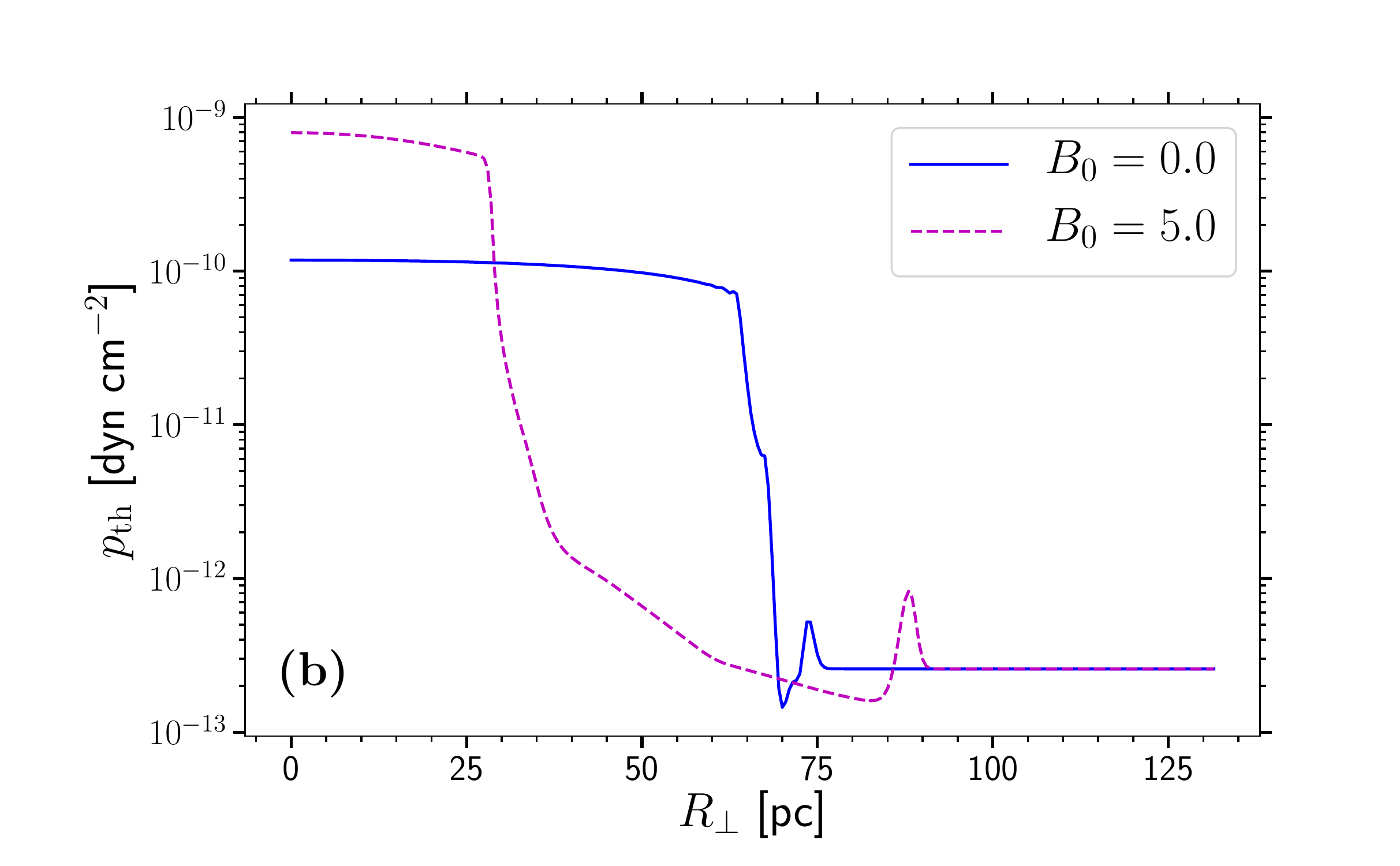}\\
  \caption{\edits{Radial profiles
	at {\bf{(a)}}\,1\,Myr and {\bf{(b)}}\,2\,Myr perpendicular to the magnetic \edits{field} of gas temperature
        \edits{(top)  and thermal pressure (bottom).}}}
  \label{fig:tt_rad}
\end{figure*}

Most unexpected is the large proportion of gas in the MHD model persisting 
around 1000\,K.} 
In the HD model {below the local maximum at} $10^4$\,K {is} thermally 
unstable{, with gas cooling rapidly to and accumulating} at about 90,\K,
where the cooling function{, $\Lambda$,} smoothly truncates. 
However with a strong magnetic field much of the gas in the remnant persists
in {this} thermally unstable range, the cold peak sustained about 200\,K and 
$n\simeq 2$\,cm$^{-3}$.
\edits{filling fraction $\rightarrow$ fractional volume, we went to some lengths
in paper 1 to define the three methods of measuring fraction by phase, so let's
keep the label precise. Delete when OK}
\edits{The high fractional volume} of thermally unstable higher density
gas in the MHD model \edits{is explained by the revised} balance of heating,
$\Gamma$, and cooling processes\edits{, in the inter-shock region illustrated in
Figure\,\ref{fig:netheat}\,{\bf{(b)}}}.

{Magnetic confinement of hot gas in the remnant is indicated strongly by the
increasing mean hot gas density plotted \edits{in Figure\,\ref{fig:rho_hot}\,{\bf{(a)}}} and
decreasing fractional volume \edits{from Figure\,\ref{fig:rho_hot}\,{\bf{(b)}}} of hot
gas in strongly magnetized remnants. This effect has also been found in
larger-scale simulations of the ISM \citep{EGSFB17,EGSFB19}.}
\edits{The residual total mass of hot gas deposited into the ISM may be 20--40\%
greater in the strong MHD remnant as depicted in 
Figure\,\ref{fig:rho_hot}\,{\bf{(c)}}}.

\edits{In Figure\,\ref{fig:tt_rad} we show radial profiles perpendicular to the
magnetic field at 1\,Myr {\bf{(a)}} and 2\,Myr {\bf{(b)}}, for 
temperature (top) and thermal pressure (bottom) for the HD and strong MHD
remnants. 
\edits{This illustrates the higher temperature in the MHD core, but smaller fractional
volume.}}

\section{Residual SN energy injection}\label{sec:energy}

\edits{Inspection of the momentum injection indicates the magnetic field 
can reduce the residual momentum appled to the ISM by an SN
remnant.
The increased mass of hot gas, on the other hand and reduced effect 
of non-adiabatic cooling within an MHD remnant invites the question how
are the energetics of the ISM affected by a strong magnetic field?
The time evolution of residual SN total energy and each 
contribution to the energy are plotted in Figure\,\ref{fig:en_ret}.} 

\edits{Panel}\,{\bf{(a)}} shows \edits{that a significant
increase in residual total energy 
occurs with strong MHD even within 1\,Myr of} 
the supernova explosion. 

\edits{From} Panel\,{\bf{(b)}} \edits{we see} that the presence
of large-scale magnetic fields \edits{retains more residual}
thermal energy,
while Panel\,{\bf{(c)}} shows that kinetic energy 
in the MHD remnants \edits{is lost more quickly}.
\edits{Beyond 3\,Myr the strong MHD residual kinetic energy 
can persist more effectively than HD, would likely be subsumed
by the ambient ISM turbulence by this time.}
\edits{For all MHD models there is a modest linear amplification of
the magnetic field trhough compression and tangling (Panel\,{\bf{(d)}}).
So the net contributions of thermal and magne sutic energy mean that the
MHD shocks in fact induce a greater total residual energy into the
ISM than HD alone.}

\begin{figure}
  \centering
  \includegraphics[trim=0.00cm 2.3cm 0.2cm 0.15cm, clip=true,width=0.93\linewidth]{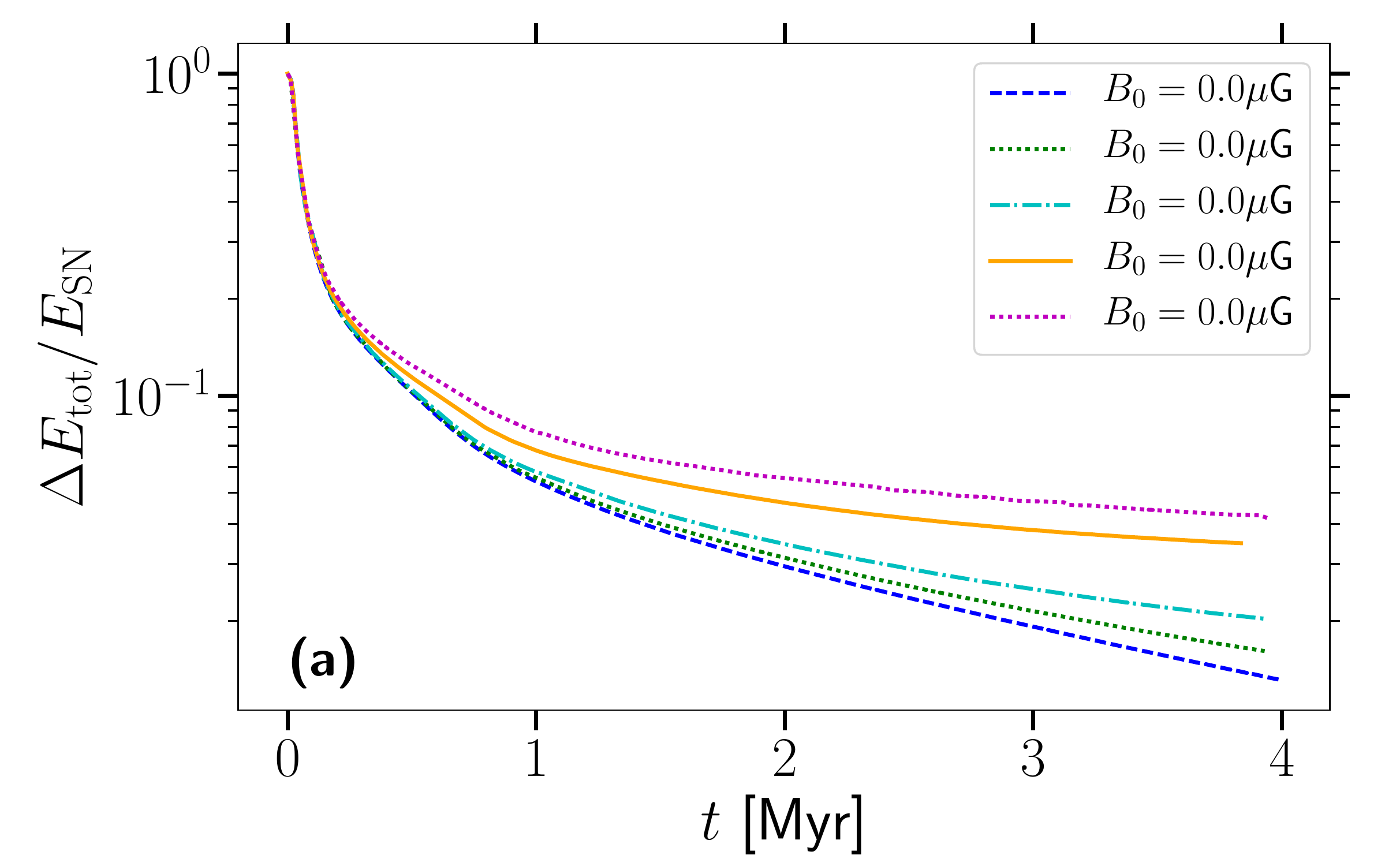}
  \includegraphics[trim=0.50cm 2.3cm 0.2cm 0.15cm, clip=true,width=0.93\linewidth]{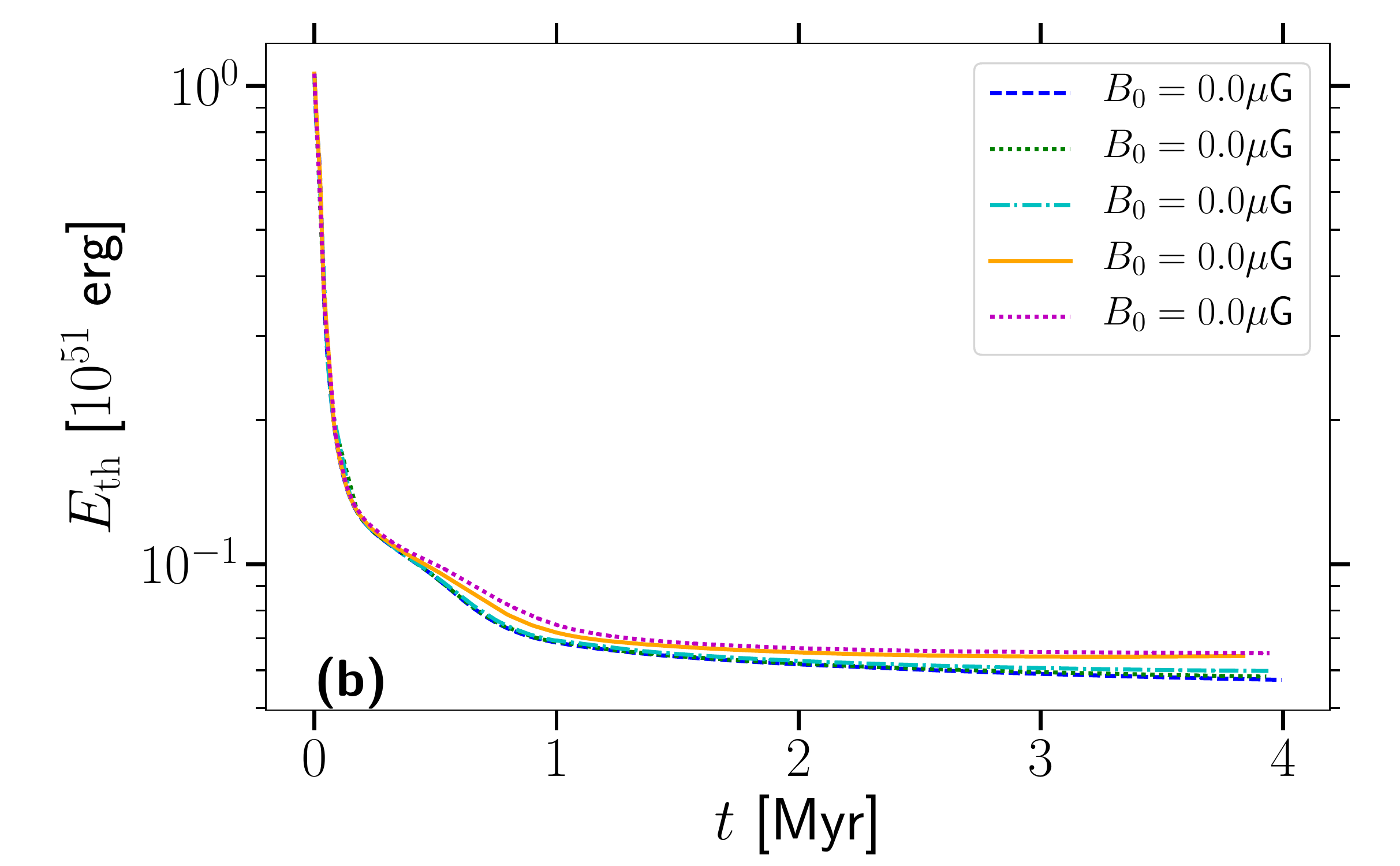}
  \includegraphics[trim=0.50cm 2.3cm 0.2cm 0.15cm, clip=true,width=0.93\linewidth]{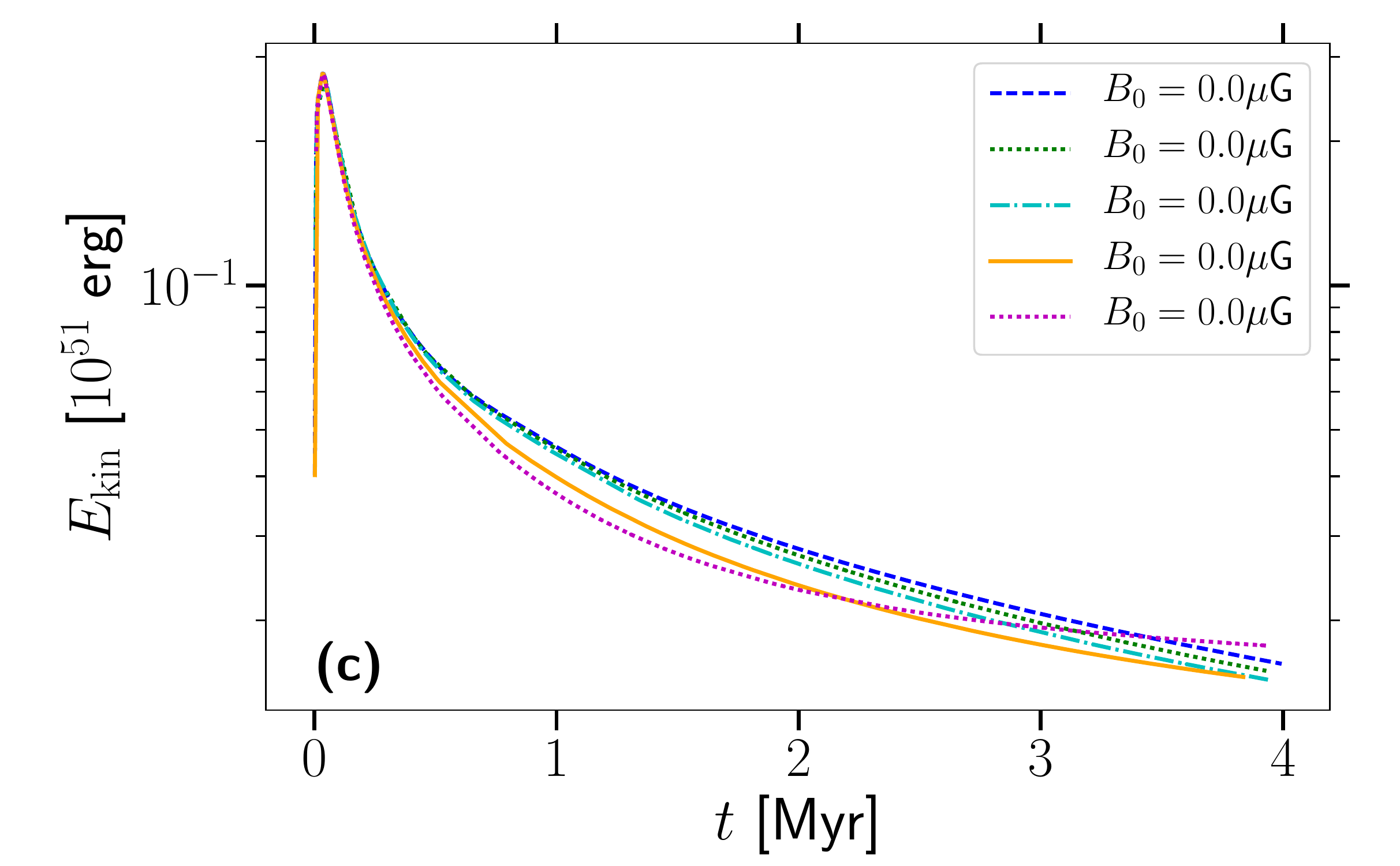}
  \includegraphics[trim=0.50cm 0.1cm 0.2cm 0.15cm, clip=true,width=0.93\linewidth]{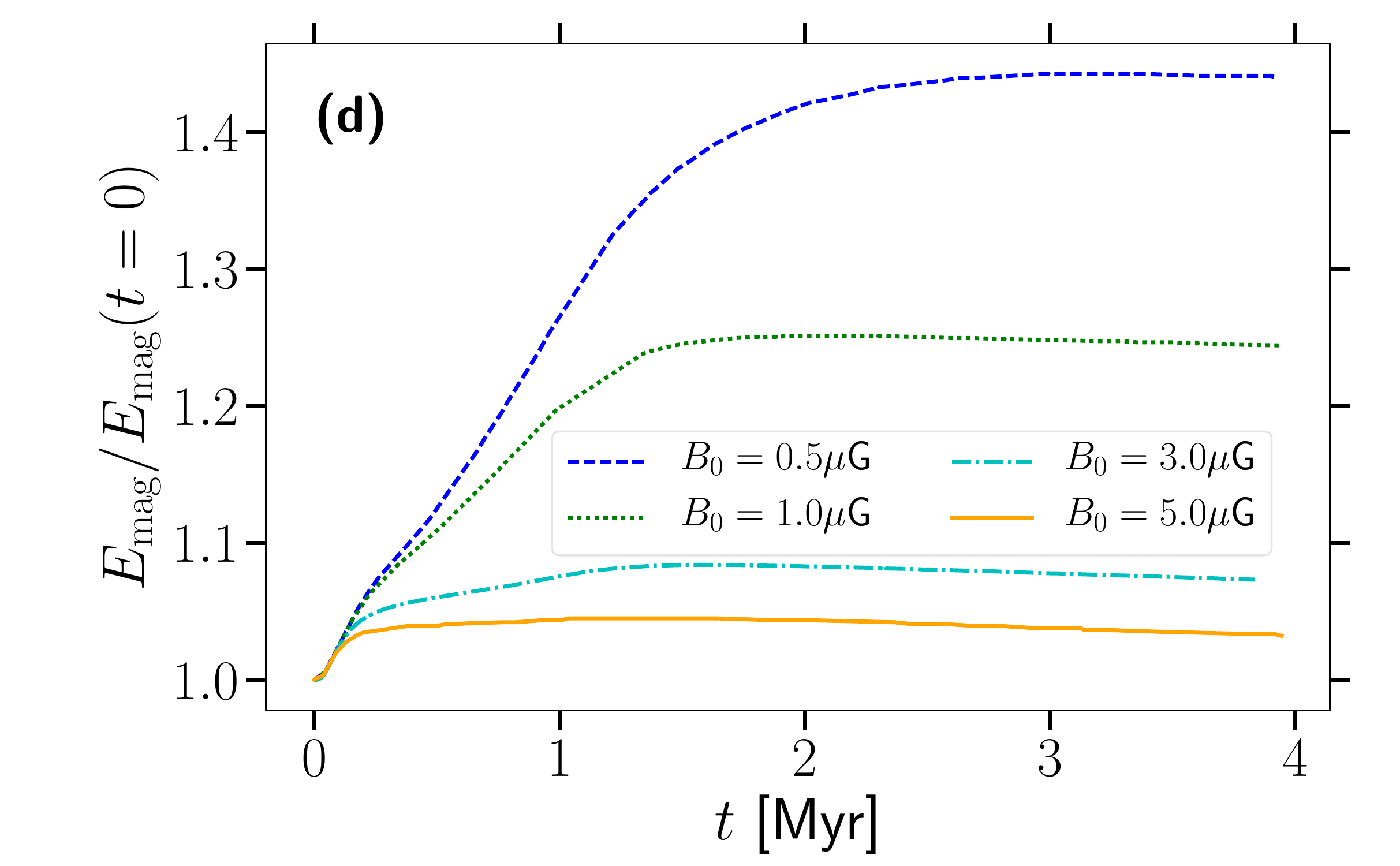}
  \caption{\edits{{\bf{(a)}}\,Energy retention by HD and MHD ($5\upmu$G remnants, where
	$\Delta E_{\rm{tot}}(t)=E_{\rm{tot}}(t)-E_{\rm{tot}}(t=0)$.
	$E_{\rm{tot}}(t=0)$ is the total energy in the ambient gas
	prior to the SN explosion, {\bf{(b)}}\, thermal energy
	{\bf{(c)}}\,kinetic energy and {\bf{(d)}}\,magnetic energy profiles.}}
  \label{fig:en_ret}
\end{figure}

\begin{table}
\caption{Energy retention as a percentage of $E_{\rm{SN}}$, given
	for each model.}
\label{tab:retention}
\begin{center}
\begin{tabular}{cccccc}
\hline
\hline
Time [Myr]           & \multicolumn{5}{c}{$B_0$ [$\upmu$G]}\\
                     & \cline{1-5}
                     & 0 & 0.5 & 1.0 & 3.0 & 5.0\\
\hline
0.25 & 16.2 & 16.2 & 16.5 & 17.2 & 18\\
0.5  & 10.2 & 10.3 & 10.4 & 11.4 & 12.4\\
1.0  & 5.4  & 5.5  & 5.9  & 6.7  & 7.7\\
2.0  & 2.9  & 3.1  & 3.5  & 4.6  & 5.5\\
4.0  & 1.3  & 1.6  & 2.0  & 3.5  & 4.2\\
\hline
\end{tabular}
\end{center}
\end{table}
\edits{The effects are tabulated for each model at various times
throughout their evolution in Table\,\ref{tab:retention}.}
Differences emerge with HD within 250\,kyr
of the SN explosion for strong MHD models. The $3\upmu$G and $5\upmu$G
models retain 6\% and 11\% more energy\edits{, rising to 
10\% and 20\% within 500\,kyr}.
\edits{The HD simulation retains $1.3\%$ of SN
energy by the end of the simulation. As magnetic field strength
increases, the SN energy retained by the system increases,
reaching $4.2\%$ for $B_0=5\upmu$G.}

\section{Critical magnetic field strength}\label{sec:beta}

{Any qualitative differences between the HD and MHD remnants 
must arise from effects due to the Lorentz force. 
Inspection of the evolution of magnetic tension reveals this to be negligible,
relative to the magnetic and thermal pressure gradients.
The maximal tension forces apply in the remnant shell nearest the polar axis 
of the remnant.}

{Therefore, modification to momentum injection 
relies on the interacting radial pressure gradients, 
particularly where $\partial_\perp \left(p\right)\simeq
\partial_\perp \left(|\vect{B}|^2/2\mu_0\right)$.
In the blast wave the magnetic gradient will grow due to compression.
The stronger the ambient field, the earlier the magnetic gradient in 
the shell and in its wake will acquire a critical strength sufficient to
significantly modify the remnant evolution.
The MHD shell begins to shed mass into its wake, compared to HD as early
as 200\,kyr for $B_0=5\upmu$G, 320\,kyr for 3$\upmu$G and 
920\,kyr for 1$\upmu$G. 
The leading shock outstrips the HD model at 440, 560 and 1560\,kyr, respectively.
From both panels in Figure\,\ref{fig:force_terms} throughout the 
inter-shock region we have

\[
\dfrac{\partial_\perp \left(|\vect{B}|^2/2\mu_0\right)|_{-}}{\partial_\perp \left(p\right)|_+}\gtrsim 1,
\]
but when the MHD shock profiles start to diverge from HD this ratio locally is still much less than 1.
A reasonable criteria for identifying the critical large-scale magnetic field
strength may be to assume plasma-$\beta<$1.
Hence, $B_{\mathrm{crit}}\sim\sqrt{2\mu_0 p}$,
where $p=k_{\mathrm{B}}nT$; $k_{\mathrm{B}}$ is the
Boltzmann constant, and $T$ denotes temperature.}

For these simulations the ambient ISM has $n=1$\,cm$^{-3}$ and $T\simeq 260$\,K.
Thus, the gas is in thermal equilibrium with radiative cooling balancing 
UV heating.
\edits{We show in Appendix\,\ref{app:amb_tt} that the results are not sensitive to 
the initial ambient temperature, which cools very quickly to the equilibrium
temperature to evolve with the same remnant properties.}
We obtain $B_{\rm crit}\sim 1\,\upmu$G.
{$B_0=1\upmu$G does lead to marginal MHD effects.}
In agreement with \citet{HT06}, we find that significant change to the nature of the remnant
is seen for $B_0\geq 3\upmu$G.
{However, the effect of cooling and heating processes are highly nonlinear functions
of density, so we shall explore how to explain critical field strength with various ambient
ISM densities and  magnetic field configurations.}

\section{\edits{Summary}}\label{sec:summary}

\edits{The merger of the remnant shell with the surrounding gas
depends on a number of factors, including but not limited to,
ambient sound speed, homogeneity (or inhomogeneity) of the ambient gas
density structures, and interaction with other shocks in a turbulent,
highly compressible and typically inhomogeneous interstellar medium.
However, the emergence of differences in energy retention, directly
linked to the presence of large-scale magnetic fields, suggests that
the efficiency with which the surrounding gas extracts energy from
SN remnants is modified by magnetic fields.}

A plane-parallel, micro-gauss strength magnetic field changes the aspect
ratio of SN remnants. 
The shockwave through the ambient {ISM propagates faster} perpendicular
to the magnetic field.
{Figure\,15 of \citet{CK01} show a similar MHD remnant expanding faster
perpendicular to the plane-parallel field.
This is also identified by \citet{FMZ91} in the context of a superbubble
blast wave.
They also report that the MHD blast is faster in the parallel direction 
than for HD.}
Conversely, the hot diffuse {remnant core is magnetically confined into
a prolate spheroid with pole aligned to the field.}
\citet{HT06,KO15} also find that expansion of the core is inhibited
perpendicular to a plane-parallel magnetic field.

Both effects are caused by {the} magnetic pressure gradient{s}
perpendicular to the magnetic field.
In the adiabatic and pressure driven phases of the blast wave expansion, while
magnetic forces are weak relative to the explosive forces, magnetic field is
expelled from the core and compressed into the shell.
{Some  energy from the thermally driven outward shock is stored in
the highly compressed field at the remnant shell, to be released through
the magnetic pressure gradient in the late MHD shock.
Compression also generates an increasingly negative magnetic pressure gradient
just behind the shell, ultimately injecting momentum inwards.}

{In similar simulations, which focus on the early (up to 400\,kyr) stage of
SN evolution},
\citet{KO15} conclude that momentum injection by an SN remnant is unaffected
by the presence of a plane-parallel magnetic field.
They suggest magnetic fields will not affect momentum injection, {because}
they do not become comparable to thermal effects before the remnant enters
the momentum-conserving phase.
\edits{\edits{In contrast we \emph{do} find for $B_0\geq 3\upmu$G a substantial effect on}
 outward momentum injection 
\edits{already within 500\,kyr, which} could be reduced by up to 10\%
in the presence of a large-scale magnetic field. Given the role of SN in driving
both turbulent and large-scale flows (such as galactic winds), such a reduction
in momentum injection could have a noticeable effect on gas dynamics within
galaxies.}

\edits{MHD effects also lead to confinement of the hot core and a more dense
inter-shock region. \edits{UV-heating \edits{exceeds} radiative losses in this region and, 
\edits{combined with the magnetic broadening of the shock front,}
inhibits the formation of cold gas, compared to HD.
This results in a region of thermally unstable warm gas, which
will eventually merge with the surrounding gas.}
In contrast to the reduced momentum injection the residual energy of 
an MHD SN remnant can be up to 40\% greater within 1\,Myr.
This, however, corresponds to a reduced fractional volume of hot gas,
\edits{an effect also noted
by \citet{EGSFB19}.}}
\edits{Even though we obtain this result in an idealized numerical setup, \citet{EGSFB19} find that
hot gas becomes more dense in numerical simulations of the local Galaxy, as the magnetic field
reaches $B_{\rm{rms}}\sim 3\upmu$G strength with a strong large-scale component.
These numerical simulations feature highly inhomogeneous gas density, compressible turbulence
and a multi-phase ISM structure. Thus, it is plausible that our remnant-scale simulations
may \edits{relate to} magnetic effects found in more sophisticated numerical models.}

{In its early stages the presence of a magnetic field has negligible
effect on SN remnant evolution}, and this is consistent with the 
results of \citet{KO15}. 
However, as the magnetic shell forms increasingly strong local pressure
gradients the dynamics {alter markedly}.
For sufficiently weak magnetic fields this may evolve so late 
that the blast wave has already merged with the ISM and it remains
dynamically insignificant.
For stronger fields the magnetic retrograde shock and late MHD shock occur
sufficiently early to alter remnant {momentum injection and structure}. 
For gas density $n=1$\,cm$^{-3}$, we find the critical field strength to
correspond to plasma-$\beta=1$ in the ambient ISM.
The magnetic confinement and late MHD shock occur earlier and are stronger
as $B_0$ increases within the range $1\leq B_0\leq5\upmu$G considered.
{With respect to structure, t}his is consistent with \citet{HT06}, who find notiecable magnetic confinement
of the remnant interior for $B_0\geq 3\upmu$G, and \cite{KO15} 
for $B_0= 7.2\upmu$G.

{Magnetic fields in the ISM are not uniform as applied in these models
and planned studies shall include the effects of turbulent structure of the
magnetic field and a range of  ambient ISM density.
Nevertheless,} it is widely observed that spiral galaxies have a large-scale, coherently
structured magnetic field {of} strength in the range $1$--$30\,\upmu$G
\citep{Beck01,TKFB08,FBSB11}.
Thus, it is plausible
that SN remnants evolve subject to the magnetic effects described here, which could have significant implications on pressure support,
mass loading of the galactic halo, and generally the multiphase and vertical structure of the ISM.

\section*{Acknowledgements}
\edits{We would like to thank the referee for their insightful feedback, which has
helped improve both the scientific content and narrative of this Paper.}

The work has been performed under the Project HPC-EUROPA3 (INFRAIA-2016-1-730897), with the support of the EC Research Innovation Action under the H2020 Programme; in particular, CCE gratefully acknowledges the support of the Department of Computer Science at Aalto University and the computer resources and technical support provided by the CSC HPC Centre in Finland.
  FAG acknowledges financial support of the Grand Challenge project SNDYN
  and GDYNS,
  CSC-IT Center for Science Ltd. (Finland) and the Academy of Finland 
  Project 272157.

\bibliography{refs}

\appendix

\section{\edits{Identifying numerical effects through 1D shock-tube tests}}\label{app:riemann}

\edits{To verify that the significant qualitative effects we identify from the MHD blast waves are attributable to the
physics rather than the numerics, we examine the impact of employing artificial
diffusivities compared to the impact of a magnetic field on the solutions. 
For numerical economy we use a set of 1D shock-tube tests.} 
\edits{The numerical solutions for adiabatic shocks with parameters relevant to SN blast waves are compared to the
system} \edits{originally described by \citet{Sod78},} \edits{and its} \edits{exact analytical solution} \edits{derived} \edits{by
\citet{HSW84}.}
\edits{We extend the analysis applied to the HD solutions of \citet{GMKSH19} to consider specifically whether MHD or 
numerical effects account for the differences in the MHD solutions}

\begin{figure}
  \centering
  \includegraphics[trim=0.25cm 0.1cm 0.3cm 0.7cm, clip=true,width=0.45\linewidth]{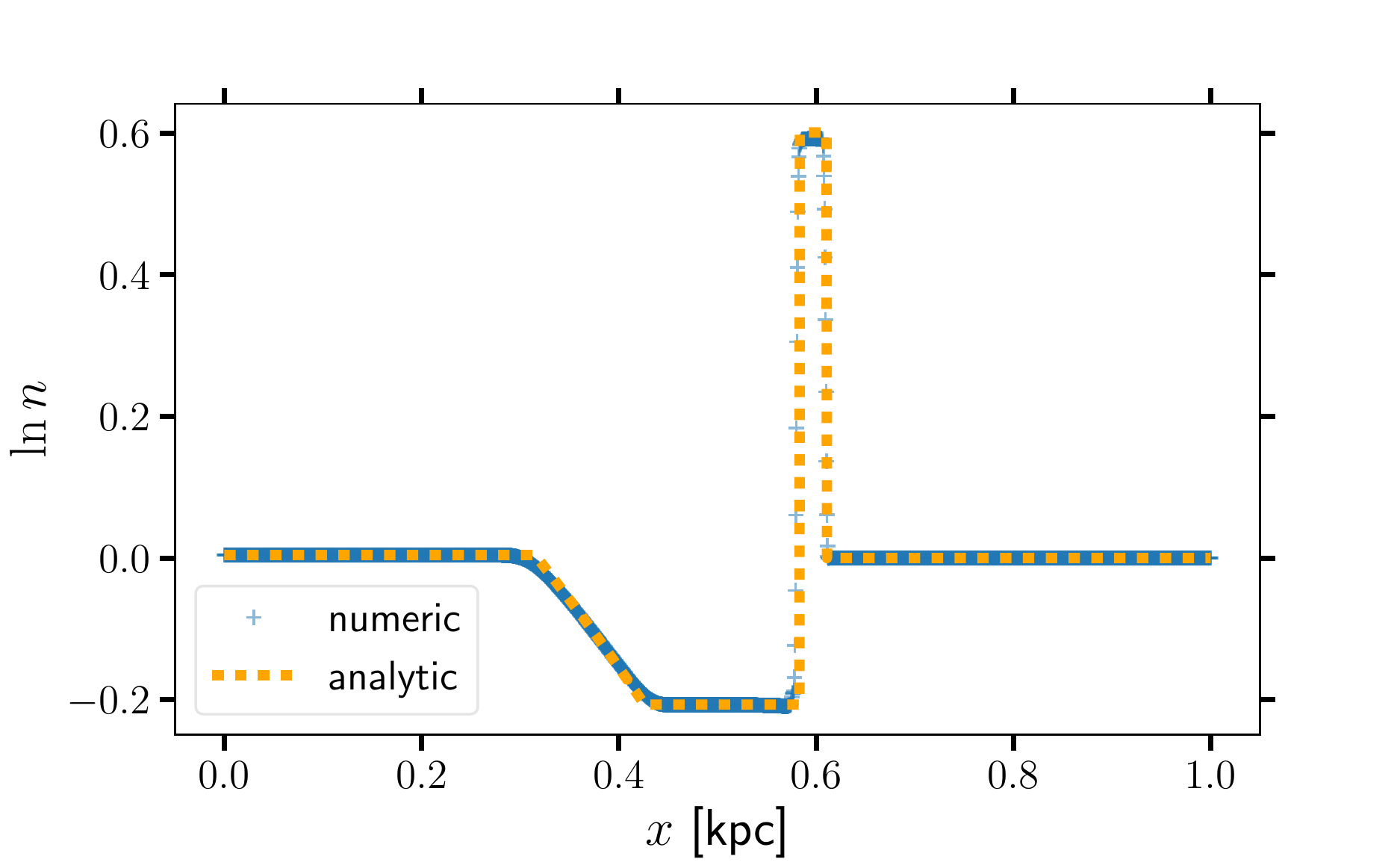}
  \includegraphics[trim=0.25cm 0.1cm 0.3cm 0.7cm, clip=true,width=0.45\linewidth]{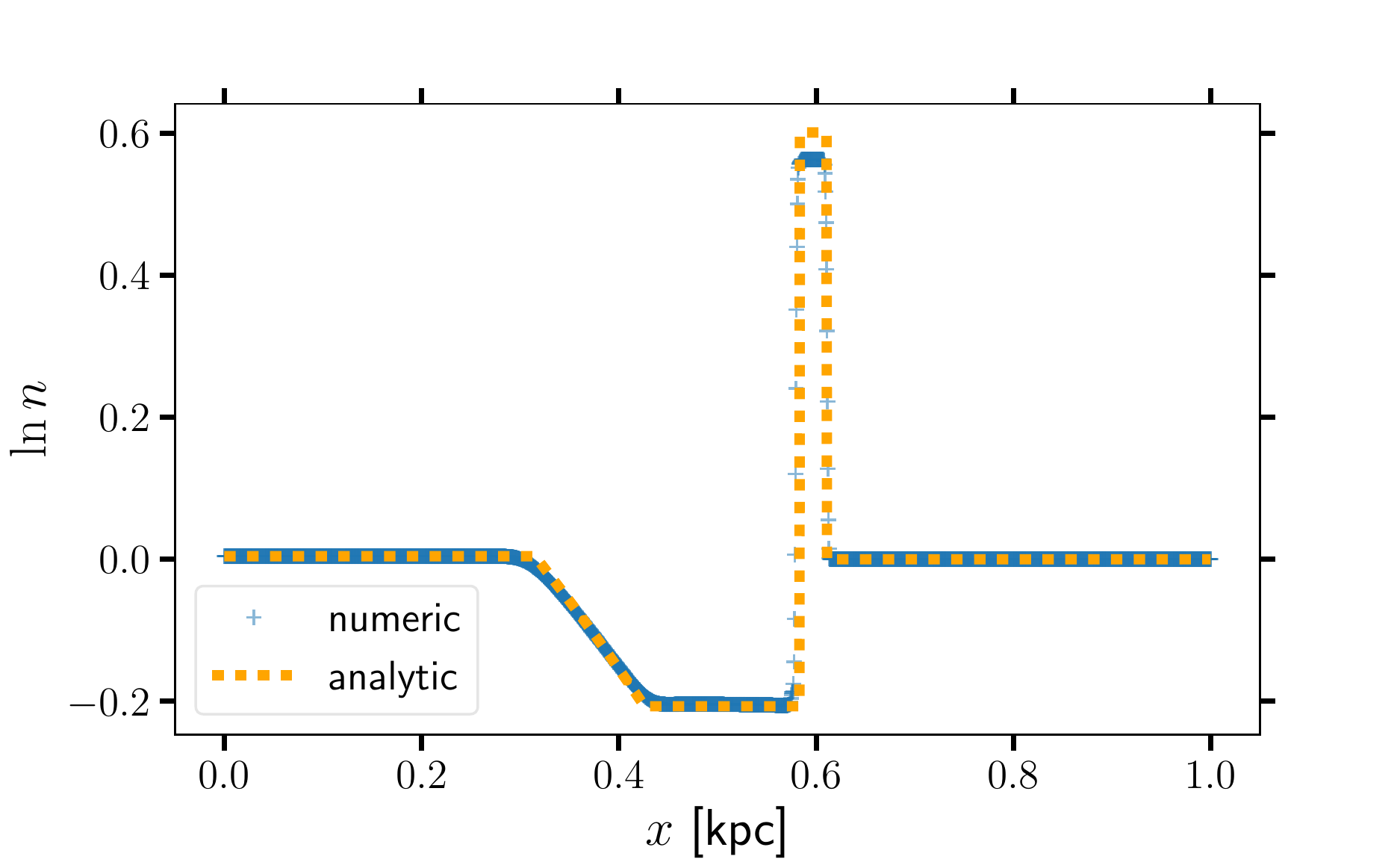}\\
  \includegraphics[trim=0.25cm 0.1cm 0.3cm 0.7cm, clip=true,width=0.45\linewidth]{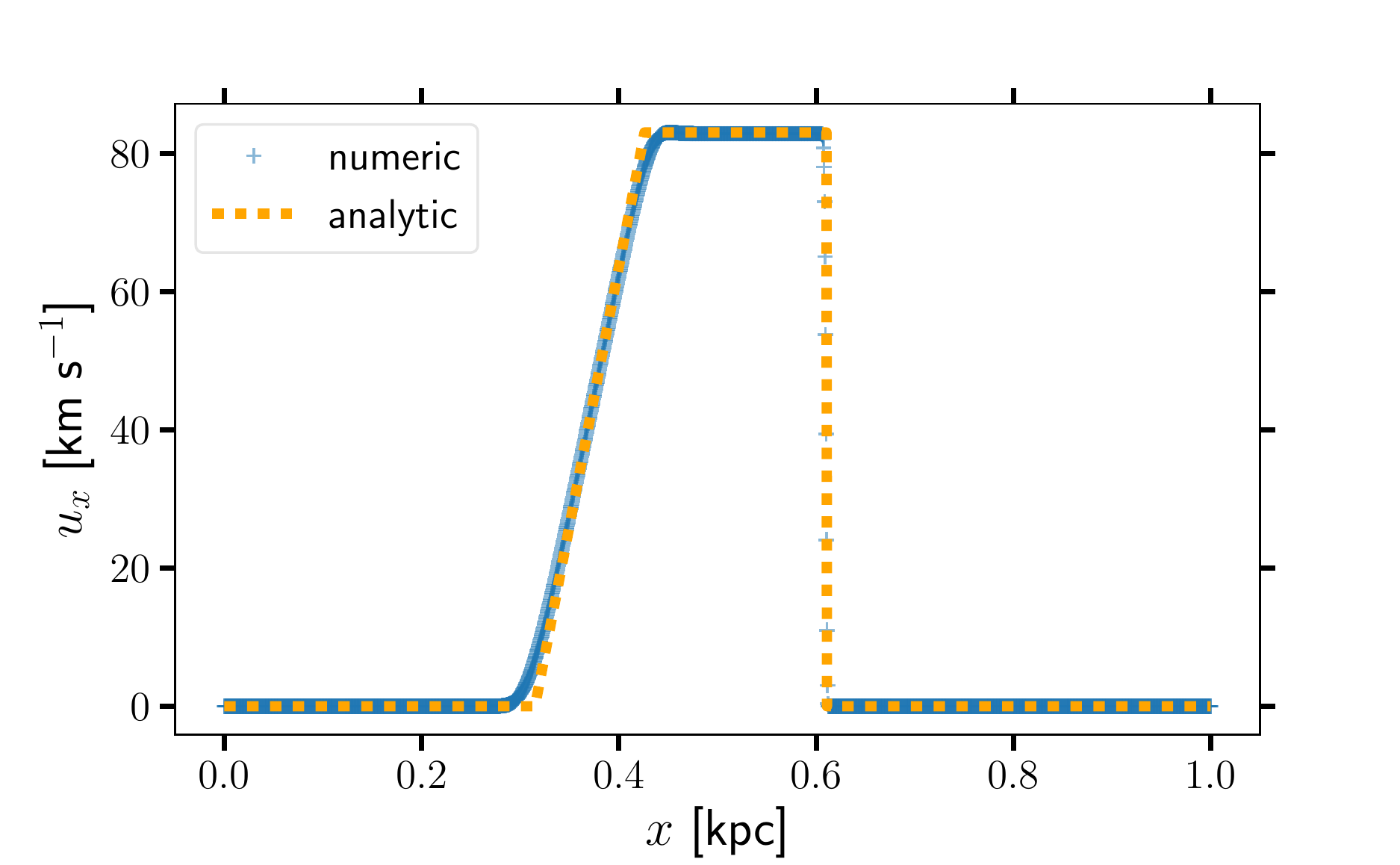}
  \includegraphics[trim=0.25cm 0.1cm 0.3cm 0.7cm, clip=true,width=0.45\linewidth]{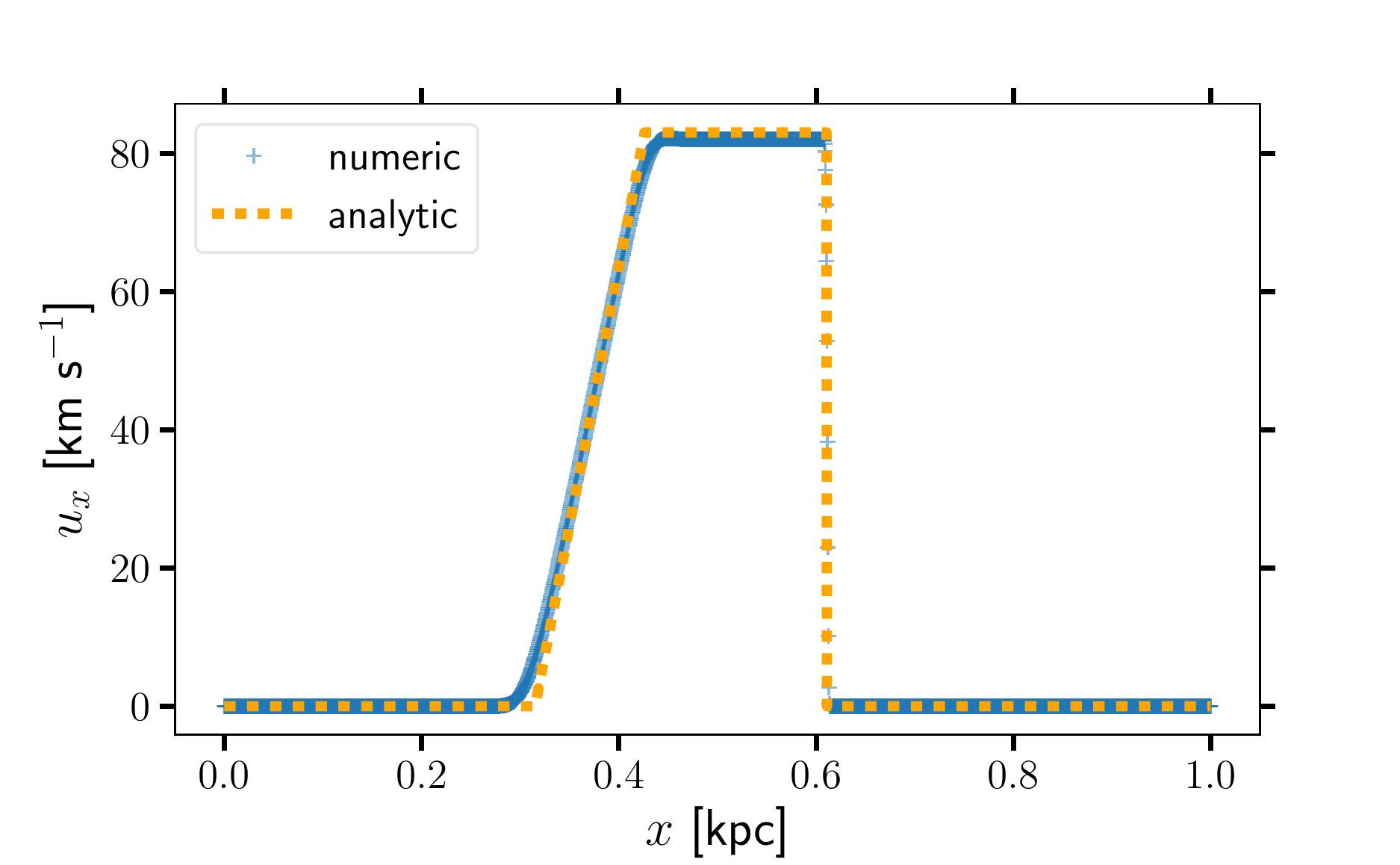}
  \caption{\edits{Shock-tube simulations for
        HD \edits{(left panels)} and
        MHD \edits{(right panels)} with $\nu$ and $\eta$ \edits{as applied to
        } the 3D simulations. The spatial resolution
	is 0.5\,pc. The MHD simulation features a $5\upmu$G uniform magnetic field perpendicular
	to the shock tube.} 
        \edits{Upper panels show log gas number density and lower panels gas velocity at $t=1$\,Myr.}}
  \label{fig:shock_standard}
\end{figure}

\edits{Figure\,\ref{fig:shock_standard} shows that \edits{for an adiabatic shock the HD numerical solution reliably 
produces the analytic solution. Even the MHD solution} has \edits{only} marginally enhanced gas density in the shock.}
\edits{Hence, neither numerical parameters nor the presence of the magnetic field have a significant effect on the
solution in the adiabatic system, and this is consistent with most previous assessments.}

\edits{We next consider the effect radiative cooling and UV heating on the shock-tube solution.
In Figure\,\ref{fig:shock_cool_rho} the non-adiabatic HD and MHD numerical solutions are contrasted to the adiabatic
analytic solution.
We do not have a non-adiabatic analytic shock-tube solution.}
%
\begin{figure}
  \centering
  \includegraphics[trim=0.25cm 0.1cm 0.3cm 0.7cm, clip=true,width=0.45\linewidth]{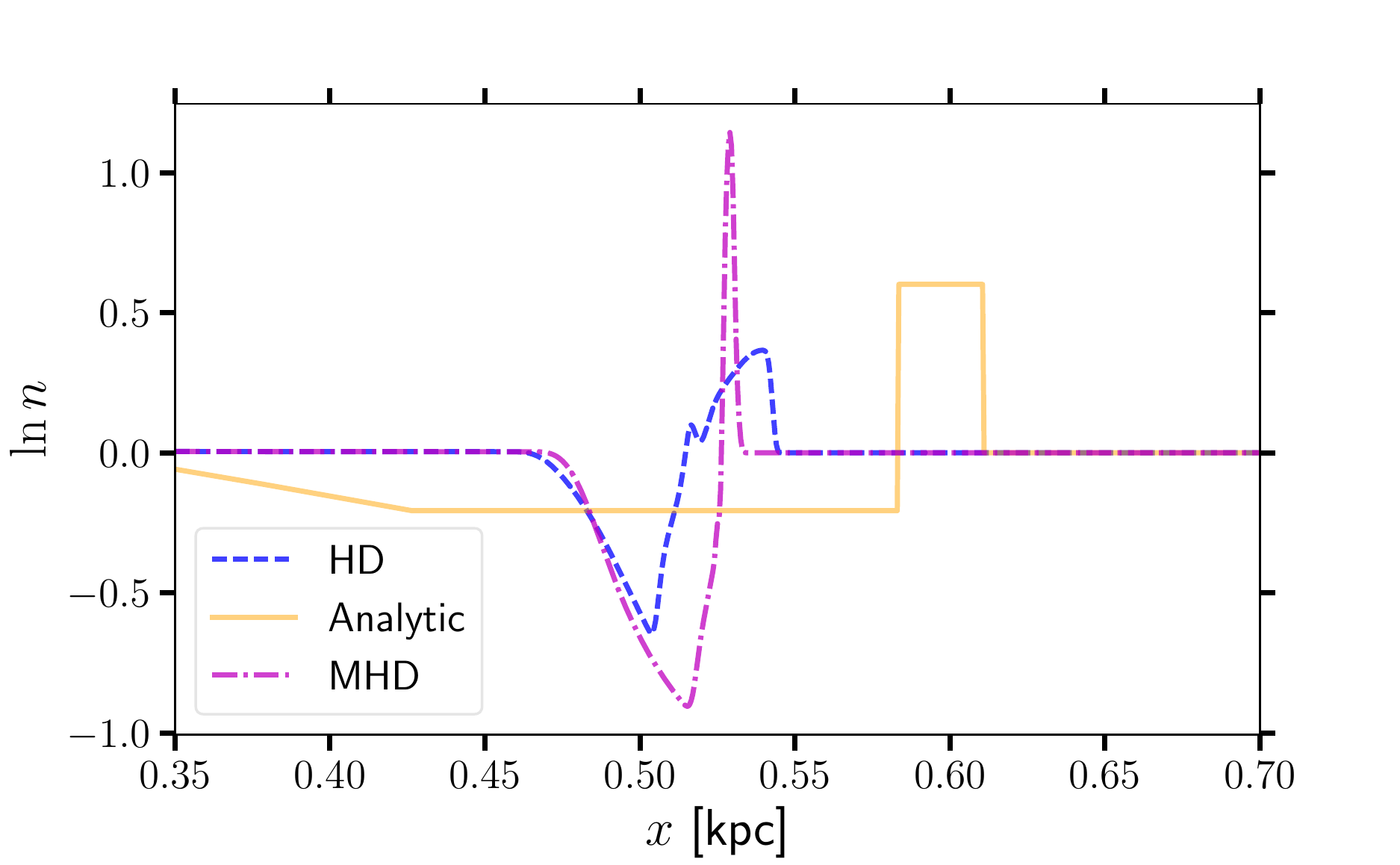}
  \caption{Comparison of \edits{the numerical non-adiabatic} HD and MHD shock\edits{-}tube \edits{solutions, alongside
           the adiabatic analytic solution.}}
  \label{fig:shock_cool_rho}
\end{figure}
%
\edits{The inclusion of cooling and heating processes has a noticeable effect on both
HD and MHD shock tubes; both shocks propagate more slowly through the ambient gas.
This is not unexpected, since these processes extract energy which would otherwise
be used up in the propagation of the shock. We note that the MHD shock propagates
faster than the HD shock, as also seen in the 3D simulations presented in this Paper.
\edits{The} gas density behind the shock tube is higher than the standard set up for both HD and MHD shock tubes with
cooling and heating.
However, as seen in the 3D simulations, the gas density is higher behind the shock for the
MHD shock tube. In addition, the shock density is lower in the MHD shock.}
\edits{This is evidence that the nonlinear interaction between the MHD and non-adiabatic effects is a significant
 factor in the divergence of the HD and MHD solutions.}

\edits{For the 1D shock-tube tests an artificial shock-dependent mass diffusion is unnecessary, but we include it
in these experiments to verify that its inclusion in SN-driven turbulence simulations does not induce excessive 
numerical diffusion.
We consider a range of mass shock diffusivity $\zeta_{\rm D}\in[0,10]f_{\rm shock}$.}
\begin{figure}
  \centering
  \includegraphics[trim=0.25cm 0.1cm 0.3cm 0.7cm, clip=true,width=0.45\linewidth]{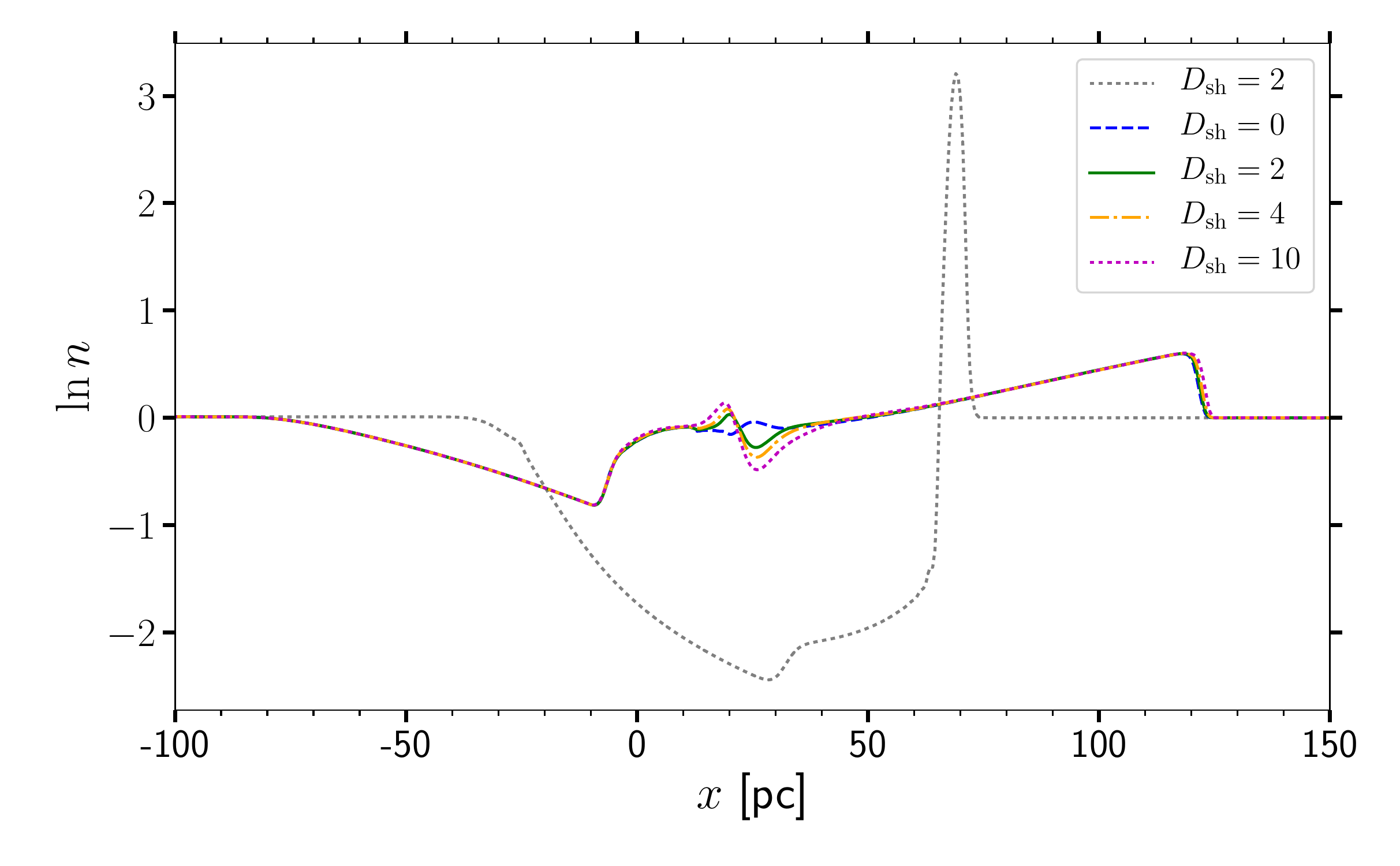}
  \caption{\edits{The} non-adiabatic MHD shock\edits{-}tube \edits{solution is contrasted with its HD solution for a 
           range of} mass diffusion rates\edits{, $\zeta_{\rm{D}}=[0,2,4,10]f_{\rm shock}$.}}
  \label{fig:mass_diff}
\end{figure}
\edits{\edits{From} Figure\,\ref{fig:mass_diff} \edits{we} show 
that \edits{adding mass diffusion, while providing numerical stability,} produces a minor quantitative difference in the 
gas density \edits{shock} profile.
The profiles are otherwise \edits{qualitatively the same and the contrast with the HD profile is clearly down to the 
presence of the magnetic effects rather than the mass diffusion}.}
\edits{\edits{A remaining concern is that the broadening of the mass profile in the MHD blast wave may arise from
      the high value of the viscosity, $\nu=\nu_0$ with $\nu_0=0.0005c_s$\,kpc\,km\,s$^{-1}$ in the 3D SN blast waves. 
      Although the estimates of microscopic viscosity in the real ISM are orders of magnitude lower, the viscosity
      in the model represents instead a turbulent viscosity.
      To model turbulence we require this dissipation scale applies above the grid scale, but sufficiently below
      the SN forcing scale to accurately capture the energy spectrum down to the smallest scales of interest in the
      model.}
      We \edits{vary} $\nu=[\nu_0,2\nu_0,4\nu_0,8\nu_0,16\nu_0]$ \edits{with
      $\zeta_{\rm D}=2f_{\rm shock}$\,kpc\,km\,s$^{-1}$ and
      $\eta=0.0008$\,kpc\,km\,s$^{-1}$ for the non-adiabatic 1D shock-tube tests.}
      This also effectively changes the magnetic Prandtl number in the MHD simulations, defined as
      $\rm{P_m}=\nu/\eta$\edits{, which yields $\rm{P_m} = [0.625,1.25,2.5,5,10]\,c_s$.}
      \edits{Given that in these solutions $c_s\gg1$\,km\,s$^{-1}$, we also have ${\rm P_m}\gg1$}}

\begin{figure}
  \centering
  \includegraphics[trim=0.25cm 0.1cm 0.3cm 0.7cm, clip=true,width=0.45\linewidth]{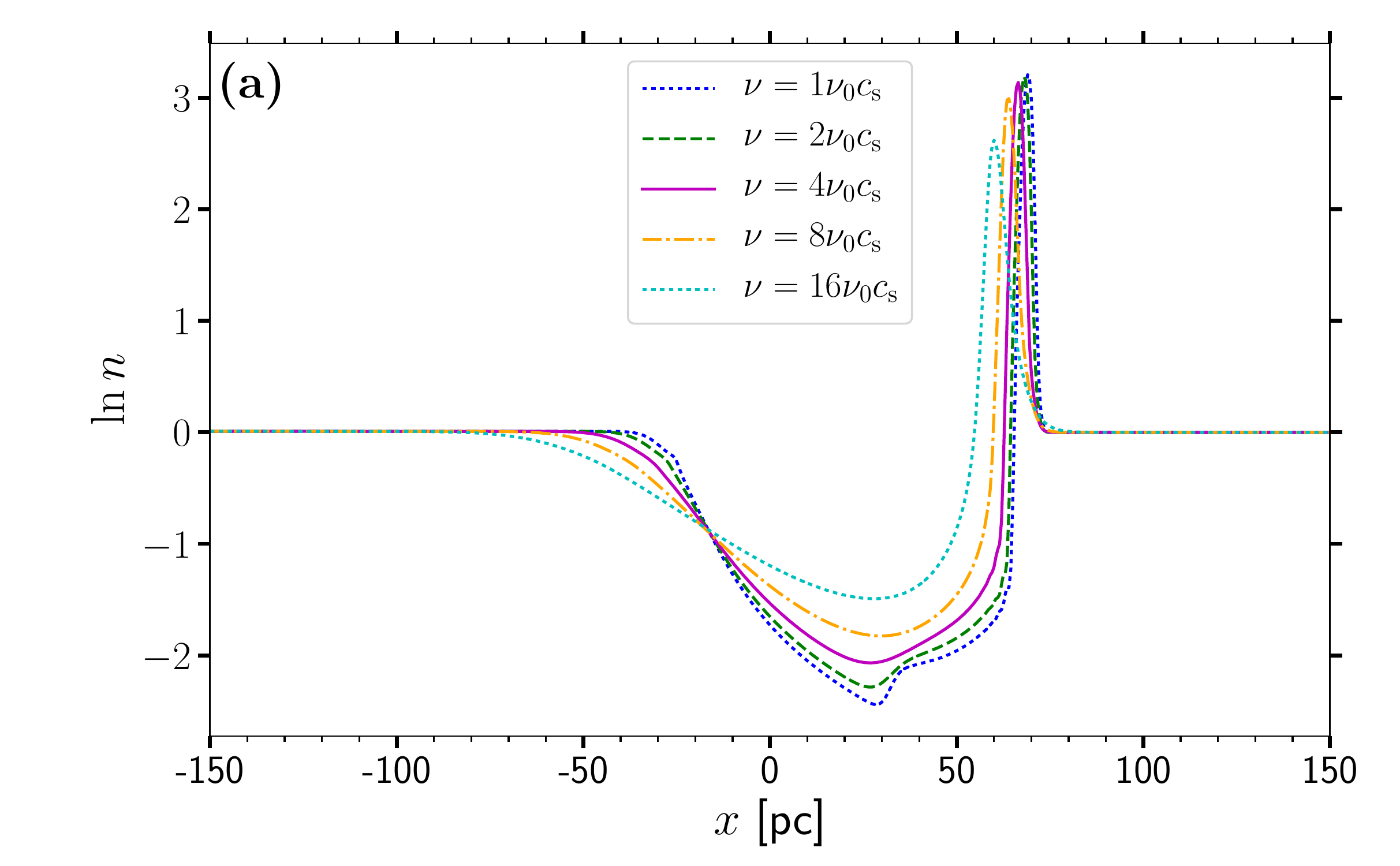}
  \includegraphics[trim=0.25cm 0.1cm 0.3cm 0.7cm, clip=true,width=0.45\linewidth]{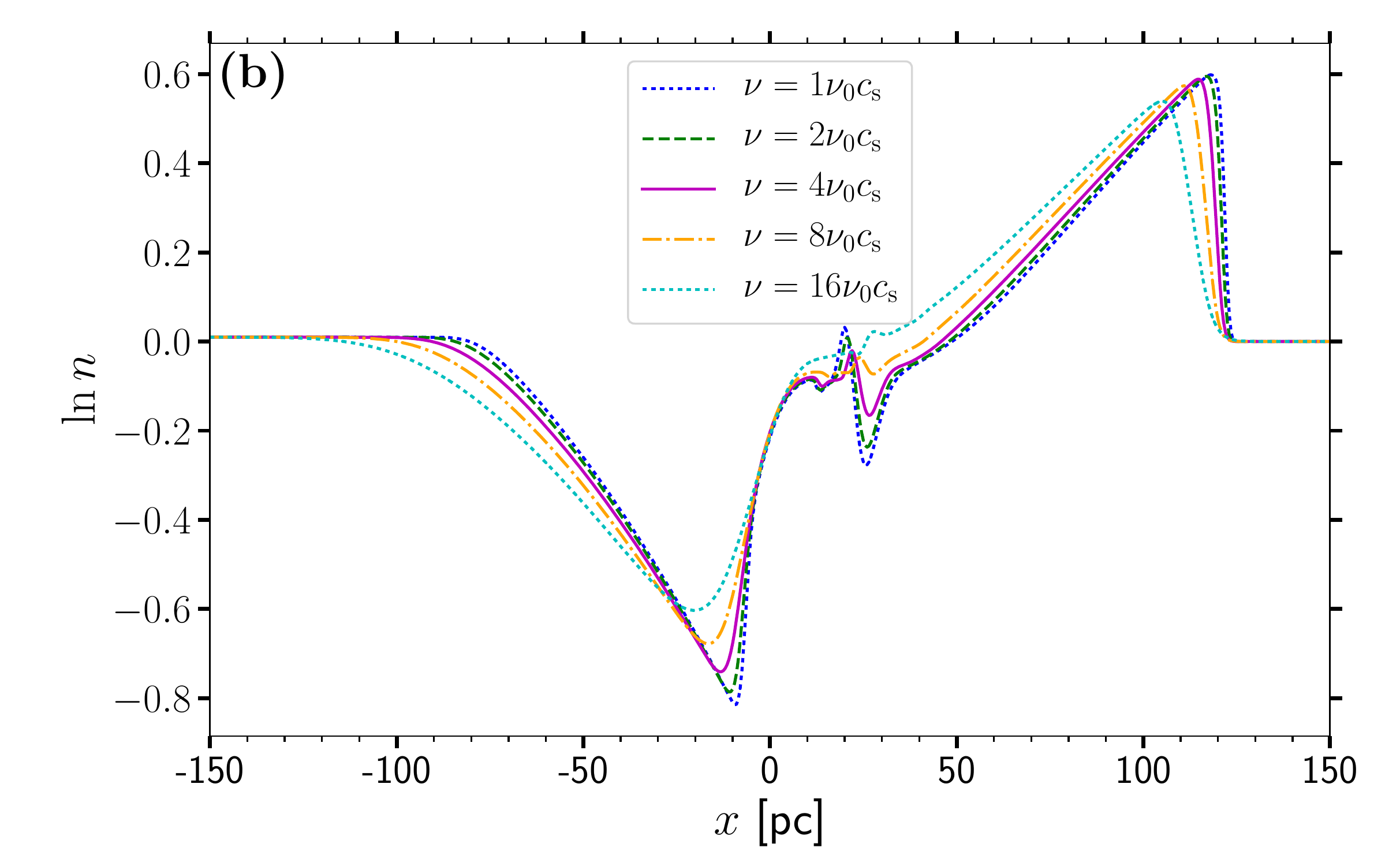}
  \caption{HD (left) and MHD (right) \edits{non-adiabatic} shock\edits{-}tube \edits{solutions for
           $\nu\in[\nu_0,2\nu_0,4\nu_0,8\nu_0,16\nu_0]$}.}
  \label{fig:visc}
\end{figure}

\edits{\edits{Some resulting density profiles are shown in Figure\,\ref{fig:visc}.
B}oth HD and MHD are affected similarly by increasing viscosity.
The shock \edits{front} does not propagate as fa\edits{st} for higher values of $\nu$ and the region behind the shock \edits{is 
smoothed}.
However both HD and MHD shocks \edits{are not affected} qualitatively as viscosity is increased.
The difference between the HD and MHD \edits{solutions cannot be explained by high diffusivity}.}

\section{Changing the ambient gas density}\label{app:amb_rho}

\edits{There is a significant change to the HD SN evolution when a moderately strong uniform magnetic field is embedded
in the ambient ISM.
We conclude that this is in part due to the alteration of the radiative cooling and UV-heating characterstics in
response to the retrograde flow of gas into the remnant.
The cooling and heating profile of the ISM is highly sensitive to the gas density and temperature.
We would like to explore the response to varying ambient ISM density in the 3D MHD SN remnant evolution in future
work, but here we perform a preliminary low budget experiment with 1D shock-tube tests of varying ambient gas density.}
\edits{
\edits{We apply ambient gas number density, }
$n_0=10^{-2},10^{-1},1,10,100$\edits{\,cm$^{-1}$} to assess whether the magnetic effects described in this Paper
may \edits{be generalised beyond 1\,cm$^{-3}$}.
The range of densities chosen reflect the range from diffuse, hot gas to dense, cold gas.
The initial entropy of the \edits{ambient gas}
is adjusted such that the \edits{ambient}
temperature remains constant \edits{with net heating at zero appropriate to} 
the initial gas density.}
\edits{Specfic gas entropy is expressed as
\[
s = c_{\rm{V}}\left[\ln T - (\gamma-1)\ln\rho\right],
\]
\edits{with $T$ and $\rho$ expressed} in dimensionless code units,
$\gamma=5/3$ is the adiabatic index, and $c_{\rm{V}}$ is the specific heat capacity.
Some initial configuration is represented by $T_0$, $\rho_0$ and,
\[
s_0 = c_{\rm{V}}\left[\ln T_0 - (\gamma-1)\ln\rho_0\right].
\]
We change the gas density such that $\rho_1=k\rho_0$, where $k>0$, but keep the
temperature constant, such that $T_0=T_1$.}
\begin{figure}
  \centering
  \includegraphics[trim=0.25cm 0.1cm 0.3cm 0.7cm, clip=true,width=0.45\linewidth]{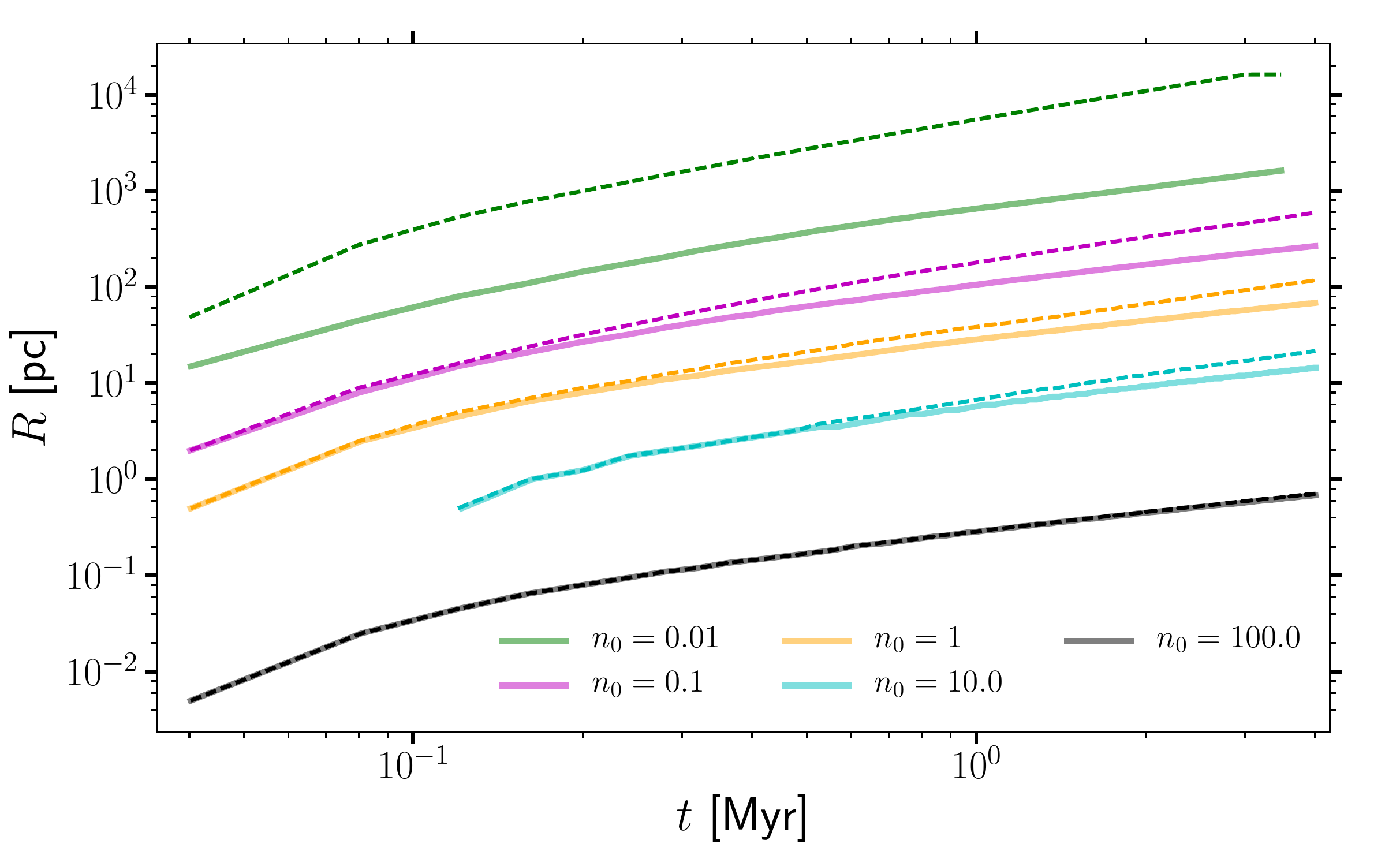}
  \caption{\edits{Shock radius for 1D non-adiabatic MHD shock-tube simulations ($B_0=5\upmu$G),
	for a range of ambient gas densities. Solid lines indicate the profiles for HD models
	at the given background density, while dashed lines of the same colour represent MHD
	models of the same background density.}}
  \label{fig:amb_rho}
\end{figure}
\edits{The two entropies, the new entropy $s_1$ can be related to the old value $s_0$: 
\begin{align}
s_1 &= s_0 + c_{\rm{V}}(\gamma-1)\left[\ln\rho_0-\ln k\rho_0\right],\nonumber\\
s_1 & = s_0 + c_{\rm{V}}(1-\gamma)\ln k.
\end{align}
\edits{We use this expression to change gas density while keeping initial temperature constant.}
In Figure\,\ref{fig:amb_rho}, we present the time profiles of distance travelled by the non-adiabatic shock (analogous to remnant shell radius) for a range of ambient gas densities. The effect of the magnetic field is measured by the difference between the profiles for HD and MHD shocks of identical ambient gas density. As in the 3D SN simulations, divergence of the MHD shock profile from the HD profile indicates the effect of magnetic field. We find that the effect of the magnetic field increases as the ambient density becomes more diffuse, due to decreasing \edits{plasma-}$\beta$. Magnetic effects are seen for all ambient densities apart from $n=100$ cm$^{-3}$. Magnetic effects appear earlier in the lower ambient gas densities.}

\section{\edits{Temperature of the ambient gas}}\label{app:amb_tt}
\edits{In our simulations, we set the ambient gas temperature $T_0=260$\,K, to ensure that
the SN explosion is set off with an ambient medium in thermal equilibrium. This particular
temperature is chosen as the radiative cooling and UV-heating balance each other, yielding
zero net heating in the ambient gas.
However, $T_0\sim10^4$\,K is a more typical temperature for the ISM, particularly for ambient
gas number density , $n_0\sim \edits{0.}1$\,cm$^{-3}$.
\edits{Using this low density would substantially increase the numerical
domain size and computational expense.}
We\edits{, therefore,} run a 3D SN simulation identical to the MHD remnant simulation with $B_0=5\upmu$G 
\edits{and $n_0=1$\,cm$^{-3}$,} but with $T_0=10^4$\,K.}

\begin{figure}
  \centering
  \includegraphics[trim=0.25cm 0.1cm 0.3cm 0.7cm, clip=true,width=0.45\linewidth]{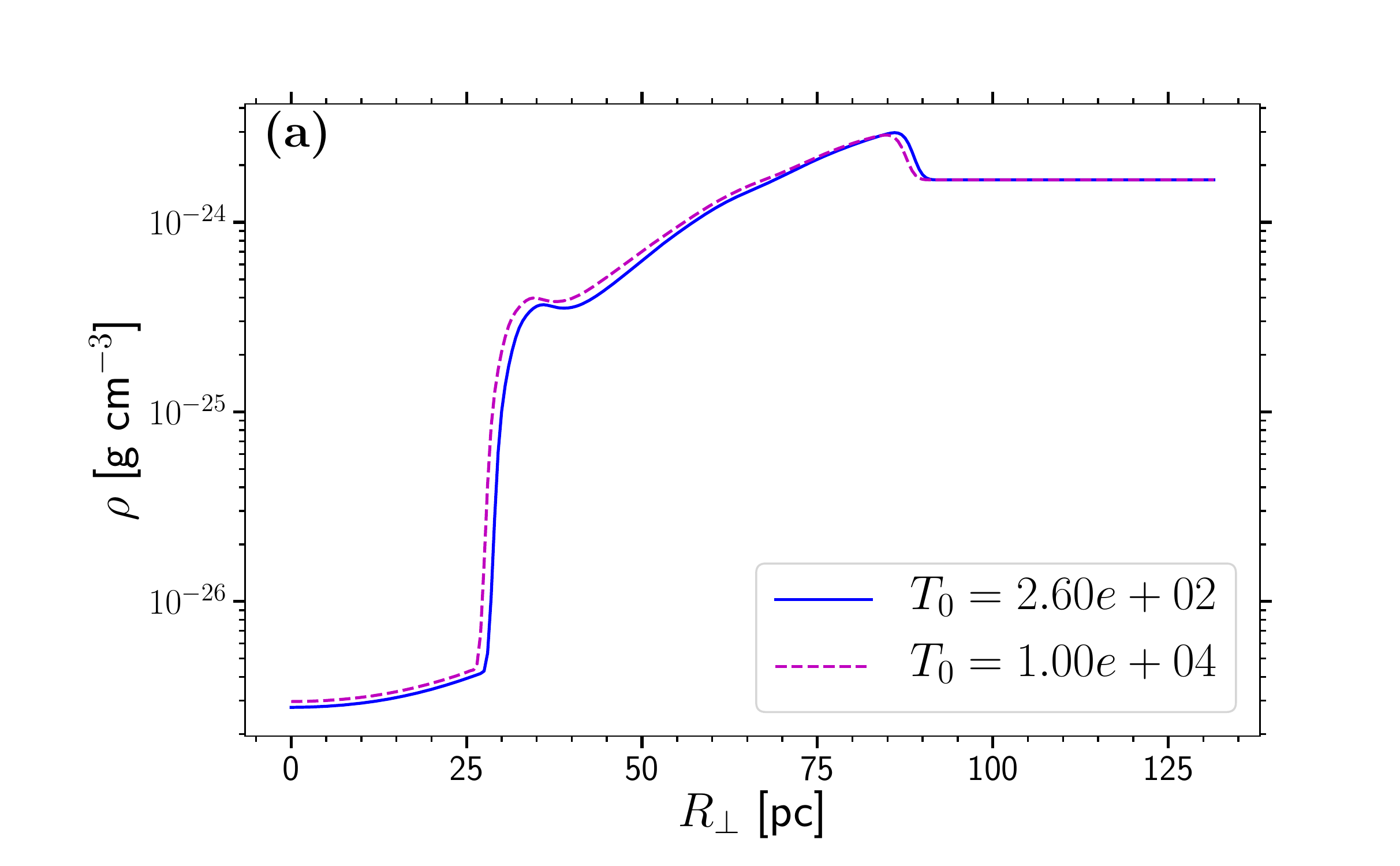}
  \includegraphics[trim=0.25cm 0.1cm 0.3cm 0.7cm, clip=true,width=0.45\linewidth]{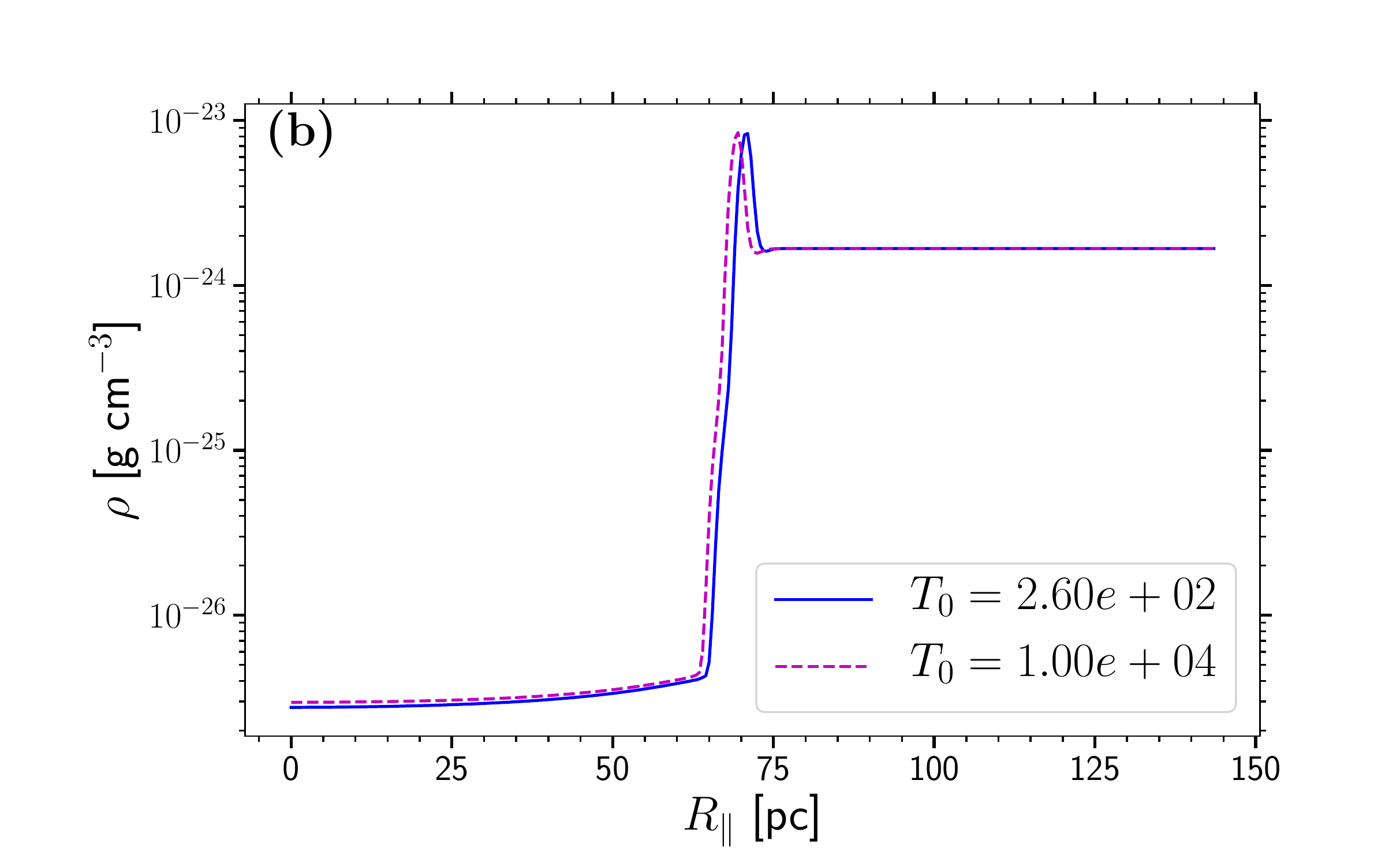}
  \includegraphics[trim=0.25cm 0.1cm 0.3cm 0.7cm, clip=true,width=0.45\linewidth]{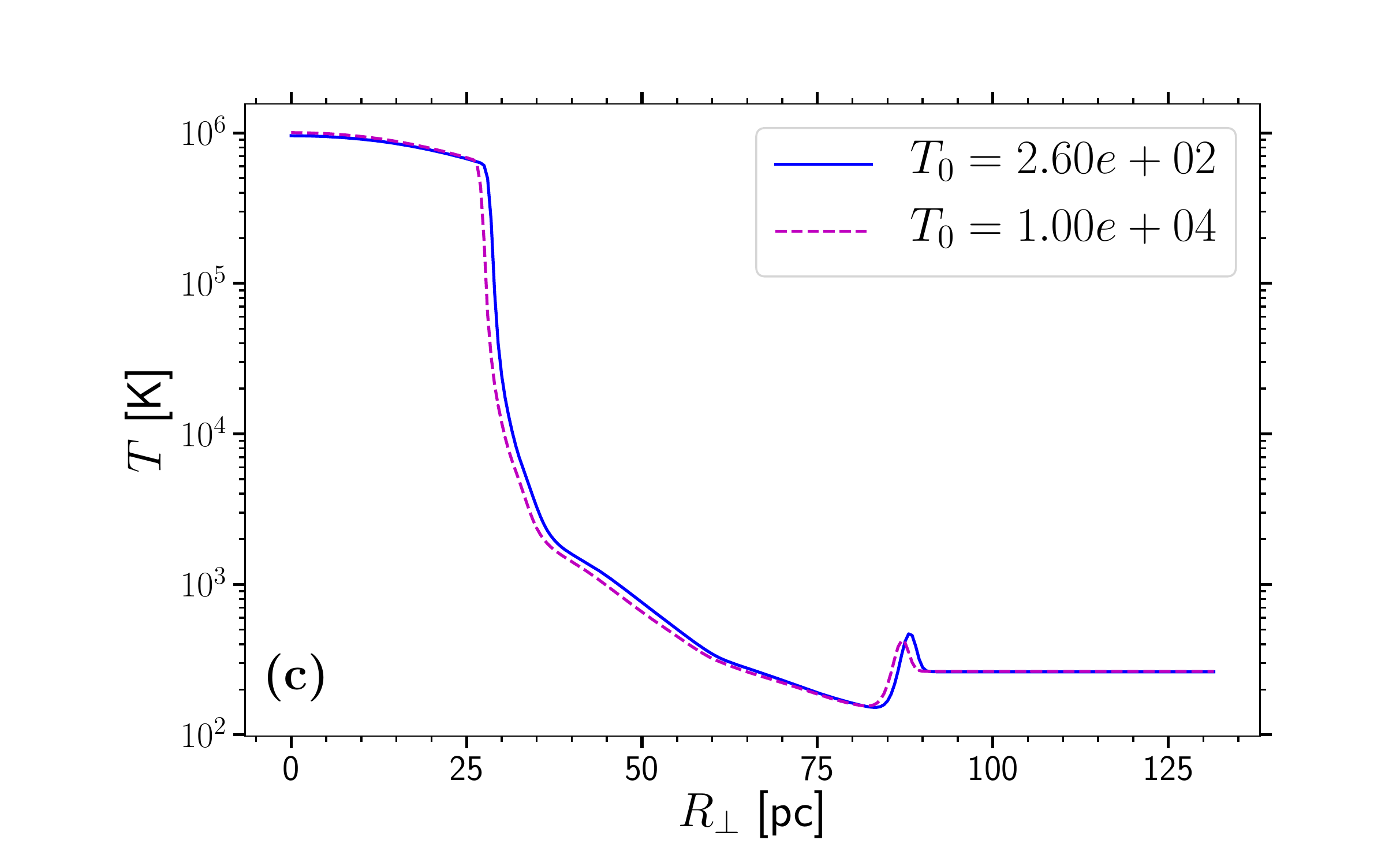}
  \includegraphics[trim=0.25cm 0.1cm 0.3cm 0.7cm, clip=true,width=0.45\linewidth]{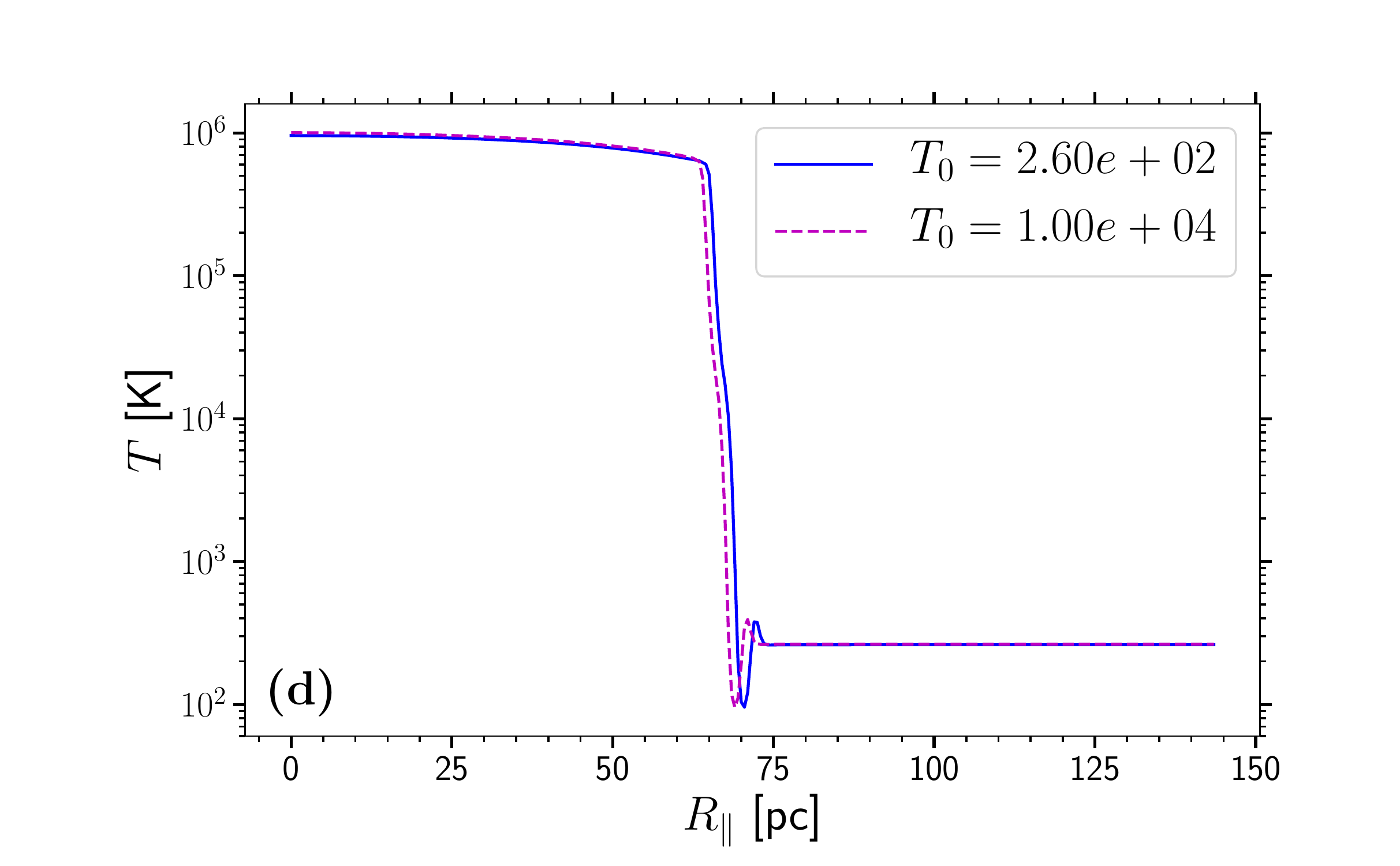}
  \caption{Radial profiles of gas temperature (top row) and density (bottom row)
	perpendicular (left column) and parallel (right column) to the magnetic
	field. The radial profiles shown here are taken at $t=2$\,Myr.}
  \label{fig:comp_prof}
\end{figure}

\edits{Figure\,\ref{fig:comp_prof} shows that the radial profiles of gas density and temperature are qualitatively identical and quantitatively very similar at 2\,Myr. The re-distribution of mass
from the shell to the interior of the remnant occurs in both models, forming the inter-shock
region between the magnetically confined diffuse core and the shell, in the direction perpendicular
to the magnetic field.}

\begin{figure}
  \centering
  \includegraphics[trim=0.25cm 0.1cm 0.3cm 0.7cm, clip=true,width=0.93\linewidth]{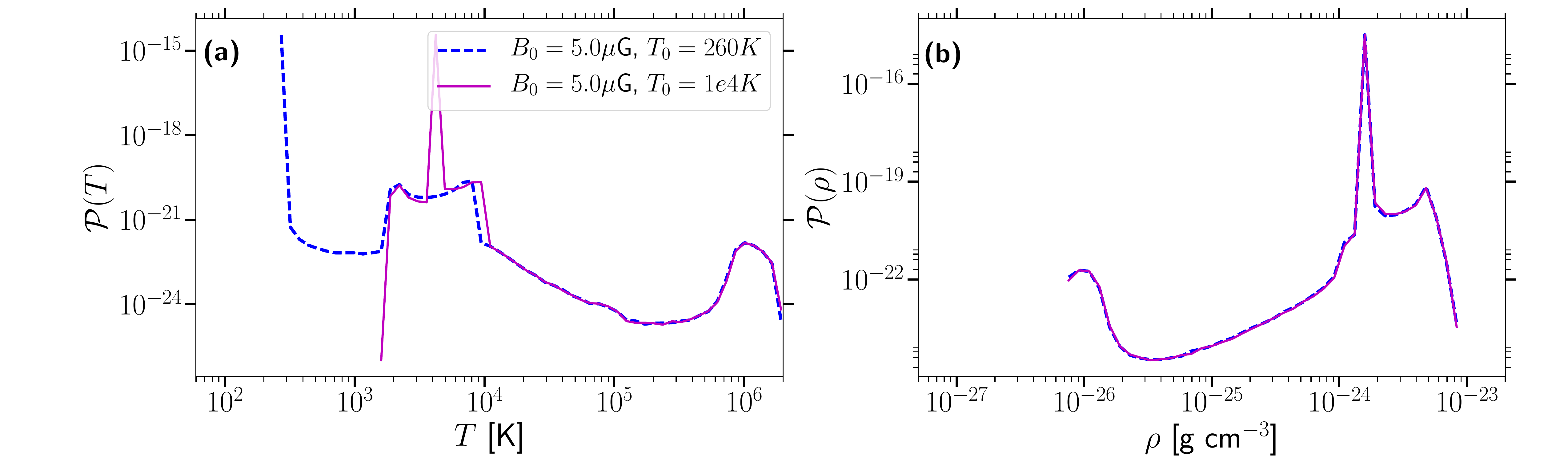}
  \includegraphics[trim=0.25cm 0.1cm 0.3cm 0.7cm, clip=true,width=0.93\linewidth]{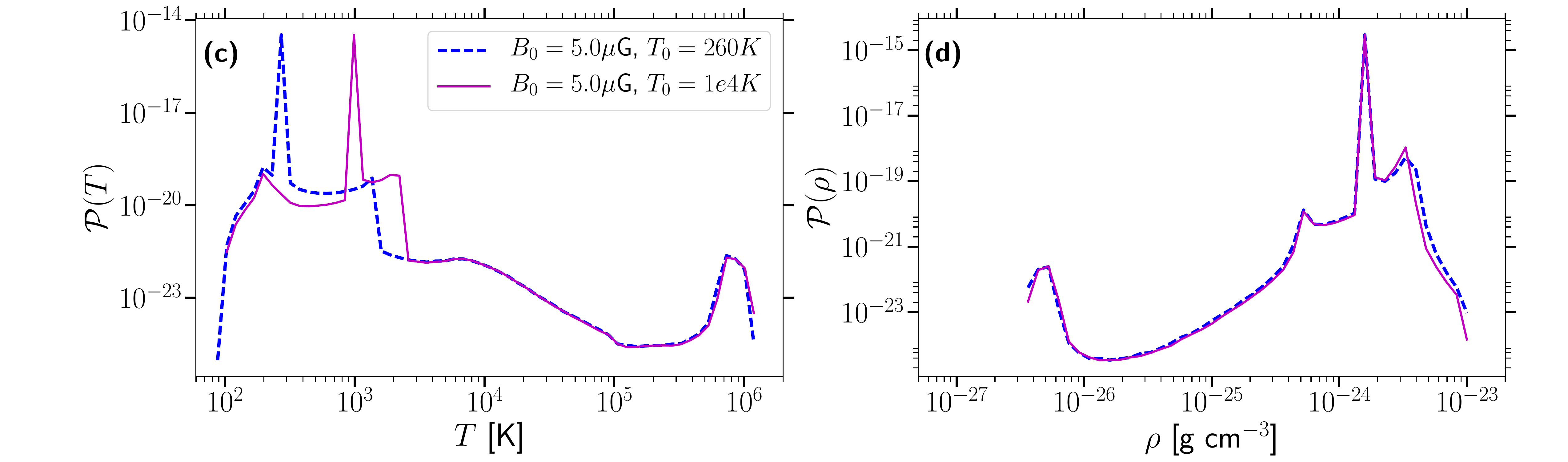}
  \includegraphics[trim=0.25cm 0.1cm 0.3cm 0.7cm, clip=true,width=0.93\linewidth]{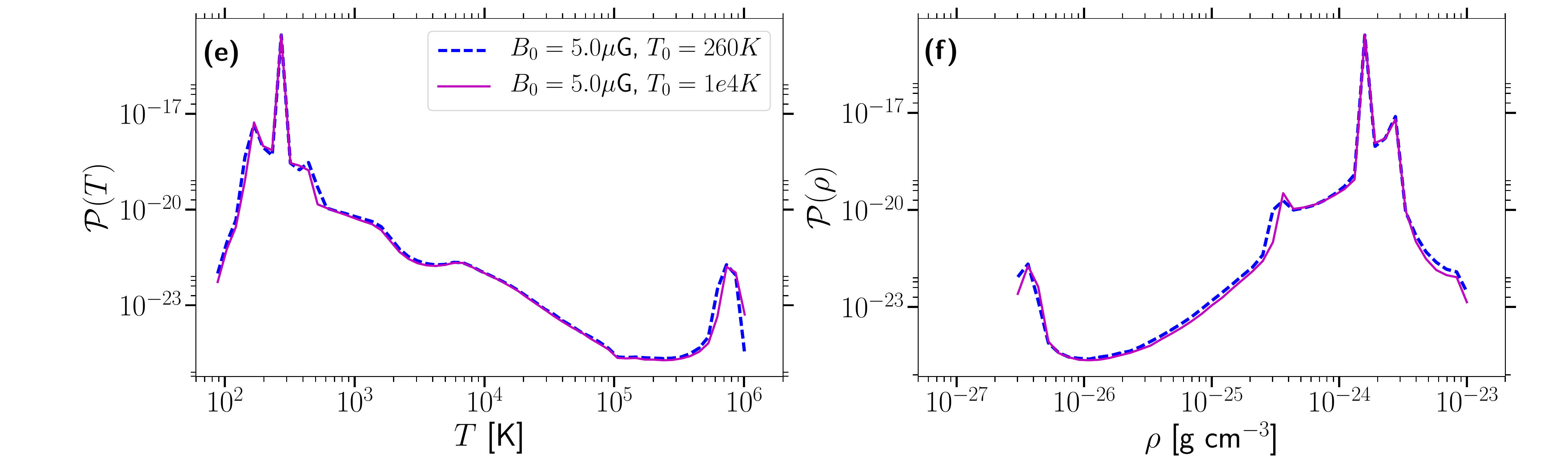}
  \caption{PDFs of gas temperature (left column) and gas density (right column) for
	the MHD model ($B_0=5\upmu$G) with initial ambient gas temperatures of
	$260$\,K and $10^4$\,K,
	at 400\,kyr (top row), 1\,Myr (middle row), and 2\,Myr (bottom row)}
  \label{fig:comp_pdfs}
\end{figure}

\edits{Moreover, the PDFs of gas temperature and density provided in Figure\,\ref{fig:comp_pdfs},
which are given at 400\,kyr, 1\,Myr and 2\,Myr, show that the temperature distributions converge
despite the different initial temperatures. The hottest gas is identical from an early stage,
as it is injected by the SN remnant, irrespective of the initial ambient gas temperature.
In the cooler range, the modal temperature of the gas, for the $10^4$\,K model, gradually
converges to the $260$\,K model, for which the ambient gas is in thermal equilibrium from the 
beginning of the simulation.}
\edits{We see that the lower peak in the temperature PDF, corresponding to the ambient ISM,
cools over time, reaching thermal equilibrium peak by 2\,Myr.
Evidently, the difference to the thermal pressure gradient form the higher temperature
is negligible.}
 
\begin{figure}
  \centering
  \includegraphics[trim=0.25cm 0.1cm 0.3cm 0.7cm, clip=true,width=0.45\linewidth]{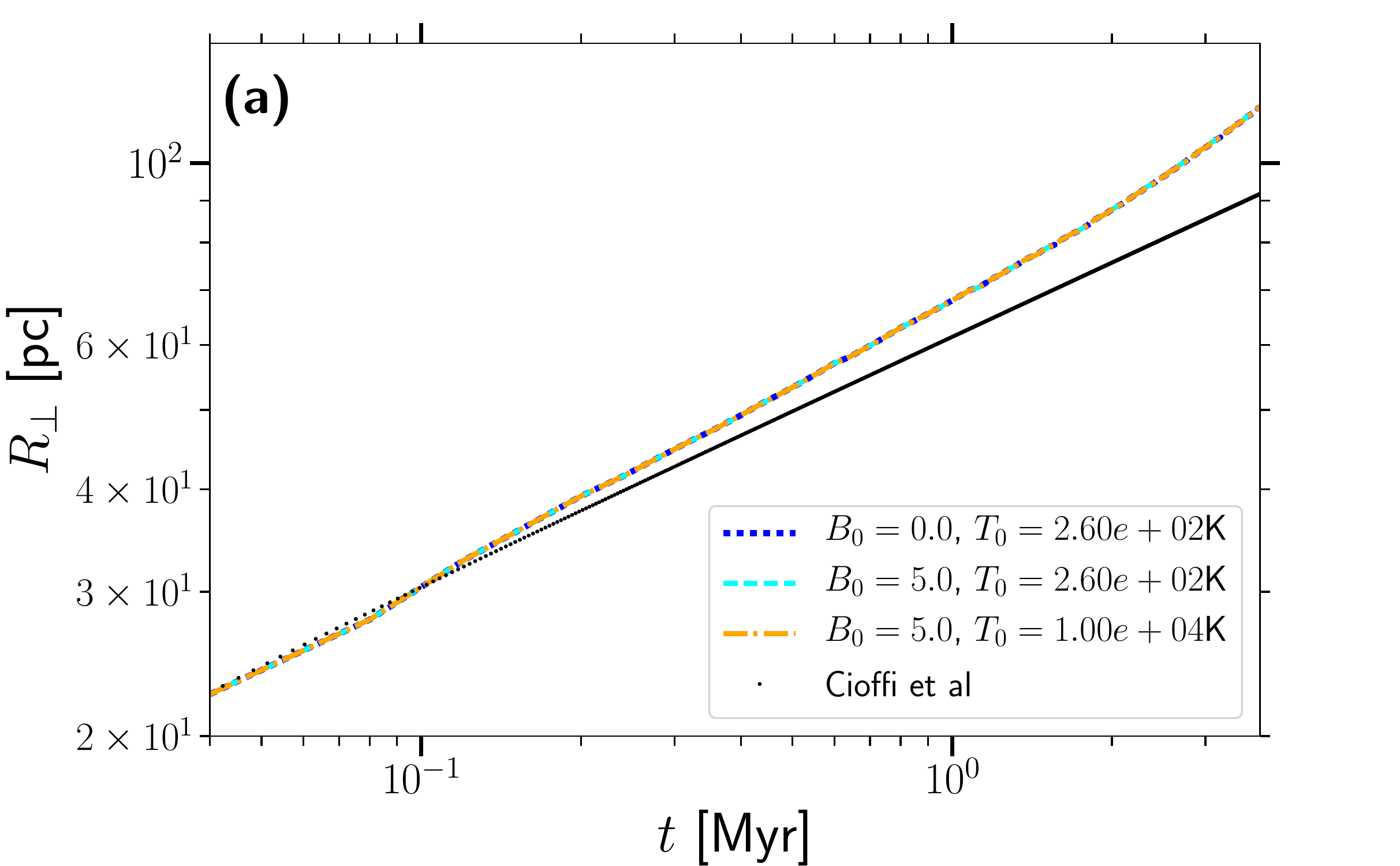}
  \includegraphics[trim=0.25cm 0.1cm 0.3cm 0.7cm, clip=true,width=0.45\linewidth]{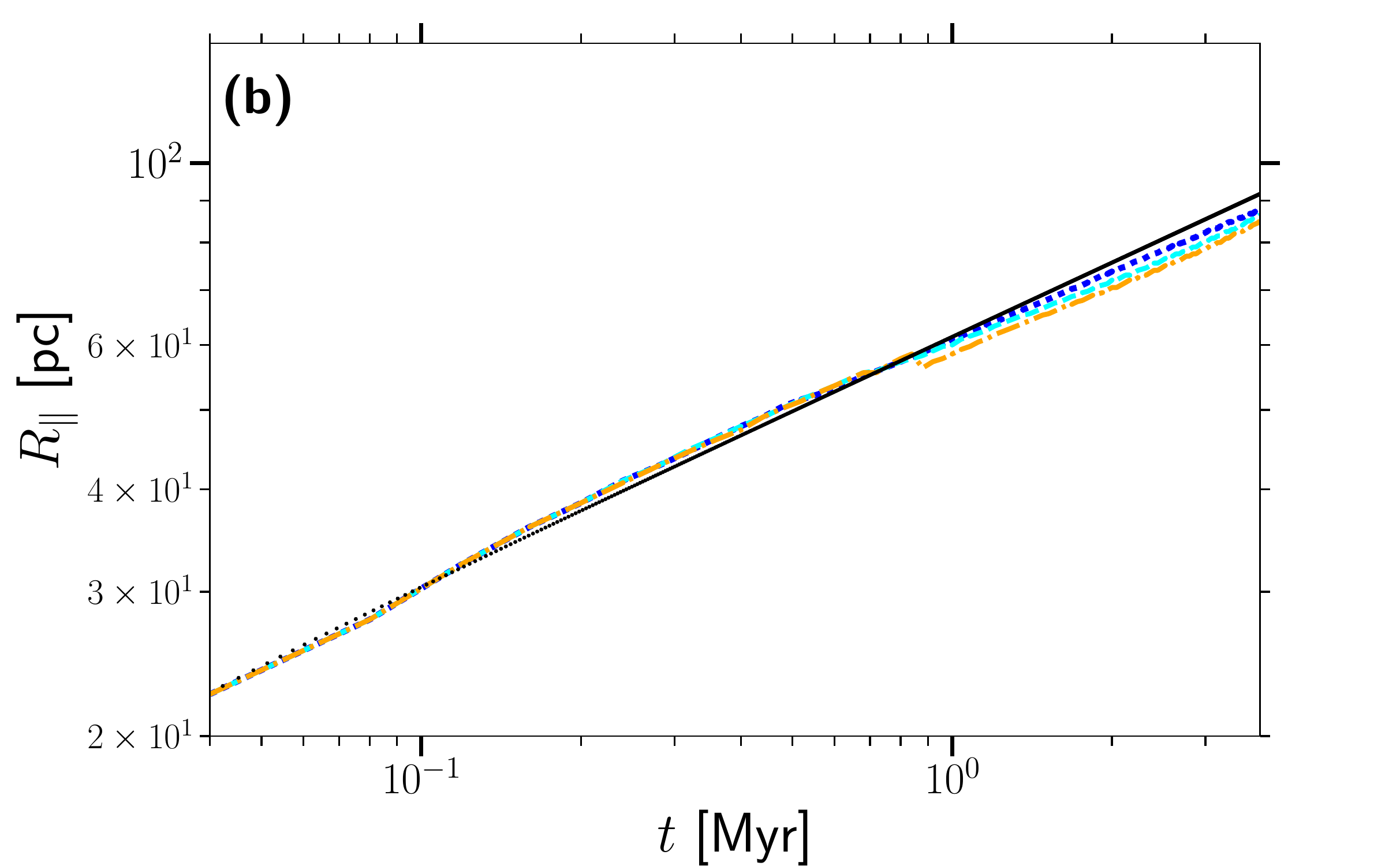}
  \caption{Time profile of the remnant radius {\bf{(a)}}\,perpendicular 
	and {\bf{(b)}}\,parallel to the magnetic field. We compare
	the HD remnant, MHD remnant with both $T_0=260$\,K and $T_0=10^4$\,K,
	with the \citet{Cioffi88} analytical solution.}
  \label{fig:comp_rads}
\end{figure}

\edits{Figure\,\ref{fig:comp_rads} shows the evolution of remnant radius perpendicular and parallel to the
magnetic field. Panel (b) shows that the HD model and both MHD models agree with the \citet{Cioffi88}
analytical solution. The radius of the $10^4$\,K MHD model is marginally (within 2\,pc) less than the
$260$\,K model. Panel (a) shows that both MHD models have an identical time profile for the radius
perpendicular to the magnetic field.}

\end{document}